\documentclass[fleqn, usenatbib, useAMS]{mnras}
\usepackage{newtxtext,newtxmath}
\usepackage{graphicx}
\usepackage{amsmath}
\usepackage{paralist,xcolor,xspace}
\usepackage{times}
\usepackage{comment}
\usepackage{orcidlink}
\usepackage[super]{nth}
\usepackage[T1]{fontenc}
\usepackage{bold-extra}

\newcommand{\codename}[1]{\textcolor{black}{\sc #1}\xspace}        
\newcommand{\simulationname}[1]{\textcolor{black}{\sc #1}\xspace}  
\newcommand{\sphflavour}[1]{\textcolor{black}{\sc #1}\xspace}      
\newcommand{\libraryname}[1]{\textcolor{black}{\tt #1}\xspace}     
\newcommand{\techjargon}[1]{\textcolor{black}{\tt #1}\xspace}      
\newcommand{\computername}[1]{\textcolor{black}{\tt #1}\xspace}    

\newcommand{\swift}{\codename{Swift}}
\newcommand{\gadget}{\codename{Gadget}}
\newcommand{\eagle}{\simulationname{Eagle}}
\newcommand{\flamingo}{\simulationname{Flamingo}}
\newcommand{\gear}{\codename{Gear}}
\newcommand{\sphenix}{\sphflavour{Sphenix}}
\newcommand{\anarchy}{\sphflavour{Anarchy}}
\newcommand{\phantomSPH}{\sphflavour{Phantom}}
\newcommand{\gasoline}{\codename{Gasoline}}
\newcommand{\swiftsimio}{\libraryname{SWIFTsimIO}}
\newcommand{\velociraptor}{\libraryname{VELOCIraptor}}
\newcommand{\woma}{\libraryname{WoMa}}
\newcommand{\seagen}{\libraryname{SEAGen}}
\newcommand{\nbody}{$N$-body\xspace}
\newcommand{\Lag}{\mathcal{L}}
\newcommand{\monofonic}{\codename{MonofonIC}}
\newcommand{\healpix}{\libraryname{HEALPix}}

\newcommand{\metis}{\libraryname{Metis}}
\newcommand{\parmetis}{\libraryname{ParMetis}}
\newcommand{\MPI}{\libraryname{MPI}}

\DeclareRobustCommand{\VAN}[3]{#2}

\title[The modern astrophysics code {\itshape \scshape Swift}]{{\bfseries \scshape Swift}: A modern highly-parallel gravity and smoothed particle hydrodynamics solver for astrophysical and cosmological applications}
\author[M. Schaller et al.]{
Matthieu Schaller\,\textsuperscript{\orcidlink{0000-0002-2395-4902}}$^{\,1,2}$\thanks{E-mail: \url{mschaller@lorentz.leidenuniv.nl}},
Josh Borrow\,\textsuperscript{\orcidlink{0000-0002-1327-1921}}$^{\,3,4,5}$,
Peter W. Draper\,\textsuperscript{\orcidlink{0000-0002-7204-9802}}$^{\,5}$,
Mladen Ivkovic\,\textsuperscript{\orcidlink{0000-0002-3539-3831}}$^{\,6,7,8}$,
Stuart McAlpine\,\textsuperscript{\orcidlink{0000-0002-8286-7809}}$^{\,9,10}$,
\newauthor
~Bert Vandenbroucke\,\textsuperscript{\orcidlink{0000-0001-7241-1704}}$^{\,2,11}$,
Yannick Bah\'{e}\,\textsuperscript{\orcidlink{0000-0002-3196-5126}}$^{\,5,2,6}$,
Evgenii Chaikin\,\textsuperscript{\orcidlink{0000-0003-2047-3684}}$^{\,2}$,
Aidan B. G. Chalk$^{\,12}$,
\newauthor
~Tsang Keung Chan\,\textsuperscript{\orcidlink{0000-0003-2544-054X}}$^{\,13,14,5}$,
Camila Correa\,\textsuperscript{\orcidlink{0000-0002-5830-8070}}$^{\,15,16}$,
Marcel van Daalen\,\textsuperscript{\orcidlink{0000-0002-8801-4911}}$^{\,2}$,
Willem Elbers\,\textsuperscript{\orcidlink{0000-0002-2207-6108}}$^{\,5}$,
Pedro Gonnet\,\textsuperscript{\orcidlink{0000-0003-4509-7291}}$^{\,17}$,
\newauthor
~Lo\"{i}c Hausammann\,\textsuperscript{\orcidlink{0000-0002-4687-4948}}$^{\,6,18}$,
John Helly\,\textsuperscript{\orcidlink{0000-0002-0647-4755}}$^{\,5}$,
Filip Hu\v{s}ko\,\textsuperscript{\orcidlink{0000-0002-1510-1731}}$^{\,5}$,
Jacob A. Kegerreis\,\textsuperscript{\orcidlink{0000-0001-5383-236X}}$^{\,19,5}$,
Folkert S. J. Nobels\,\textsuperscript{\orcidlink{0000-0002-0117-7495}}$^{\,2}$,
\newauthor
~Sylvia Ploeckinger\,\textsuperscript{\orcidlink{0000-0002-1965-1650}}$^{\,1,20}$,
Yves Revaz\,\textsuperscript{\orcidlink{0000-0002-6227-0108}}$^{\,6}$,
William J. Roper\,\textsuperscript{\orcidlink{0000-0002-3257-8806}}$^{\,21}$,
Sergio Ruiz-Bonilla\,\textsuperscript{\orcidlink{0000-0003-0925-9804}}$^{\,5}$,
\newauthor
~Thomas D. Sandnes\,\textsuperscript{\orcidlink{0000-0002-4630-1840}}$^{\,5}$,
Yolan Uyttenhove\,\textsuperscript{\orcidlink{0000-0002-0124-618X}}$^{\,11}$,
James S. Willis$^{\,22}$, and
Zhen Xiang\,\textsuperscript{\orcidlink{0009-0004-5467-872X}}$^{\,1,23,24}$

\\ \small ~\\
\emph{\normalsize \textit{Author affiliations are listed at the end of the paper}}
\vspace{-0.2cm}
}

\begin{document}  

\pagerange{\pageref{firstpage}--\pageref{lastpage}} 

\pubyear{2024}
\date{Accepted 2024 March 28. Received 2024 March 27; in original form 2023 May 22}

\maketitle

\label{firstpage}

\begin{abstract}
Numerical simulations have become one of the key tools used by theorists in all
the fields of astrophysics and cosmology. The development of modern tools that
target the largest existing computing systems and exploit state-of-the-art
numerical methods and algorithms is thus crucial. In this paper, we introduce
the fully open-source highly-parallel, versatile, and modular coupled
hydrodynamics, gravity, cosmology, and galaxy-formation code \swift. The
software package exploits hybrid shared- and distributed-memory task-based
parallelism, asynchronous communications, and domain-decomposition algorithms
based on balancing the workload, rather than the data, to efficiently exploit
modern high-performance computing cluster architectures. Gravity is solved for
using a fast-multipole-method, optionally coupled to a particle mesh solver in
Fourier space to handle periodic volumes. For gas evolution, multiple modern
flavours of Smoothed Particle Hydrodynamics are implemented. \swift also evolves
neutrinos using a state-of-the-art particle-based method. Two complementary
networks of sub-grid models for galaxy formation as well as extensions to
simulate planetary physics are also released as part of the code.  An extensive
set of output options, including snapshots, light-cones, power spectra, and a
coupling to structure finders are also included. We describe the overall code
architecture, summarise the consistency and accuracy tests that were performed,
and demonstrate the excellent weak-scaling performance of the code using a
representative cosmological hydrodynamical problem with $\approx$$300$ billion
particles. The code is released to the community alongside extensive
documentation for both users and developers, a large selection of example test
problems, and a suite of tools to aid in the analysis of large simulations run
with \swift.
\end{abstract}


\begin{keywords}
\vspace{-0.1cm}
software: simulations, methods: numerical, software: public release
\vspace{-0.3cm}
\end{keywords}

\section{Introduction}
\label{sec:introduction}

Over the last four decades, numerical simulations have imposed themselves as the
key tool of theoretical astrophysics. By allowing the study of the highly
non-linear regime of a model, or by allowing \emph{in-silico} experiments of
objects inaccessible to laboratories, simulations are essential to the
interpretation of data in the era of precision astrophysics and cosmology. This
is particularly true in the field of galaxy evolution and non-linear structure
formation, where the requirements of modern surveys are such that only large
dedicated campaigns of numerical simulations can reach the necessary precision
and accuracy targets. Hence, it is no surprise that this field has seen a recent
explosion in numerical tools, models, analysis methods and predictions
\citep[for reviews, see][]{Somerville2015, Naab2017, Vogelsberger2020,
  Angulo2022, Crain2023}.\\

Meeting this growing demand and complexity of numerical simulations requires
increasingly efficient and robust tools to perform such calculations. For
instance, these softwares involve more and more coupled differential equations
to approximate, themselves coupled to increasingly complex networks of sub-grid
models. At the same time, the evolution of computer architectures towards
massively parallel systems further complicates the software development
task. The details of the machine used, as well as an intimate knowledge of
parallelisation libraries, are often required to achieve anywhere near optimal
on these the systems. This, however, often puts an additional burden on
scientists attempting to make small alterations to the models they run and is
often a barrier to the wider adoption of software packages.  Nevertheless, the
significant ecological impact of large astrophysical simulations
\citep{Stevens2020, SPZ2020} make it imperative to address these technical
challenges.

Jointly, all these needs and sometimes orthogonal requirements make constructing
such numerical software packages a daunting task. For these reasons, developing
numerical software packages that are both efficient and sufficiently flexible
has now become a task undertaken by large teams of contributors with mixed
expertise, such as our own. This, in turn, implies that better code development
practices need to be adopted to allow for collaborative work on large code
bases.

Despite all this, the community has seen the arrival of a number of simulation
software packages that rise to these challenges, many of which have also been
released publicly. This recent trend, guided by open-science principles, is an
important development allowing more scientists to run their own simulations,
adapt them to their needs, and modify the code base to solve new problems. The
public release of software is also an important step towards the reproducibility
of results. Whilst some packages only offer the core solver freely to the
community, some other collaborations have made the choice to fully release all
their developments; we follow this latter choice here. This is an essential step
that allows for more comparisons between models (as well as between models and
data) to be performed and to help understand the advantages and shortcomings of
the various methods used. The characterisation and inclusion of uncertainty on
model predictions, especially in the field of non-linear structure formation, is
now becoming common practice \citep[for examples targeted to the needs of large
  cosmology surveys see][]{Heitmann2008, Schneider2016, Grove2021}. \\

\begin{figure*}
\includegraphics[width=1.95\columnwidth]{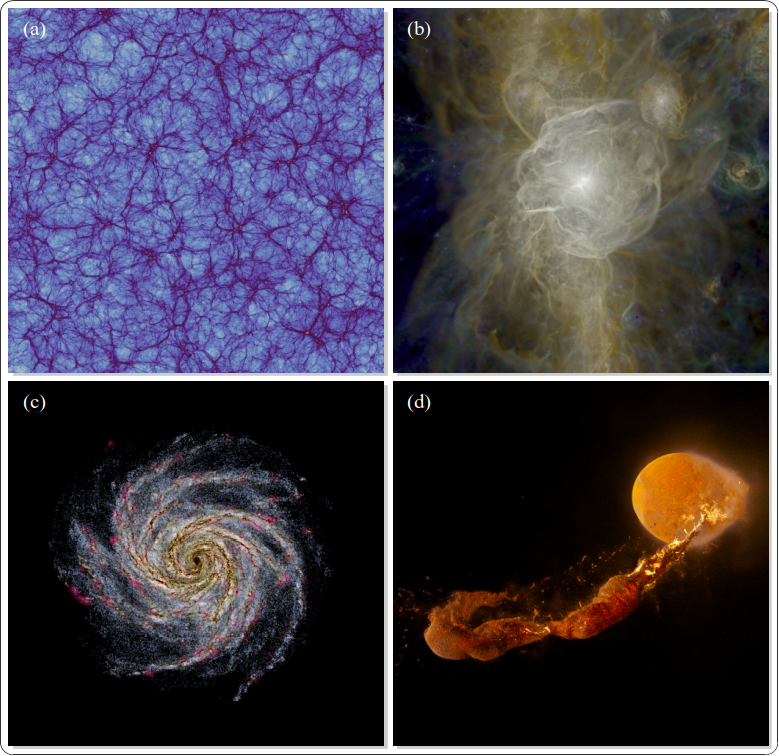}
\vspace{-0.05cm}
\caption{ A selection of simulation results obtained with the \swift code,
  illustrating the huge range of problems that have already been targeted and
  the flexibility of the solver. The panels show: \textit{(a)} a projection of
  the large-scale distribution of dark matter from a $10~{\rm Mpc}/h$ slice of
  the $(500~{\rm Mpc}/h)^3$ benchmark simulation of
  \citet[\S\,\ref{ssec:cosmology:validation}]{Schneider2016}; \textit{(b)} the
  temperature of the gas weighted by its velocity dispersion in a zoom-in
  simulation of a galaxy cluster using the \swift-\eagle galaxy formation model
  (\S\,\ref{ssec:eagle}) extracted from the runs of \citet{Altamura2022};
  \textit{(c)} an idealised isolated galaxy from the \emph{Agora}-suite
  \citep{Kim2016} simulated using the \gear model (\S\,\ref{ssec:gear}) rendered
  using \textsc{pNbody} \citep{pNbody}; and \textit{(d)} a snapshot extracted
  from a Moon-forming giant impact simulation of \citet{Kegerreis2022} using the
  planetary physics extension of the code (\S\,\ref{ssec:planetary}) and
  rendered using the \textsc{Houdini} software.}
\label{fig:publicity}
\end{figure*}

In this paper, we introduce the fully open-source code
\swift\footnote{\textbf{S}PH \textbf{W}ith \textbf{I}nter-dependent
\textbf{F}ine-grained \textbf{T}asking} designed to solve the coupled equations
of gravity and hydrodynamics together with multiple networks of extensions
specific to various sub-fields of astrophysics. The primary applications of the
code are the evolution of cosmic large-scale structure, cluster and galaxy
formation, and planetary physics. A selection of results obtained with the code
is displayed in Fig.~\ref{fig:publicity}.

\swift was designed to be able to run the largest numerical problems of interest
to the large-scale structure, cosmology \& galaxy formation communities by
exploiting modern algorithms and parallelisation techniques to make efficient
use of both existing and the latest CPU architectures. The scalability of the
code was the core goal, alongside the flexibility to easily alter the physics
modules. Our effort is, of course, not unique and there is now a variety of
codes exploiting many different numerical algorithms and targeted at different
problems in the ever-growing field of structure formation and galaxy evolution.
Examples in regular use by the community include \codename{Art}~\citep{ART},
\codename{Falcon}~\citep{Dehnen2000}, \codename{Flash}~\citep{flash},
\codename{Ramses}~\citep{Ramses}, \codename{Gadget-2}~\citep{Springel2005},
\codename{Arepo}~\citep{Springel2010}, \codename{Greem}~\citep{Greem},
\codename{Pluto}~\citep{pluto},
\codename{CubeP$^3$M}~\citep{CUBEPM}, \codename{2HOT}~\citep{2HOT},
\codename{Enzo}~\citep{ENZO}, \codename{Nyx} \citep{Nyx},
\codename{Changa}~\citep{CHANGA}, \codename{Gevolution}~\citep{gevolution},
\codename{HACC}~\citep{HACC}, \codename{Gasoline-2}~\citep{Wadsley2017},
\codename{Pkdgrav-3}~\citep{pkdGrav3}, \codename{Phantom}~\citep{Price2018},
\codename{Athena++}~\citep{athena}, \codename{Abacus}~\citep{Abacus}, and
\codename{Gadget-4}~\citep{Springel2021} as well as many extensions and
variations based on these solvers. They exploit a wide variety of numerical
methods and are designed to target a broad range of astrophysics, galaxy
formation, and cosmology problems.

Besides exploiting modern parallelisation concepts, \swift makes use of
state-of-the-art implementations of the key numerical methods. The gravity
solver relies on the algorithmically-ideal fast-multipole method \citep[see
  e.g.][]{Greengard1987, Cheng1999, Dehnen2014} and is optionally coupled to a
particle-mesh method using the Fourier-space representation of the gravity
equations to model periodic boundary conditions (See \citet{Springel2021} for a
detailed discussion of the advantages of this coupling over a pure tree
approach). The hydrodynamics solver is based on the Smoothed Particle
Hydrodynamics (SPH) method \citep[see e.g.][]{Price2012, SpringelSPHreview} with
multiple flavours from the literature implemented as well as our own version
\citep[\sphenix;][]{Borrow2022}. The code is also being extended towards other
unstructured hydrodynamics methods (such as moving mesh \citep[see
  e.g.][]{Springel2010, shadowfax}, renormalised mesh-free techniques or SPH-ALE
\citep[see e.g.][]{Hopkins2015}), which will be released in the future. For
cosmological applications, \swift was extended to use the particle-based
``delta-f'' method of \citet{Elbers2020} to evolve massive neutrinos, allowing
us to explore variations of the $\Lambda$CDM model. On top of these core
components, the software package was extended to provide models for galaxy
formation. We make two such models available: one based on that used for the
\eagle project \citep{Schaye2015,Crain2015} and a second one based on the \gear
code \citep{Revaz2018, LoicThesis}. These were designed to target very different
scales and resolution ranges--massive galaxies and their large-scale environment
for \eagle, and dwarf galaxies for \gear--and are hence highly
complementary. The \eagle model is additionally and optionally extended with the
implementation of jet feedback from active galactic nuclei by
\citet{Husko2022b}.

Although \swift was originally developed for large-scale structure cosmology and
galaxy formation applications, it quickly became clear that the benefits of the
improved parallelisation of the coupled gravity--hydrodynamics solver could also
be extended to other areas in astrophysics. In particular, the code has been
extended to support planetary simulations by adding equations of state for the
relevant materials. These extensions have been designed by expanding the
existing SPH schemes to allow for multiple materials to interact, hence opening
the window to simulate the collisions and interactions of planets and other
bodies made of various layers of different materials.

Another, and to our knowledge unique, feature of \swift is the extent of the
material distributed as part of the public release\footnote{See
\url{www.swiftsim.com}.}. We do not only distribute the core gravity and
hydrodynamics solver but also offer the multiple modules for galaxy formation
mentioned and other applications above, as well as the different flavours of
SPH, the full treatment of cosmological neutrinos, and more than $100$ ready-to-run example
problems. All these elements are documented in detail, including developer
instructions for extending the code.  We emphasise too that the code is in
active development and we expect future releases to further extend the physics
modules presented here. ~\\

\noindent This paper is arranged as follows. In Section \ref{sec:design} we
present the overall \swift code design philosophy and core principles. The
equations of SPH that the code solves are summarised in Section
\ref{sec:sph}. In Section \ref{sec:gravity} and \ref{sec:cosmo}, we introduce
the equations for gravity, neutrinos, and the cosmology framework used by the
code. Sections \ref{sec:io} and \ref{sec:stf} are dedicated to the input \&
output strategy and cosmological structure finding respectively. In Section
\ref{sec:extensions}, we present some extensions including galaxy formation
(sub-grid) models and planetary physics models. We complete the code
presentation in Section \ref{sec:details} with some implementation details and
performance results. Finally, some conclusions are given and future plans are
presented in Section \ref{sec:conclusion}.

\section{Code design and implementation choices}
\label{sec:design}

We begin by laying out the core design principles of \swift, in particular its
strategy for making efficient use of massively parallel (hybrid shared and
distributed memory) high-performance computing systems.

\subsection{The case for a hydrodynamics-first approach}
\label{ssec:design:hydro_first}

Astrophysical codes solve complex networks of coupled differential equations,
often acting on a large dynamic range of temporal and spatial scales. Over time
these pieces of software frequently evolve from their original baseline, through
the addition of increasingly complex equations and physical processes, some of
them treated as ``sub-grid'' models. This process is often repeated multiple
times with each new iteration of the code, leading to multiple layers of
additions on top of one another. In many cases these layers do not use the most
appropriate algorithms or parallelisation strategies, but rather rely on the
decisions made for the previous layers' implementations.

A particularly relevant example of this issue is the generalised use of a
tree-code infrastructure \citep[e.g.][]{Barnes1986}, originally designed to
solve the equations of gravity, to also perform a neighbour-finding search for
SPH \citep[see e.g.][for a review]{Monaghan1992, Price2012}. Similarly, this gas
neighbour-finding code is then sometimes reused to find neighbours of star
particles (for feedback or enrichment), although the two species are clustered
very differently. These kinds of infrastructure re-use are ubiquitous in
contemporary simulation codes \citep[e.g.][]{Hernquist1989, Couchman1995,
  Dave1997, Springel2001, Wadsley2004, Springel2005, Springel2010, Hubber2011,
  Wadsley2017, Price2018, Springel2021}. Although appealing for its reduced
complexity, and successful in the past, this approach can in some cases result
in noticeable sub-optimal computational efficiency, in particular for modern
computing hardware. The data structure itself (a nested set of grids) is not the
culprit here, the way it is traversed is the limitation. For example, tree walks
typically involve frequent jumps in memory moving up and down the tree, a
pattern that is not ideal for modern CPUs or GPUs. Such a pattern is
particularly sub-optimal to make efficient use of the hierarchy of memory caches
as most of the data read will be discarded. Instead, modern hardware prefers to
access memory linearly and predictably, which also allows for a more efficient
utilisation of the memory bandwidth and caches, but also enables vector
instructions (SIMD). To exploit vector instructions, we need all the elements of
the vector (e.g. particles) to follow the same branching path. Thus, if an
independent tree-walk has to be performed for each particle, and there is no
obvious way to meaningfully group the particles into batches that will follow
the same path in the tree, then it will seriously hinder our ability to use such
vector instructions in our algorithms. Such an approach would hence, from the
outset, forfeit 7/\nth{8}\footnote{On a computer using \techjargon{AVX2}
  instructions (i.e. a SIMD vector size of 8), which is typical of current
  hardware. We note however that such peak performance is rarely achieved in
  actual production simulations.}  of the available computing performance of a
modern system. The loss of performance due to a tree-walk's inability to make
use of the various cache levels is more difficult to quantify. However, the
recent trend in computing hardware to add more layers of caches is a clear sign
that their use ought to be maximised in order to extract performance out of the
computing units. To back up this intuition, we performed a detailed analysis of
the large cosmological simulations from the \eagle project \citep{Schaye2015},
based on a heavily modified version of the \codename{Gadget}-3 code. It showed
that the majority ($>65\%$) of the computing time was spent in the
neighbour-finding operations (both for gas and stars)
performed via a tree walk. \\

\begin{figure}
\centering
\includegraphics[width=0.8\columnwidth]{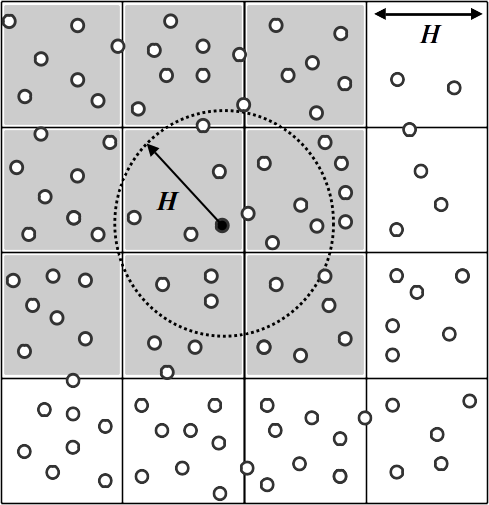}
\vspace{-0.1cm}
\caption{The Verlet-list method. By constructing a mesh structure with cell
  sizes matching the search radius $H$ of particles, the neighbour-finding
  strategy is entirely set by the geometry of the cells and the list of
  potential candidates is thus exactly known. The particle in black only has
  potential neighbours in the cell where it resides or any of the 8 (26 in 3D)
  directly neighbouring cells (in grey). The smoothly varying nature of SPH
  leads to particles having similar $H$ in nearby regions, with this scale only
  varying slowly over the whole simulated domain.}
\label{fig:design:cells_sph}
\end{figure}

All these considerations suggest that a simulation code designed with a
hydrodynamics-first approach could achieve substantial performance gains. In
SPH-like methods, the neighbourhood is defined entirely by demanding a certain
number $N_{\rm{ngb}}\sim50$--$500$ of particles around the particle of interest
from which to compute physical quantities and their derivatives. Similarly, many
sub-grid implementations (See e.g. \S\,\ref{ssec:eagle}, \S\,\ref{ssec:gear}, and
\S\,\ref{ssec:spin_jets}) rely on the same neighbourhoods for most of their
calculations. Hence, grouping particles in cells that contain a number of
particles $\gtrsim N_{\rm{ngb}}$ will naturally construct neighbourhoods of the
required size. This will lead to the construction of a Cartesian grid with cells
whose size is similar to the size of the search radius of the particles. The
neighbour-finding algorithm can then be greatly simplified. Each particle only
needs to search for particles in the cell where it lies and any of the directly
adjacent cells (Fig. \ref{fig:design:cells_sph}). To ensure this property is
always fulfilled, we force the cell sizes to not be smaller than the search
radii of the particles in a given region. If the condition is violated, this
triggers a reconstruction of the grid. This so-called \emph{Verlet-list} method
\citep{Verlet1967} is the standard way neighbour-finding is performed in
molecular dynamics simulations. Once the cell structure has been constructed,
all the required information is known. There is no need for any speculative
tree-walk and the number of operations, as well as the iteration through memory,
are easily predictable.

\begin{figure}
\centering
\includegraphics[width=0.85\columnwidth]{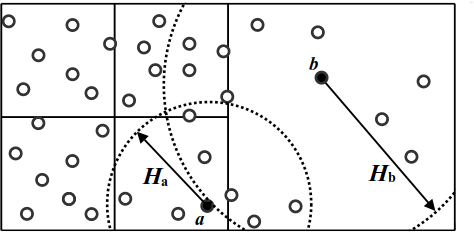}
\vspace{-0.1cm}
\caption{An example of interactions between regions of different densities,
  i.e. particles with different search radii. Particle $a$ will interact with
  the particles on the left and above using the smaller cells. It will interact
  with the particles on the right using the larger cell. The particle $b$ will
  only interact using the cells at the coarser level. Thanks to the nested
  grids, interactions happen at different levels in the hierarchy depending on
  the local search radius. Once the grid is constructed, all the possible
  interactions at the different levels are known without the need of a
  speculative tree-walk.}
\label{fig:design:cells_resolution}
\end{figure}

In the case of SPH for astrophysics, the picture is slightly more complex as the
density of particles and hence the size of their neighbourhoods can vary by
orders of magnitude. The method can nevertheless be adapted by employing a
series of nested grids (Fig. \ref{fig:design:cells_resolution}). Instead of
constructing a single grid with a fixed cell size, we recursively divide them,
which leads to a structure similar to the ones employed by
adaptive-mesh-refinement codes (See \S\,\ref{ssec:parallel:cells}). As we split
the cells into eight children, this entire structure can also be interpreted as
an oct-tree. We emphasise, however, that we do not walk up and down the tree to
identify neighbours; this is a key difference with respect to other packages.

With the cells constructed, the entire SPH neighbour-related workload can then
be decomposed into two sets of operations (or two sets of \emph{tasks}): the
interactions between all particles within a single cell and the interactions
between all particles in directly adjacent cells. Each of these operations
involves $\sim N_{\rm{ngb}}^2$ particle operations. For typical scenarios, that
is an amount of work that can easily be assigned to one single compute core with
the required data fitting nicely in the associated memory cache. Furthermore,
since the operations are straightforward (no tree-walk), one can make full use
of vector instructions to parallelise the work at the lowest level.

This approach, borrowed from molecular dynamics, was adapted for
multi-resolution SPH and evaluated by \cite{Gonnet2015} and
\cite{Schaller2016}. It forms the basis of the \swift code described here. We
emphasise that such an approach is not restricted to pure SPH methods; other
mesh-free schemes, such as the arbitrary Lagrangian-Eulerian (ALE) renormalised
mesh-free schemes \citep{Vila1999, Gaburov2011, Hopkins2015, Alonso2022}, finite
volume particle methods \citep[e.g.][]{Hietel2001, Hietel2005, Ivanova2013}, or
moving mesh \citep{Springel2010, shadowfax} also naturally fit within this
paradigm as they also rely on the concepts of neighbourhoods and localised
interactions.

As it turns out, the series of nested grids constructed to accommodate the
distribution of particles also forms the perfect structure on which to attach a
gravity solver.  We argued against such re-use at the start of our presentation;
the situation here is, however, slightly different. Unlike what is done for the
hydrodynamics, the gravity algorithm we use requires a tree-walk and some amount
of pointer-chasing (jumps in memory) is thus unavoidable. We eliminated the
tree-walk for the identification of SPH neighbourhoods, which was our original
goal. We can now use a much more classic structure and algorithm for the gravity
part of the \swift solver.  Viewing the grid cells as tree nodes and leaves, we
implement a \emph{Fast-Multipole-Method} \citep[FMM;
see][]{Greengard1987,Cheng1999,Dehnen2002, Dehnen2014, Springel2021} algorithm
to compute the gravitational interactions between particles. Here again, the
work can be decomposed into interactions between particles in the same cell
(tree-leaf), particles in neighbouring cells, or in distant cells. Once the tree
is constructed, all the information is available and no new decision making is
in principle necessary. The geometry of the tree and the choice of opening angle
entirely characterises all the operations that will need to be performed. All
the arithmetic operations can then be streamlined with the particles treated in
batches based on the tree-leaves they belong to.

\subsection{Parallelisation strategy: Task-based parallelism}
\label{ssec:design:parallel_strategy}

All modern computer architectures exploit multiple levels of parallelism.  The
trend over the last decade has been to increase the number of computing units
(CPUs, GPUs, or other accelerators) in a single system rather than to speed up
the calculations performed by each individual unit. Scientific codes that target
modern high-performance computing systems must thus embrace and exploit this
massive parallelism from the outset to get the most out of the underlying
hardware.

\begin{figure*}
\includegraphics[width=2\columnwidth]{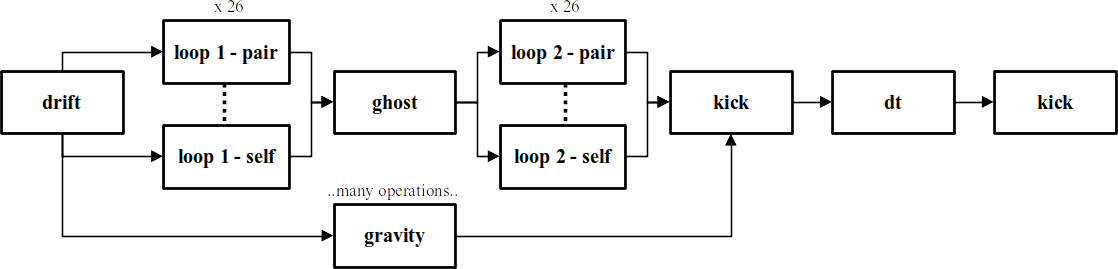}
\caption{A simplified graph of the tasks acting on a given cell for SPH and
  gravity during one time step in \swift. Dependencies are depicted as arrows
  and conflicts by dotted lines. Once the particles have been drifted to the
  current point in time, the first loop over neighbours can be run. The
  so-called ``ghost'' task serves mainly to reduce the number of dependencies
  between successive loops over the neighbours. Once the second loop has run,
  the time integration (\S\,\ref{ssec:design:multi_dt}) can be performed. In
  parallel to the SPH operations, the gravity tasks (condensed into a single one
  here for clarity) can be run as they act on different subsets of the data. To
  prevent different threads from over-writing each others' data, the various SPH
  loop tasks (1 self and 26 pairs) are prevented from running concurrently via
  our conflict mechanism. Additional loops over neighbours, used for instance in
  more advanced SPH implementations, in sub-grid models or for radiative
  transfer, can be added by repeating the same pattern. They can also be placed
  after the time integration tasks if they correspond to terms entering the
  equations in an operator splitting way.}
\label{fig:design:tasks}
\end{figure*}

As discussed in the previous section, the construction of a cell-based
decomposition of the computational volume leads to natural units of work to be
accomplished by the various compute cores. In principle, no ordering of these
operations is required: as long as all the internal (\emph{self}
i.e. particle-particle interactions of particles within a single cell) and
external (\emph{pair} i.e. particle-particle interactions of particles residing
in two different cells) interactions of these cells have been performed, all
particles will have iterated over all their neighbours. One can therefore list
all these cell-based units of work or \emph{tasks} and use a piece of software
that simply lets the different compute threads on a node fetch a task, execute
it, and indicate its successful completion. Such tasks can e.g.~take all the
particles in a cell and compute the $N_{\rm cell}^2$ SPH (or gravity)
interactions between them; or take all the particles and drift them
(i.e. integrate their positions) forward. This constitutes a very basic form of
\emph{task-based parallelism}. In astrophysics, the \codename{ChanGa} code
\citep{CHANGA} uses a similar parallel framework.

Compared to the traditional ``branch-and-bound'' approach in which all
operations are carried out in a pre-specified order and where all compute units
perform the same operation concurrently, as used by most other astrophysics
simulation codes, this task-based approach has two major performance
advantages. Firstly, it dynamically balances the work load over the available
compute cores. In most simulations, the distribution of computational work over
the simulation domain is highly inhomogeneous, with a small part of the volume
typically dominating the total cost. Decomposing this work a priori
(\textit{i.e.} statically) is a very challenging problem, and practical
solutions inevitably lead to substantial work imbalance. By not pre-assigning
regions to a specific computing unit, the task scheduler can instead naturally
and dynamically assign fewer cells to an individual computing unit if they turn
out to have a high computational cost, and vice versa.

The second advantage of the task-based approach is that it naturally allows the
gravity and hydrodynamics computations to be performed at the same time without
the need for a global synchronisation point between the two that typically leads to
(sometimes substantial) idle time. The list of tasks simply contains both kinds
of calculations and the threads can pick any of them; there is no need for the
code to wait for all the gravity operations to be done before the SPH
calculations can begin, or vice versa (Fig. \ref{fig:design:tasks}).

\begin{figure*}
\includegraphics[width=2.05\columnwidth]{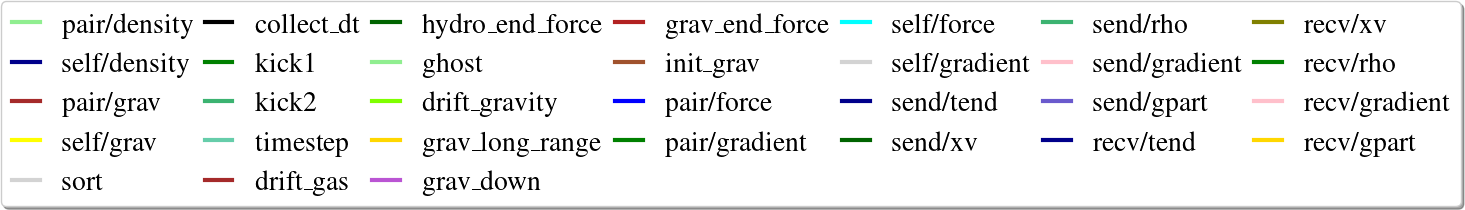}
\includegraphics[width=2.05\columnwidth]{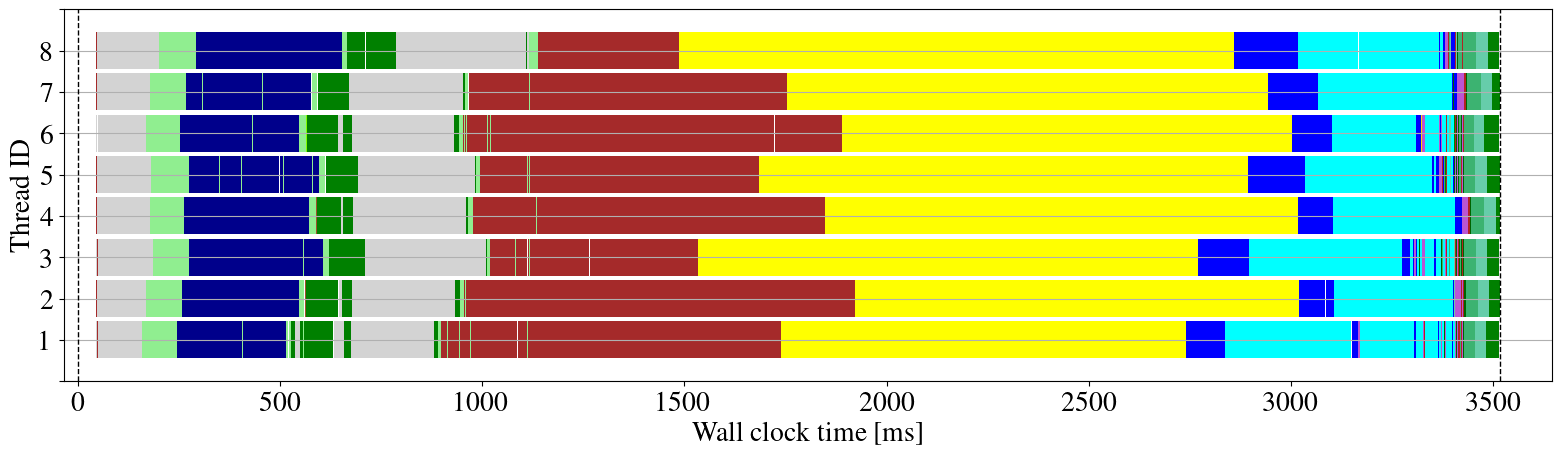}
\vspace{-0.1cm}
\caption{The execution of various tasks using 8 threads over the course of one
  time-step, extracted from a cosmological hydrodynamical simulation with
  $2\times128^3$ particles using only gravity and hydrodynamics on a
  shared-memory system. The different rows correspond to the different threads
  on the compute node. The work each thread performs is coloured to correspond
  to the task type it executes. Yellow, for instance, corresponds to a self-task
  performing gravity operation on a cell, whereas navy blue corresponds to a
  pair-task performing a \nth{3} SPH loop over two cells. Note that some tasks
  displayed in the legend do not actually run in this example. For instance, no
  \techjargon{MPI}-related \emph{send} or \emph{recv} tasks are executed
  here. We show them in the legend for consistency with
  Fig. \ref{fig:design:task_graph_mpi}. The long bands are actually a series of
  the same task acting on different cells one after the others. There are for
  instance 512 yellow tasks. As desired, the threads display essentially no idle
  time (white gaps) between operations and all end their work at very nearly the
  same time. In other words, the load balancing is near-perfect with no parallel
  performance loss. The small gap at the start corresponds to cost of deciding
  what tasks to activate for this step. Bands of a given colour can have
  different lengths, indicating that tasks can correspond to very different
  workloads depending on how many particles are present in the cell(s) on which
  they act. At a given point in time, different threads often process different
  task types, and hence solve a different set of equations. This is different
  from the traditional branch-and-bound parallelism approach where all threads
  perform the same action and have to wait until they have all completed it
  before moving to the next piece of physics.}
\label{fig:design:task_graph}
\end{figure*}

This tasking approach forms the basis of \swift. In its form discussed above,
however, it is too simple for the complex physics entering actual
simulations. Most SPH implementations require multiple loops over the particles
in their neighbourhoods. Sub-grid models often require that some hydrodynamic
quantities be computed before they can themselves operate. One could first
construct a list of all tasks related to the first loop and then distribute the
threads on it. A second list could then be constructed of all the tasks related
to the second loop and the process repeated. This would, however, re-introduce
global synchronisation points between the individual lists, leading to
undesirable idle time. Instead, we construct a single list but introduce
so-called \emph{dependencies} between operations acting on a given cell (and
hence its particles). For instance, all the first loop tasks have to be
performed on a given cell before the tasks associated with the second loop can
be performed. This transforms the list of tasks into an orientated \emph{graph}
with connections indicating the localised ordering of the physical operations to
perform. This graph can now include all the operations, even the ones not
requiring neighbour loops (e.g. time integration). Different cells can thus
naturally progress in a given time step at different rates, leading to no global
barriers between each loop (Fig. \ref{fig:design:task_graph}).  When a task has
completed, it reports this to all other tasks that depend on it. Once all
dependencies for a task are satisfied (i.e. all the other tasks that must have
run before it in the graph have completed), it is allowed to run; it is placed
in a queue from where it can be fetched by available compute threads.

In addition to this mechanism, the task scheduling engine in the \swift code
also uses the notion of \emph{conflicts} (Fig. \ref{fig:design:tasks}) to
prevent two threads from working on the same cell at the same time. This
eliminates the need to replicate data in different caches, which is detrimental
to performance. More crucially, it also ensures that all work performed inside a
single task is intrinsically thread-safe without the need to use atomic
operations. Because the code executed by a thread inside a task is guaranteed to
run on a private piece of data, developers modifying the physics kernels need
not worry about all the usual complexities related to parallel programming. This
reduces the difficulty barrier inherent to programming on modern architectures
and allows astrophysicists to easily modify and adapt the physics model in
\swift to their needs. To our knowledge, the combination of dependency and
conflict management in the tasking engine is a unique feature of
\swift\footnote{The classical alternative to conflict management is to introduce
explicit dependencies between tasks acting on the same data. This is less
desirable as it introduces an ordering of the cells where no natural one
exists.}. For a detailed description, we refer the reader to \cite{Gonnet2016},
where a stand-alone problem-agnostic version of this task scheduling engine is
introduced.

One additional advantage of this conflict mechanism is the opportunity to
symmetrize the operations. As no other compute thread is allowed to access the
data within a cell, we can update \emph{both} particles that take part in an
interaction simultaneously, effectively halving the number of interactions to
compute.  This is typically not possible in a classic tree-walk scenario as each
particle would need to independently search for its neighbours. The same
optimisation can be applied to the gravity interactions involving direct
interactions of particles, usually between two tree leaves.

\begin{figure}
\centering
\includegraphics[width=0.9\columnwidth]{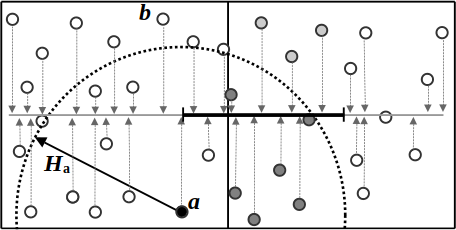}
\vspace{-0.1cm}
\caption{Pseudo-Verlet list optimisation for the interactions between all
  particles within a pair of neighbouring cells. Here the particles in the left
  cell receive contributions from the particles in the right cell. In the first
  phase, all particles are projected onto the axis linking the two cells (grey
  line) and sorted based on their projected coordinates. In the interaction
  phase, the particles iterate along this axis to identify candidates. For
  instance, the particle $a$ (in black) will identify plausible neighbours (in
  light and dark grey) on this axis up to a distance $H_a$ (indicated by the
  black ruler). These candidates are then tested for 3D distance to verify
  whether they are genuine neighbours (i.e. within the dotted circle and
  highlighted in dark grey here) or not. With this technique, the number of
  false-positives (light grey) is greatly reduced compared to the total number
  of possible candidates in the right-hand cell (here, 3 vs. 11). The advantage
  is even greater when considering the next particle (from right to left) on the
  axis. Particle $b$ knows that it will at most have to iterate on the axis up
  to the end of the ruler set by particle $a$, i.e. its list of candidates is at
  most as large as $a$'s for the same value of $H$. Moving from particle to
  particle in the left-hand cell, we can also stop the whole operation as soon
  as the distance on the axis does not reach at least the first particle in the
  right-hand cell. Because particles move only by small amounts between steps,
  the sorted list can be re-used multiple times provided a sufficient buffer is
  added to the length of the black ruler. Finally, the process is reversed to
  update the particles on the RHS with contributions from particles in the left
  cell. In 3D, even larger gains are achieved when the two cells share only an
  edge or just a corner.}
\label{fig:design:sort}
\end{figure}

Last but not least, the thread-safe nature of the work performed by the tasks,
combined with the small memory footprint of the data they act on, leads to them
being naturally cache efficient but also prime candidates for SIMD
optimization. The gravity calculations are simple enough that modern compilers
are able to automatically generate vector instructions and thus parallelise the
loops over pairs of particles. For instance, on the realistic gravity-only test
problem of \S \ref{ssec:gravity:convergence} we obtain speed-ups of 1.96x, 2.5x,
and 3.14x on the entire calculation when switching on \techjargon{AVX},
\techjargon{AVX2}, and \techjargon{AVX512} auto-vectorization on top of regular
optimization levels. This could also be the case for simple versions of the SPH
loops \citep[see discussion by][]{Willis2018}. The cut-off radius beyond which
no interactions take place does, however, allow for additional
optimizations. Borrowing, once more, from molecular dynamics, we implement
sorted interactions and pseudo-Verlet lists \citep{Gonnet2013}. Instead of
considering all particles in neighbouring cells as potential candidates for
interactions, we first sort them along the axis linking the cells' centres. By
walking along this axis, we drastically reduce the number of checks on particles
that are within neighbouring cells but outside each other's interaction range,
especially in the cases where the cells only share an edge or a corner
(Fig. \ref{fig:design:sort}). This way of iterating through the particle pairs
is much more complex and compilers are currently unable to recognize the pattern
and generate appropriate vector instructions. We therefore implemented SIMD code
directly in \swift, for some of the flavours of SPH, following the method of
\cite{Willis2018}. This approach does, however, break down when more complex
physics (such as galaxy formation models, see \S \ref{sec:extensions}) are solved, as too
many variables enter the equations.

Despite the advantages outlined above, one possible drawback to the task-based
approach, as implemented in \swift, is the lack of determinism. The ordering in
which the tasks are run will be different between different runs, even on the
same hardware and with the exact same executable. This can (and does) lead to
small differences in the rounding and truncation of floating point numbers
throughout the code, which, in turn will lead to slightly different results each
time. This is, of course, not an issue on its own as every single one of these
results was obtained using the same combination of operations and within the
same set of floating point rules. As an example, the study by
\cite{Borrow2023random} shows that the level of randomness created by the code
is consistent with other studies varying random seeds to generate different
galaxy populations. The same differences between runs can also arise in pure
\techjargon{MPI} codes or when using other threading approaches such as
\techjargon{OpenMP} as neither of these guarantee the order of operations (at
least in their default operating modes). Our approach merely exacerbates these
differences. In practice, we find that the main drawback is the difficulty this
intrinsic randomness can generate when debugging specific math-operation related
problems. We note that nothing prevents us from altering the task scheduling
engine to force a specific order. This would come at a performance cost, but
could be implemented in a future iteration of the code to help with the
aforementioned debugging scenario.

\subsection{Beyond single-node systems}
\label{ssec:design:mpi}

So far, we have described the parallelisation strategy within single
shared-memory compute nodes. To tackle actual high-performance computing (HPC)
systems and run state-of-the-art calculations, mechanisms must be added to
extend the computational domain to more than one node. The classic way to
achieve this is to decompose the physical volume simulated into a set of
discrete chunks and assign one to each compute node or even each compute
thread. Communications, typically using an \MPI implementation, must then be
added to exchange information between these domains, or to perform reduction
operations over all domains.

\swift exploits a variation of this approach, with two key guiding principles:
first, \MPI communication is only used between different compute nodes, rather
than between individual cores of the same node (who use the previously-described
tasking mechanism to share work and data between each other). Second, we base
the \MPI domain decomposition on the same top-level grid structure as used for
the neighbour finding, and aim to achieve a balanced distribution of work,
rather than data, between nodes.

The base grid constructed for neighbour finding
(\S\,\ref{ssec:design:hydro_first}) is split into regions that get assigned to
individual compute nodes. The algorithm used to decide how to split the domain
will be described in \S\,\ref{ssec:dd}; we focus here on how the exchange of
data is integrated into the task-based framework of \swift.

As the domain decomposition assigns entire cells to compute nodes, none of the
tasks acting on a single cell require any changes; all their work is, by
definition, purely local. We only need to consider operations involving pairs of
particles, and hence pairs of cells, such as SPH loops, gravitational force
calculation by direct summation (see \S\,\ref{ssec:gravity:walk}), or sub-grid
physics calculations (see \S\,\ref{sec:extensions}).

Consider a particle needing information from a neighbour residing on another
node to update its own fields. There are generally two possible approaches
here. The first one is to send the particle over the network to the other node,
perform a neighbour finding operation there, update the particle, and send the
particle back to its original node. This may need to be repeated multiple times
if the particle has neighbours on many different nodes. The second approach
instead consists of importing all foreign neighbours to the node and then only
updating the particles of interest local to the node once the foreign neighbour
particle data is present. We use this second approach in \swift and construct a
set of \emph{proxy} cells to temporarily host the foreign particles needed for
the interactions. The advantage of this approach is that it requires only a
single communication, since no results have to be reported back to the node
hosting the neighbour particle. Also, since we constructed the grid cells in
such a way that we know a priori which particles can potentially be neighbours,
and since we attach the communications to the cells directly, we also know which
particles to communicate. We do not need to add any walk through a tree to
identify which cells to communicate.

As \swift exploits threads within nodes and only uses \MPI domains and
communications between nodes, we actually construct relatively large domains
when compared to other \MPI-only software packages that must treat each core as
a separate domain. This implies that each node's own particle (or cell) volume
is typically much larger than any layer of proxy cells surrounding it. In
typical applications, the memory overhead for import buffers of foreign
particles is therefore relatively small. Furthermore, the trend of the last
decade in computing hardware is to have an ever larger number of cores and
memory on each node, which will increase the volume-to-surface ratio of each
domain yet further. Note, however, that some of these trends are not followed by
a proportional raise in memory bandwidth and some architectures also display
complex \techjargon{NUMA} designs. On such systems it may be beneficial to use a
few \MPI domains per node rather than a single one.

\begin{figure}
\centering
\includegraphics[width=0.9\columnwidth]{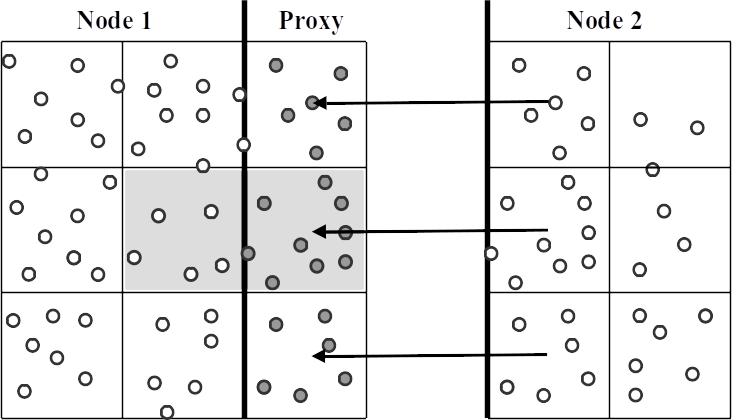}
\vspace{-0.1cm}
\caption{A pair interaction taking place over a domain boundary. The cell pair
  interaction in grey involves cells residing on either side of the domain
  boundary (thick black line), on two separate nodes.  To allow for the
  interaction to happen, we create a set of proxy cells on the first node and
  create communication tasks (arrows) that import the relevant particles (in
  grey) from the second node.  We also create a dependency between the
  communication and the pair task to ensure the data have arrived before the
  pair interaction can start. The pair task can then update the particles
  entirely locally, i.e. by exploiting exactly the same piece of code as for
  pairs that do not cross domain boundaries. A similar proxy exists on the other
  node to import particles in the opposite direction in order to process the
  pair also on that node and update its local particles.}
\label{fig:design:proxies}
\end{figure}

Once the proxy cells have been constructed, we create communication tasks to
import their particles (see Fig.~\ref{fig:design:proxies}). When the import is
done, the work within the pair task itself is identical to a purely local
pair. Once again, users developing physics modules need therefore not be
concerned with the complexities of parallel computing when writing their code.

\begin{figure}
\centering
\includegraphics[width=0.85\columnwidth]{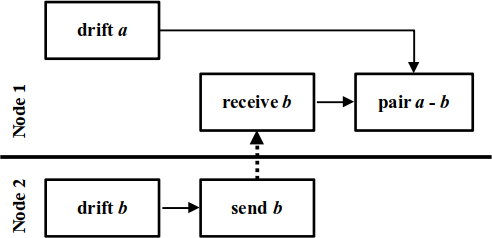}
\vspace{-0.1cm}
\caption{Extra communication tasks. The pair $a$--$b$ task (SPH or gravity)
  corresponds to the grey pair in Fig. \ref{fig:design:proxies}. Each compute
  node has a task to drift its own local cell. The foreign node (here below the
  thick black line) then executes a \emph{send} operation. On the local node, a
  \emph{receive} task is run to get the data before unlocking the dependency
  (solid arrow) and letting the scheduler eventually run the pair $a$--$b$
  interaction task. The communication itself (dotted arrow) implicitly acts as a
  dependency between the nodes.  The converse set of tasks exists on the other
  compute node to allow the pair $b$--$a$ to also be run on that node.}
\label{fig:design:tasks_mpi}
\end{figure}

The particles need to be communicated prior to the start of the pair
interactions. After all, the correct up-to-data particle data needs to be
present before the computation of the interactions for them to be correct. The
commonly adopted strategy is to communicate all particles from each boundary
region on all nodes to their corresponding proxy regions before the start of the
calculations. This can be somewhat inefficient, for two reasons. Firstly, it
typically saturates the communication network and the memory bandwidth of the
system, leading to poor performance, especially on smaller, mid-range, computing
facilities where the commuication hardware is less powerful than in big national
centres. Secondly, no other operations are performed by the code during this
phase, even though particles far from any domain boundaries require no foreign
neighbours at all and could therefore, in principle, have their interactions
computed in the meantime. The traditional branch-and-bound approach prevents
this, but \swift treats the communications themselves as tasks that can
naturally be executed concurrently with other types of calculation (see above).

\begin{figure*}
\includegraphics[width=2.05\columnwidth]{design/task_graph/Legend.png}
\includegraphics[width=\columnwidth]{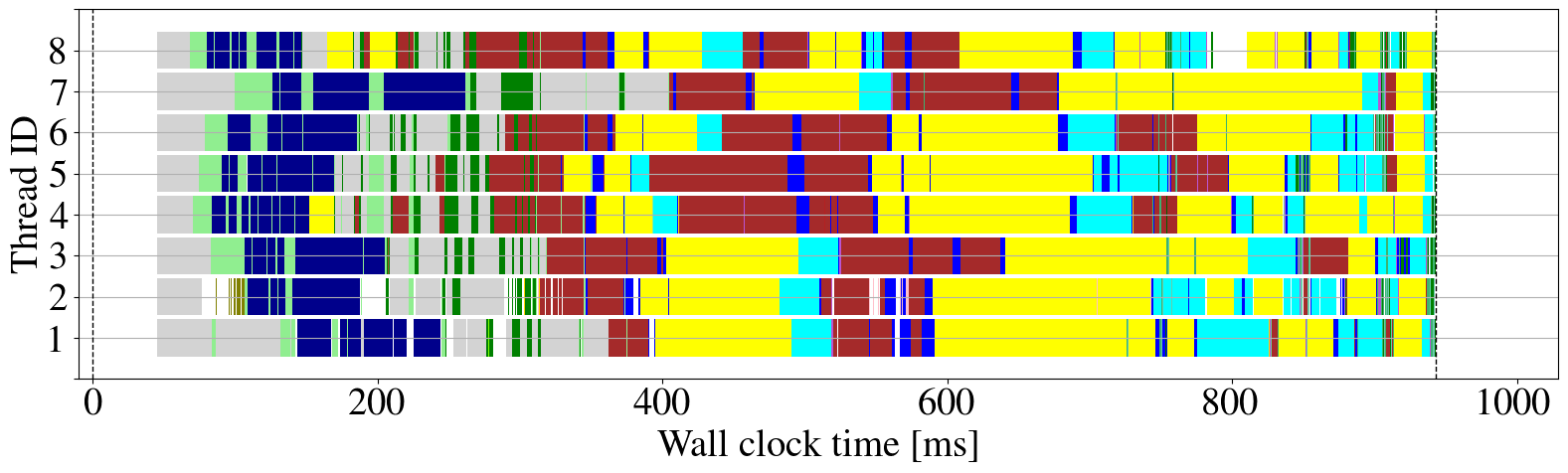}~
\includegraphics[width=\columnwidth]{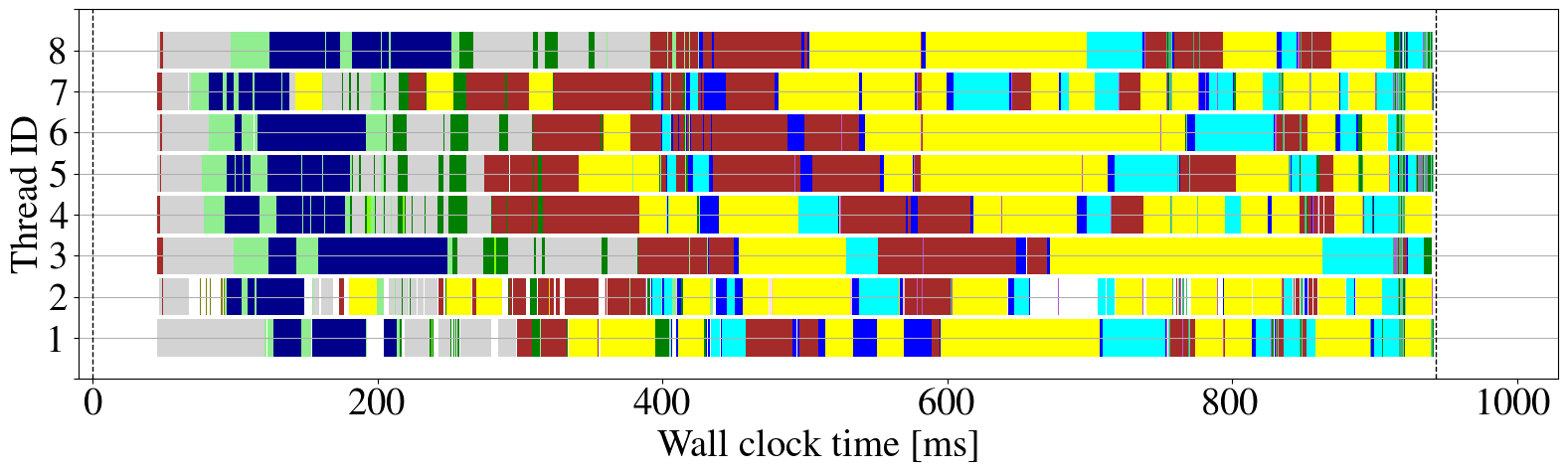}
\includegraphics[width=\columnwidth]{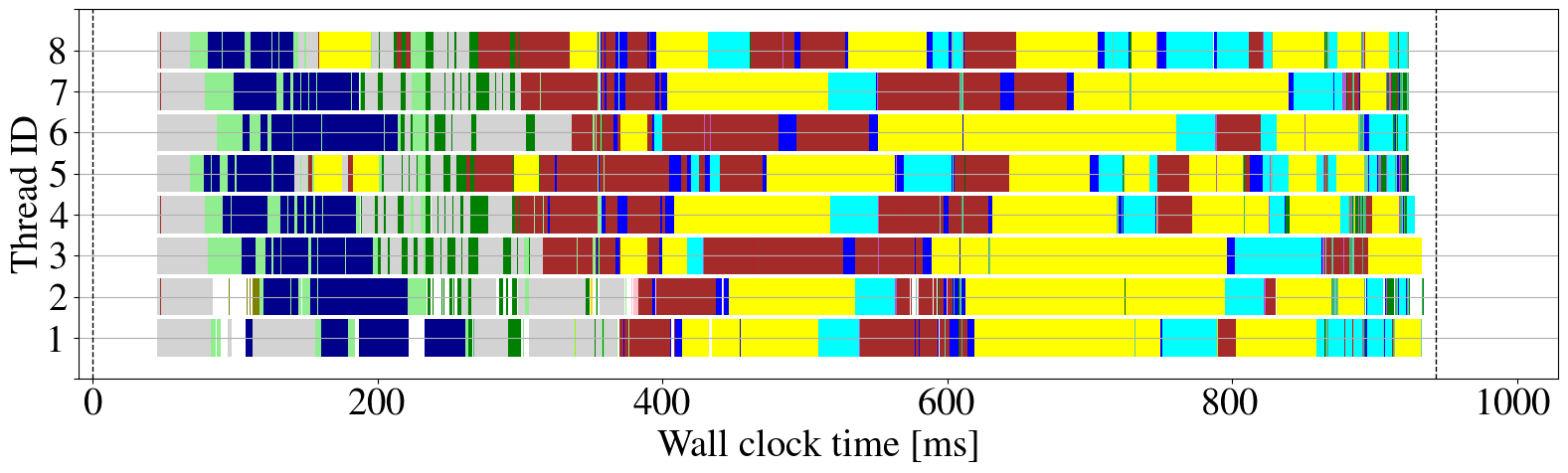}~
\includegraphics[width=\columnwidth]{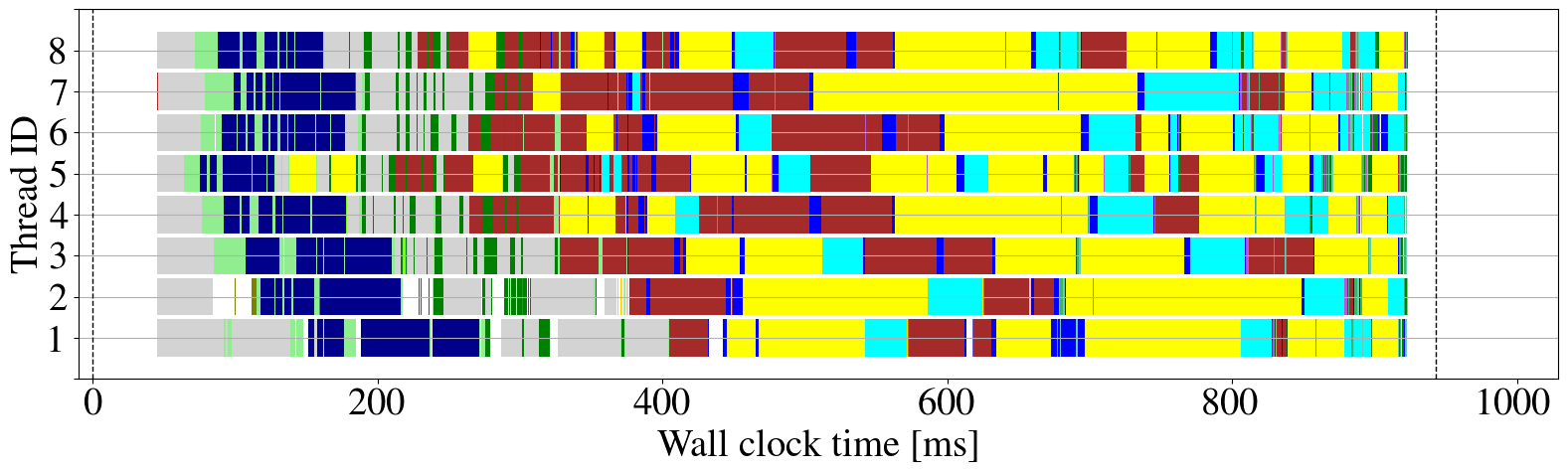}
\vspace{-0.1cm}
\caption{The same physics problem ($2\times128^3$ particles cosmological
  simulation) as displayed on Fig.~\ref{fig:design:task_graph} but now split
  across $4$ nodes, each using $8$ threads, i.e. a combination of distributed
  and shared parallelism. This is the hybrid mode in which \swift is run for
  large calculations that do not fit on a single node. Each panel corresponds to
  a different compute node. Within each panel the different rows correspond
  to the different threads on the compute node. The work each thread performs is
  coloured to correspond to the task type it executes using the same scheme as
  on Fig.~\ref{fig:design:task_graph}. The vertical dashed line on the right of
  each panel indicates the end of the time-step, which is determined by the
  point where the last compute node finishes. As can be seen, the node-to-node
  balance is not perfect; some nodes complete their work slightly earlier. This
  is due to the \techjargon{MPI} library requiring some time to process messages
  in an unpredictable way, which the domain decomposition algorithm (\S
  \ref{ssec:dd}) can thus not compensate for. This leads to small gaps in the
  execution (white gaps in the coloured bands). All required communication for
  the tasks occurs within this same figure, and overlaps (asynchronously) with
  work that only has local or already satisfied dependencies. All the exchanges
  happen whilst other tasks are running. The communications are overlapping with
  actual work. Note also that with less work per node overall compared to the
  shared-memory case, shown in Fig.~\ref{fig:design:task_graph}, it is easier to see here that a given point in time
  different threads often process different task types, and hence solve a
  different set of equations.}
\label{fig:design:task_graph_mpi}
\end{figure*}

At a technical level, we achieve this concurrency by exploiting the concept of
non-blocking communications offered by the \MPI standard\footnote{See \S\,3.7 of
\citet{mpi_standard}.}. This allows one compute node to mark some data to be
sent and then return to process other work. The data are silently transferred in
the background. On the receiving end, the same can be done and a receive
operation can be posted before the execution returns to the main code. One can
then probe the status of the communication itself, i.e. use the facilities
offered by the \MPI standard to know whether the data have arrived or are still
in transit. By using such a probe, we can construct \emph{send} and
\emph{receive} communication tasks that can then be inserted in the task graph
where needed and behaving like any of the other (computing) tasks. Once the data
have arrived on the receiving side, the receive task can simply unlock its
dependencies and the work (pair tasks) that required the foreign data can now be
executed (Fig. \ref{fig:design:tasks_mpi}). By adding the communications in the
tasking system, we essentially allow computational work to take place \emph{at
the same time} as communications. Note that the communication operations can be
performed by any of the running threads. We do not reserve one thread for
communications. The tasks not requiring foreign data can run as normal while the
data for other pairs is being exchanged, eliminating the performance loss
incurred from waiting for all exchanges to complete in the traditional
approach. The large volume-to-surface ratio of our domains (see above) implies
that there are typically many more tasks that require no foreign data than ones
that do. There is, hence, almost always enough work to perform during the
communication time and overheads.

An example of task execution over multiple nodes is displayed on
Fig.~\ref{fig:design:task_graph_mpi}. This is running the same simulation as was
shown on Fig.~\ref{fig:design:task_graph} but exploiting $4$ nodes each using
$8$ threads. We show here the full hybrid distributed and shared memory
capability of \swift. Here again, tasks of different kind are executed
simultaneously by different threads. No large data exchange operation is
performed at the start of the step; the threads immediately start working on
tasks involving purely local data whilst the data is being transferred. The work
and communication are thus effectively overlapping. The four nodes complete
their work at almost the same time and so do the threads within each node, hence
showing near perfect utilisation of the system and thus the ability to scale well.

The ability of \swift to perform computations concurrently with \MPI
communications reduces idle time, but the actual situation is somewhat more
complex. In reality, the \MPI library as well as the lower software layers
interacting with the communication hardware also need to use CPU cycles to
process the messages and perform the required copies in memory, so that a
complete overlap of communications and computations is not feasible. This is
often referred to as the \MPI progression problem. Such wasted time can for
instance be seen as blank gaps between tasks on
Fig.~\ref{fig:design:task_graph_mpi}. The extra cost incurred can vary dramatically
between different implementations of the \MPI protocol and depending on the
exact hardware used. A similar bottleneck can occur when certain sub-grid models
requiring many neighbour loops are used \citep[e.g.][]{Chaikin2023}. These may
generate many back-and-forth communications with only little work to be done
concurrently.

We remark, however, that whilst the communications taking place during a
time-step are all formally asynchronous, we still have a synchronisation point
at the end of a step where all the compute nodes have to wait. This is necessary
as we need all nodes to agree what the next time-step size is for instance.
This can be detrimental in the cases where the time-step hierarchies become very
deep (see below) and when only a handful of particles require updates every
step.  A strategy akin to the one used by the \codename{Dispatch} code
\citep{dispatch}, where regions can evolve at independent rates, would remove
this last barrier. In practice, thanks to our domain decomposition aiming to
balance the work not the data (see \S \ref{ssec:dd}), this barrier is typically
not a bottleneck for steps with a lot of work as the nodes all take a similar
amount of time to reach this end-of-step barrier.

\subsection{Local time-step optimisations}
\label{ssec:design:multi_dt}

In most astrophysical simulations, not only do the length-scales of interest
span several orders of magnitude, but so too do the time-scales. It would
therefore, typically, be prohibitively expensive to update all particles at
every step; localised time-step sizes or even per-particle time-steps are
essential. For a system governed by a Hamiltonian, it is possible to rewrite the
classic leapfrog algorithm and consider \emph{sub-cycles} where only a fraction
of the particles receive acceleration updates (a kick operation) whilst all
other particles are only moved (drifted) to the current point in time
\citep{Duncan1998, Springel2005}. \swift exploits this mechanism by first
creating long time-steps for the long-range gravity interaction
(\S\,\ref{ssec:gravity:mesh_summary}), where all the particles are updated, and
then creating a hierarchy of smaller steps using powers-of-two subdivisions,
where only the short-range gravity and hydrodynamic forces are updated
\citep{Hernquist1989}. This hierarchy is implemented by mapping the physical
time from start to end of a simulation to the range of values representable by
an integer. A jump of one thus represents the minimum time-step size reachable
by a particle (e.g. $(t_{\rm end} - t_{\rm begin}) / 2^{32}$ for a 32-bit
integer.). Each actual time-step size is then a power-of-two multiple of this
base quantum of time, hence ensuring exactly the hierarchy of time-steps we
expected. Using a 64-bit integer, we get a maximal possible number of steps in a
run of $2^{64}~\approx10^{19}$, much more than will be necessary.

In real applications, this hierarchy can be more than 10 levels deep, meaning
that the longest time-step sizes can be $>$1000$\times$ larger than the base
time-step length \citep[see e.g.][]{Borrow2018}.

The speed gains obtained by updating only a small fraction of the particles are
immense. However, at the level of code implementation and parallelisation, this
concept creates complicated challenges. Firstly, it requires added logic
everywhere to decide which particle and hence which cell needs updating. This
can be detrimental on some architectures (e.g. GPUs or SIMD vector units) where
more streamlined operations are required. Secondly, and most importantly, it
leads to global simulation steps where less computing time is spent moving the
system forward than is spent in overheads. This challenge cannot simply be
overcome by making the software more parallel; there will be steps where there
are fewer particles to update than there are CPU threads running. As small steps
(i.e. steps with a low number of particles to update) are orders of magnitude
more frequent than the base step, they can actually dominate the overall
simulation run time. It is hence of paramount importance to minimize all
possible overheads.

One of the key overheads is the time spent communicating data across the
network. The domain decomposition algorithm used in \swift (see
\S\,\ref{ssec:dd}) attempts to minimize this by not placing frequently active
particles (or their cells) close to domain boundaries. If this is achieved, then
entire steps can be performed without a single message being exchanged. The
other main overhead is the drift operation. In the classic sub-cycling leapfrog
\citep[e.g.][]{Quinn1997, Springel2005}, only the active particles are kicked,
but \emph{all} particles are drifted, since they could potentially be neighbours
of the active ones. Whilst the drift is easily scalable, as it is a pure
per-particle operation, it would nevertheless be wasteful to move all particles
for only the handful of them that are eventually found in the neighbourhood of
the few active particles. In \swift, as is also done in some other modern codes,
we alleviate this by first identifying the regions of the domain that contain
active particles and all their neighbours. We then activate the drift task for
these cells and only them. We thus do not drift all the particles just the
required ones, which is, to our knowledge, not an approach that is discussed in
the literature by other authors. This additional bit of logic to determine the
regions of interest is similar to a single shallow tree-walk from the root of
the tree down to the level where particles will be active. The benefit of this
reduced drift operation is demonstrated by \citet{Borrow2018}.  We note that
\swift can nevertheless be run in a more standard ``drift-everything'' mode to
allow for comparisons.

\subsection{Language, implementation choices, and statistics}
\label{ssec:design:language}

The design described above is, in principle, agnostic of the programming
language used and of the precise libraries exploited\footnote{With the exception
of \MPI, as its programming model drove many of the design decisions.}
to implement the physics or parallelism approach. It was decided early on to
write the code in the \techjargon{C} language (specifically using the \techjargon{GNU99}
dialect) for its ease of use, wide range of available libraries, speed of
compilation, and access to the low level threads, vector units, and memory
management of the systems.

The task engine exploited by \swift is available as a stand-alone tool,
\libraryname{QuickSched} \citep{Gonnet2016}, and makes use of the standard
\techjargon{POSIX} threads available in all \techjargon{UNIX}-based systems. The
advantage of using our own library over other existing alternative
(e.g. \libraryname{Cilk} \citep{Cilk}, \libraryname{TBB} \citep{TBB},
\libraryname{SMPSs} \citep{SMPSs}, \libraryname{StarPU} \citep{StarPU}, or the
now standard \libraryname{OpenMP} tasks) is that it is tailored to our specific
needs and can be adapted to precisely match the code's structure. We also
require the use of task conflicts (see \S\,\ref{ssec:design:parallel_strategy})
and the ability to interface with \MPI calls (see \S\,\ref{ssec:design:mpi}),
two requirements not fulfilled by other alternatives when the project was
started.

By relying on simple and widely available tools, \swift can be (and has been)
run on a large variety of systems ranging from standard x86 CPUs, ARM-based
computers, BlueGene architecture, and IBM Power microprocessors.

The entirety of the source code release here comprises more than 150\,000 lines
of code and 90\,000 lines of comments. These large numbers are on the one hand
due to the high verbosity of the \techjargon{C} language and on the other hand due to the
extent of the material released and the modular nature of the code.  The
majority of these lines are contained in the code extensions and i/o
routines. Additionally, about 30\,000 lines of python scripts are provided to
generate and analyse examples.  The basic \libraryname{Cocomo} model \citep{cocomo}
applied to our code base returns an estimate of 61 person-years for the
development of the package.

\swift was also designed, from the beginning, with a focus on an open and
well-documented architecture both for ease of use within the development team
but also for the community at large. For that reason, we include fifteen
thousand lines of narrative and theory documentation\footnote{Documentation is
available at \url{http://www.swiftsim.com/docs}}, a user onboarding guide, and large
open-source, well-documented, and well-tested analysis tools\footnote{These
tools are all available on the \swift project GitHub page
\url{http://www.github.com/swiftsim}}.

\section{Smoothed Particle Hydrodynamics Solver}
\label{sec:sph}

Having discussed the mechanism used by \swift to perform loops over neighbouring
particles, we now turn to the specific forms of the equations for hydrodynamics
evolved in the code.

Smoothed particle hydrodynamics \citep[SPH;][]{Lucy1977, Monaghan1977} has been
prized for its adaptivity, simplicity, and Lagrangian nature. This makes it a
natural fit for simulations of galaxy formation, with these simulations needing
to capture huge dynamic ranges in density (over 4 orders of magnitude even for
previous-generation simulations), and where the coupling to gravity solvers is
crucial. Future releases of \swift will also offer more modern hydrodynamics
solver options (see \S \ref{ssec:future}).

\swift implements a number of SPH solvers, all within the same neighbour-finding
and time-stepping framework. These solvers range from a basic re-implementation
of equations from \citet{Monaghan1992} in \S \ref{ssec:sph:intro} \& \S
\ref{ssec:sph:basic}, to newer models including complex switches for artificial
conductivity and viscosity.  We introduce our default scheme \sphenix in \S
\ref{ssec:sph:sphenix} and present our implementation of a time-step limiter and
of particle splitting in \S \ref{ssec:sph:limiter} and \S
\ref{ssec:sph:splitting} respectively. For completeness, we give the equations
for the additional flavours of SPH available in \swift in Appendix
\ref{appendix:SPH}. Note also that in this section, we limit ourselves to the
equations of hydrodynamics in a non-expanding frame. Information on comoving
time integration is presented later in \S\,\ref{ssec:operators}.

As comparing hydrodynamic models is complex, and often a significant level of
investigation is required even for a single test problem
\citep[e.g.][]{Agertz2007, Braspenning2022}, we do not directly compare the
implemented models in \swift here. We limit our presentation to the classic
``nIFTy cluster'' problem \citep[][\S \ref{ssec:sph:nifty}]{Sembolini2016},
which is directly relevant to galaxy formation and cosmology applications. For
our fiducial scheme, \sphenix, the results of many of the standard hydrodynamics
tests were presented by \cite{Borrow2022}. The initial conditions and parameters
for these tests, and many others, are distributed as part of \swift and can be
run with all the schemes introduced below.

\subsection{A brief introduction to SPH}
\label{ssec:sph:intro}

SPH is frequently presented from two lenses: the first, a series of equations of
motion derived from a Lagrangian with the constraint that the particles must
obey the laws of thermodynamics
\citep[see e.g.][]{Nelson1994, Monaghan2001, Springel2002, Price2012, Hopkins2013}; or
a coarse-grained, interpolated, version of the Euler equations
\citep[as in][]{Monaghan1992}.

As the implemented methods in \swift originate from numerous sources, there are
SPH models originally derived from, and interpreted through, both of these
lenses. Here, we place all of the equations of motion into a unified framework
for easy comparison.

SPH, fundamentally, begins with the kernel\footnote{An expanded discussion of
the following is available in both \citet{Price2012} and \citet{Borrow2021}.}.
This kernel, which must be normalised, must have a central gradient of zero, and
must be isotropic, is usually truncated at a compact support radius $H$.  We
describe the kernel as a function of radius $r$ and smoothing length $h$, though
all kernels implemented in \swift are primarily functions of the ratio between
radius and smoothing length $r/h$ to ensure that the function remains
scale-free. The kernel function
\begin{align}
    W(r, h) = \frac{1}{h^{n_{\rm d}}} w(r/h)
    \label{eqn:kernel}
\end{align}
where here $n_{\rm d}$ is the number of spatial dimensions and $w(r/h)$ is a
dimensionless function that describes the form of the kernel.

Throughout, \swift uses the \citet{Dehnen2012} formalism, where the smoothing
length of a particle is independent of the kernel used, with the smoothing
length given by $h=\sqrt{2\ln2} \,a$, with $a$ the full-width half maximum of a
Gaussian.  The cut-off radius $H=\gamma_{\rm K} h$ is given through a
kernel-dependent $\gamma_{\rm K}$.  We implement the kernels from that same
paper, notably the \citet{Wendland1995} C2, C4, and C6 kernels, as well as the
Cubic, Quartic, and Quintic splines \citep{Monaghan1985} using their
normalisation coefficients. Generally, we recommend that production simulations
are performed with the Wendland-C2 or Quartic spline kernels for efficiency and
accuracy reasons.

\subsubsection{Constructing the number density \& smoothing length}

The kernel can allow us to construct smoothed, volume-dependent quantities from
particle-carried quantities. Particle-carried quantities are intrinsic to
individual mass elements (e.g. mass, thermal energy, and so on), whereas
smoothed quantities (here denoted with a hat) are created from particle-carried
quantities convolved with the kernel across the smoothing scale (e.g. mass
density, thermal energy density, and so on).

The most basic smoothed quantity is referred to as the particle number density,
\begin{align}
    \hat{n}(\mathbf{r}, h) = \sum_j W(|\mathbf{r} - \mathbf{r}_j|, h),
    \label{eqn:numberdensity}
\end{align}
for a sum runs over neighbouring particles $j$. This is effectively a partition of unity
across the particle position domain when re-scaled such that
\begin{align}
    \hat{n}(h)\left(\frac{h}{\eta}\right)^{n_{\rm d}} = 1,
    \label{eqn:partition}
\end{align}
for all positions $\mathbf{r}$ and constant smoothing scale
$\eta$\footnote{Relationships between the classic `number of neighbours'
definition and the smoothing scale $\eta$ are described in \citet{Price2012}.},
assuming that the smoothing length $h$ is chosen to be large enough compared to
the inter-particle separation.

Given a disordered particle arrangement (i.e. any arrangement with non-uniform
particle spacing in all dimensions), it is possible to invert
eq. \ref{eqn:partition} with a fixed value of $\eta$ to calculate the expected
smoothing length given a measured number density from the current particle
arrangement. In principle, this is possible for all values of $\eta$, but in
practice there is a \citep[kernel dependent, see][]{Dehnen2012} lower limit on
$\eta$ which gives acceptable sampling of the particle distribution (typically
$\eta > 1.2$). Higher values of $\eta$ give a smoother field, and can provide
more accurate gradient estimates, but lead to an increase in computational
cost. For some kernels, high values of $\eta$ can also lead to occurrences of
the pairing instability \citep{Price2012,Dehnen2012}.

Given a computation of $\hat{n}_i$ at the position of a particle $\mathbf{r}_i$,
for a given smoothing length $h_i$, an expected particle number density can be
computed from eq. \ref{eqn:partition}.  In addition, we compute the derivative
\begin{align}
    \frac{\mathrm{d}\hat{n}_i}{\mathrm{d}h} = - \sum_j \left(\frac{n_{\rm d}}{h_i} W_{ij} + \frac{r_{ij}}{h_i} \nabla_i W_{ij}\right),
\end{align}
where here $r_{ij} \equiv |\mathbf{r}_i - \mathbf{r}_j|$, and $W_{ij} \equiv
W(r_{ij}, h_i)$, with $\nabla_i$ implying a spatial derivative with respect to
$\mathbf{r}_i$. This gradient is used, along with the difference between the
expected density and measured density, within a Newton--Raphson scheme to ensure
that the smoothing length $h_i$ corresponds to eq. \ref{eqn:partition} to within
a relative factor of $10^{-4}$ by default.

We calculate the mass density of the system in a similar fashion, with this
forming our fundamental interpolant:
\begin{align}
    \hat{\rho}_i = \sum_j m_j W_{ij},
\end{align}
where here $m_j$ is the particle mass. We choose to use the particle number
density in the smoothing length calculation, rather than mass, to ensure
adequate sampling for cases where particle masses may be very different, which
was common in prior galaxy formation models due to stellar enrichment sub-grid
implementations.

\swift calculates (for most implemented flavours of SPH) the pressure of particles based upon their
smoothed density and their internal energy per unit mass $u$, or adiabat $A$,
with
\begin{align}
    P_i = (\gamma - 1) u_i \hat{\rho}_i = A_i \hat{\rho}_i^\gamma,
    \label{eqn:eos}
\end{align}
where $\gamma$ is the ratio of specific heats.

\subsubsection{Creating general smoothed quantities}

Beyond calculating the density, any quantity can be convolved with the kernel to
calculate a smoothed quantity. For a general particle-carried quantity $Q$,
\begin{align}
    \hat{\mathbf{Q}}_i = \frac{1}{\hat{\rho}_i} \sum_j m_j \mathbf{Q}_j W_{ij},
\end{align}
with spatial derivatives
\begin{align}
    \nabla \cdot \hat{\mathbf{Q}}_i &= \frac{1}{\hat{\rho}_i} \sum_j m_j \mathbf{Q}_j \cdot \nabla W_{ij},\\
    \nabla \times \hat{\mathbf{Q}}_i &= \frac{1}{\hat{\rho}_i} \sum_j m_j \mathbf{Q}_j \times \nabla W_{ij},
\end{align}
provide basic estimates of smoothed quantities. Better estimators exist, and are
used in specialised cases \citep[see e.g.][]{Price2012}, but in all other cases
when we refer to a smoothed quantity these are the interpolants we rely on.

\subsubsection{SPH equations of motion}

Following \citet{Hopkins2013}, we write equations of motion for SPH in terms of
two variables describing a volume element for conserving neighbour number
($\tilde{x}$ in their formalism, here we use $a$) and a volume element for the
thermodynamical system ($x$ in their formalism, here we use $b$).  We then can
write the conservative equations of motion for SPH as derived from a Lagrangian
as follows:
\begin{align}
    \frac{\mathrm{d}\mathbf{v}_i}{\mathrm{d}t} = - \sum_j b_i b_j \left[
        \frac{f_{ij} P_i}{\hat{b}_i^2} \nabla_i W_{ij} +
        \frac{f_{ji} P_j}{\hat{b}_j^2} \nabla_j W_{ji}
    \right],
    \label{eqn:genericeom}
\end{align}
where here the factors $f_{ij}$ are given by
\begin{align}
    f_{ij} = 1 - \frac{a_j}{b_j} \left(\frac{h_i}{n_{\rm d} \hat{b}_i} \frac{\partial \hat{b}_i}{\partial h_i}\right)
    \left(1 + \frac{h_i}{n_{\rm d} \hat{a}_i} \frac{\partial \hat{a}_i}{\partial h_i}\right)^{-1}.
\end{align}
The second equation of motion, i.e. the one evolving the thermodynamic variable
($u$ or $A$) depends on the exact flavour of SPH, as described below.

\subsection{Basic SPH flavours}
\label{ssec:sph:basic}

\swift includes two so-called traditional SPH solvers, named
\sphflavour{Minimal} (based on \citet{Price2012}) and \sphflavour{Gadget2}
(based on \citet{Springel2005}), which are Density--Energy and
Density--Entropy-based solvers respectively. This means that they use the
particle mass as the variable $b$ in eq.~\ref{eqn:genericeom} and evolve the
internal energy $u$ or, respectively the adiabat $A$ (eq.~\ref{eqn:eos}), as
thermodynamic variable. These two solvers use a basic prescription for
artificial viscosity that is not explicitly time-varying. They are included in
the code mainly for comparison to existing literature and to serve as basis for
new developments.

\noindent These two solvers share the same equation of motion for velocity and
internal energy,
\begin{align}
    \frac{\mathrm{d} \mathbf{v}_i}{\mathrm{d} t} =& - \sum_j
        m_j \left[
            \frac{f_i P_i}{\hat{\rho}_i^2} \nabla_i W_{ij} + \frac{f_j P_j}{\hat{\rho}_j^2} \nabla_i W_{ij}
        \right], \\
            \frac{\mathrm{d} u_i}{\mathrm{d} t} =& \sum_j m_j \frac{f_i P_i}{\hat{\rho}_i^2} \mathbf{v}_{ij} \cdot \nabla_i W_{ij}
    \label{eqn:densityenergy}
\end{align}
but as they each track different thermodynamic variables ($u$, internal energy
per unit mass for \sphflavour{Minimal}, and entropy/adiabat $A$ for
\sphflavour{Gadget2}). In this latter flavour, the equation for the adiabat is
absent as ${\rm d}A/{\rm d}t = 0$ in the absence of additional source terms. In
the equations above we also defined, $\mathbf{v}_{ij} \equiv \mathbf{v}_i -
\mathbf{v}_j$, and
\begin{align}
    f_i = \left(1 + \frac{h_i}{n_{\rm d} \hat{\rho}_i} \frac{\partial \hat{\rho}_i}{\partial h}\right),
\end{align}
which is known as the `f-factor' or `h-factor' to account for non-uniform
smoothing lengths.\\

In addition to these conservative equations, the two basic SPH solvers include a
simple viscosity prescription, implemented as an additional equation of motion
for velocity and internal energy (entropy). The artificial viscosity
implementation corresponds to the equations 101, 103, and 104 of
\citet{Price2012}, with $\alpha_u = 0$ and $\beta = 3$.  We solve the following
equations of motion
\begin{align}
    \left.\frac{\mathrm{d} \mathbf{v}_i}{\mathrm{d} t}\right|_{\rm visc} =& - \sum_j m_j \frac{\nu_{ij}}{2}
        \left(f_{i}\nabla W_{ij} + f_{j}\nabla W_{ji}\right), \\
    \left.\frac{\mathrm{d} u_i}{\mathrm{d} t}\right|_{\rm visc} =& \sum_j m_j \frac{\nu_{ij}}{4} f_{i} \mathbf{v}_{ij} \cdot \nabla W_{ij},
    \label{eqn:basicartvisc}
\end{align}
where the interaction-dependent factor
\begin{align}
    \nu_{ij} &= - \frac{\alpha_{{\rm V}, ij} \,\mu_{ij} \,v_{{\rm
          sig}, ij}}{\hat{\rho}_i \hat{\rho}_j},  \label{eqn:alpha}  \\
    \mu_{ij} &= \begin{cases}
        \frac{\mathbf{v}_{ij} \cdot \mathbf{x}_{ij}}{|\mathbf{x}_{ij}|}  & {\rm if}~
        \mathbf{v}_{ij} \cdot \mathbf{x}_{ij} < 0,\\
        0 & {\rm otherwise}. \\
  \end{cases}
  \label{eqn:nuij}
\end{align}
These rely on the signal velocity between all particles, which is also used in
the time-step calculation, and is defined in these models as
\begin{align}
    v_{{\rm sig}, ij} = c_{{\rm s}, i} + c_{{\rm s}, j} - \beta \mu_{ij},
\end{align}
where the constant $\beta = 3$.

Finally, the viscosity is modulated using the \citet{Balsara1989} switch, which
removes viscosity in shear flows. The switch is applied to the viscosity
constants $\alpha_{{\rm V}, ij}$ is as follows:
\begin{align}
    \alpha_{{\rm V}, ij} &= \alpha_{\rm {V}, i} = \alpha_{\rm{V}} B_i, \\
    B_i &= \frac{| \nabla \cdot \mathbf v_i |}{|\nabla \cdot \mathbf v_i| + |\nabla \times \mathbf v_i| + \epsilon c_{{\rm s}, i} / h_i},
    \label{eqn:balsara}
\end{align}
where here $\alpha_{\rm V}=0.8$ is a fixed constant, $c_{{\rm s}, i}$ is the gas
sound speed, and $\epsilon = 0.0001$ is a small dimensionless constant
preventing divisions by zero.

\subsection{The {\bfseries \scshape Sphenix} flavour of SPH}
\label{ssec:sph:sphenix}

The \sphenix flavour of SPH is the default flavour in \swift, and was described
in detail by \citet{Borrow2022}. \sphenix inherits from the Density--Energy
formulation of SPH, uses similar discontinuity treatments and limiters as the
\anarchy scheme use in the \eagle cosmological simulations \citep[see][and
  Appendix \ref{ssec:sph:anarchy}]{Schaller2015, Schaye2015}, and uses a novel
limiter for feedback events. \sphenix was designed with galaxy formation
applications in mind. As the scheme uses the Density--Energy equation of motion and not a
pressure-smoothed implementation (\S \ref{ssec:sph:psph}), it must use a
comparatively higher amount of conduction at contact discontinuities to avoid
spurious pressure forces \citep[e.g.][]{Agertz2007, Price2008, Price2012}. As
such, removing the additional conduction in scenarios where it is not warranted
(in particular strong shocks) becomes crucial for accurate modelling and to not
dissipate energy where not desired. \\

As such, the major equations of motion are the same as described above in the
tradition SPH case, with the dissipationless component being identical to
eq. \ref{eqn:densityenergy}. The artificial viscosity term, however, is more
complex. We no longer use a constant $\alpha_{\rm V}$ in eq. \ref{eqn:alpha}. We
follow the framework of \cite{Morris1997} and turn it into a time-evolving
particle-carried quantity. This scalar parameter is integrated forward in time
using
\begin{equation}
    \alpha_{{\rm V},i}(t + \Delta t) = \alpha_{{\rm V},i}(t) - \alpha_{{\rm V, loc}, i}\exp
    \left(-\frac{\ell \cdot c_{{\rm s}, i}}{H_i} \Delta t\right),
    \label{eq:sphenix_visc1}
\end{equation}
with $H_{i} = \gamma_{\rm K} h_i$ the kernel cut-off radius, and where
\begin{align}
    \alpha_{{\rm V, loc}, i} &= \alpha_{\rm V, max} \frac{S_i}{v_{\rm sig, i}^2 + S_i},     \label{eq:sphenix_visc2}\\
    S_i &= H_i^2 \cdot \max\left(0, -\dot{\nabla}\cdot \mathbf{v}_i\right),     \label{eq:sphenix_visc3}
\end{align}
which ensures that $\alpha_{{\rm V},i}$ decays away from shocks. In these
expressions, $\ell = 0.05$ is the viscosity decay length, and $\alpha_{\rm V,
  max}=2.0$ is the maximal value of the artificial viscosity parameter. The
$S_i$ term is a shock indicator \citep[see][]{Cullen2010} which we use here to
rapidly increase the viscosity in their vicinity. For this detector, we
calculate the time differential of the velocity divergence using the value from
the previous time-step,
\begin{align}
    \dot{\nabla}\cdot \mathbf{v}_i(t + \Delta t) = \frac{\nabla \cdot \mathbf{v}_i(t + \Delta t) - \nabla \cdot \mathbf{v}_i(t)}{\Delta t}.\label{eq:sphenix_visc4}
\end{align}

Additionally, If $\alpha_{{\rm V, loc}, i} > \alpha_{{\rm V},i}(t)$, then
$\alpha_{{\rm V},i}(t + \Delta t)$ is set to $\alpha_{{\rm V, loc}, i}$ to ensure a
rapid increase in viscosity when a shock front approaches. The value of the
parameter entering the usual viscosity term (eq.~\ref{eqn:alpha}) is then
\begin{align}
    \alpha_{{\rm V}, ij} = \frac{\alpha_{{\rm V}, i} + \alpha_{{\rm V},j}}{2}\cdot\frac{B_i
      + B_j}{2}, 
    \label{eqn:av_ij}
\end{align}
which exploits the \citet{Balsara1989} switch so that we can rapidly
shut down viscosity in shear flows. Note that, by construction, these
terms ensure that the interaction remains fully symmetric. ~\\

In \sphenix, we also implement a thermal conduction (also known as artificial
diffusion) model following \citet{Price2008}, by adding an additional equation
of motion for internal energy
\begin{align}
  \left.\frac{\mathrm{d}u_i}{\mathrm{d}t}\right|_{\rm diff} = \sum_j \alpha_{{\rm c}, ij} v_{\mathrm{c}, ij} m_j (u_i - u_j) \frac{f_{ij} \nabla_i W_{ij} + f_{ij} \nabla_j W_{ji}}{\rho_i + \rho_j},
    \label{eqn:art_cond}
\end{align}
where here the new dimensionless parameter for the artificial conduction
strength is constructed using a pressure weighting of the contribution of both
interacting particles:
\begin{align}
    \alpha_{{\rm c},ij} = \frac{P_i \alpha_{{\rm c}, i} + P_j \alpha_{{\rm c}, j}}{P_i + P_j}.
\end{align}
with the $\alpha_{{\rm c}, i}$ evolved on a particle-by-particle basis with a
similar time dependency to the artificial viscosity parameter. The artificial
conduction uses the Laplacian of internal energy as a source term, in an effort
to remove nonlinear gradients of internal energy over the kernel width, with
\begin{equation}
    \frac{\mathrm{d} \alpha_{{\rm c}, i}}{\mathrm{d}t} = \beta_{\rm c} H_i \frac{\nabla^2 u_i}{\sqrt{u_i}} - (\alpha_{{\rm c}, i} - \alpha_{{\rm c, min}})\frac{v_{{\rm c}, i}}{H_i},
    \label{eqn:artconddt}
\end{equation}
where here $\beta_{\rm c}=1$ is a dimensionless parameter, and $\alpha_{{\rm c},
  i, {\rm min}}=0$ is the minimal value of the artificial conduction
coefficient. The artificial conduction parameter is bounded by a maximal value
of $\alpha_{{\rm c}, i, {\rm min}}=2$ in all cases. The value of $\beta_{\rm c}$
is high compared to other schemes to ensure the conduction parameter can vary on
short timescales. Note that the velocity entering the last term of
eq. \ref{eqn:artconddt} is not the signal velocity but we instead follow
\citet{Price2018} and write
\begin{align}
        v_{{\mathrm{c}}, ij} = \frac{|\mathbf{v}_{ij} \cdot \mathbf{x}_{ij}|}{|\mathbf{x}_{ij}|} + 
        \sqrt{2\frac{|P_i - P_j|}{\hat{\rho}_j + \hat{\rho}_j}}.
\end{align}
This is a combination of the signal velocities used by \citet{Price2018} for the
cases with and without gravity. As the thermal conduction term
(eq. \ref{eqn:art_cond}) is manifestly symmetric, no equation of motion for
velocity is required to ensure energy conservation.

Finally, we ensure that the conduction is limited in
regions undergoing strong shocks, limiting $\alpha_{\rm c}$ by applying
\begin{equation}
    \alpha_{{\rm c}, {\rm max}, i} = \alpha_{\rm c, \rm{max}}\left(
                 1 - \frac{\alpha_{{\rm V}, {\rm max}, i}}
                 {\alpha_{{\rm V}, \rm{max}}}\right),
    \label{eqn:condshocklimiter}
\end{equation}
with $\alpha_{{\rm c}, \rm{max}} = 1$ a constant, and
\begin{equation}
    \alpha_{{\rm c}, i} = \begin{cases}
        \alpha_{{\rm c}, i} & \alpha_{{\rm c}, i} < \alpha_{{\rm c}, \rm{max}} \\
        \alpha_{{\rm c}, \rm{max}} & \alpha_{{\rm c}, i} > \alpha_{{\rm c}, \rm{max}}.
    \end{cases}
    \label{eqn:conductionlimiter}
\end{equation}
Note the explicit appearance of the viscosity parameters $\alpha_{{\rm V},i}$ in
these expressions.  More information on the motivation behind the limiter, and
its implementation, are presented by \citet{Borrow2022}.

\subsection{Time-step limiter}
\label{ssec:sph:limiter}

For all these schemes, a necessary condition to ensure energy conservation,
especially when additional source terms such as stellar feedback are in use, is
to impose some form of limit between the time-step size of neighbouring
particles. This allows for information to be correctly propagated between
particles \citep[see][]{Durier2012}. In \swift, we use three different
mechanisms to achieve the desired outcome; these are all called ``time-step
limiters'' in different parts of the literature. We describe them here briefly.

The first limit we impose is to limit the time-step of \emph{active}
particles. When a particle computes the size of its next time-step, typically
using the CFL condition, it also additionally considers the time-step size of
all the particles it interacted within the loop computing accelerations. We then
demand that the particle of interest's time-step size is not larger than a
factor $\Delta$ of the minimum of all the neighbours' values. We typically use
$\Delta=4$ which fits naturally within the binary structure of the time-steps in
the code. This first mechanism is always activated in \swift and does not
require any additional loops or tasks; it is, however, not sufficient to ensure
energy conservation in all cases.

The time-step limiter proposed by \cite{Saitoh2009} is also implemented in
\swift and is a recommended option for all simulations not using a fixed
time-step size for all particles. This extends the simple mechanism described
above by also considering inactive particles and waking them up if one of their
active neighbours uses a much smaller time-step size. This is implemented by
means of an additional loop over the neighbours at the end of the regular
sequence (Fig. \ref{fig:design:tasks}). Once an active particle has computed its
time-step length for the next step, we perform an additional loop over its
neighbours and activate any particles whose time-step length differs by more
than a factor $\Delta$ (usually also set to $4$).  As shown by
\cite{Saitoh2009}, this is necessary to conserve energy and hence yield the
correct solution even in purely hydrodynamics problems such as a Sedov--Taylor
blast wave. The additional loop over the neighbours is implemented by
duplicating the already existing tasks and changing the content of the particle
interactions to activate the requested neighbours.

The third mechanism we implement is a synchronisation step to change the
time-step of particles that have been directly affected by external source
terms, typically feedback events. \cite{Durier2012} showed that the
\cite{Saitoh2009} mechanism was not sufficient in scenarios where particles
receive energy in the middle of their regular time-step. When particles are
affected by feedback (see \S\,\ref{ssec:eagle}, \ref{ssec:gear}, and
\ref{ssec:spin_jets}), we flag them for \emph{synchronisation}. A final pass
over the particles, implemented as a task acting on any cell which was drifted
to the current time, takes these flagged particles, interrupts their current
step to terminate it at the current time and forces them back onto the timeline
(\S~\ref{ssec:design:multi_dt}) at the current step. They then recompute their
time-step and get integrated forward in time as if they were on a short
time-step all along. This guarantees a correct propagation of energy and hence
an efficient implementation of feedback. The use of this mechanism is always
recommended in simulations with external source terms.

\subsection{Particle splitting}
\label{ssec:sph:splitting}

In some scenarios, particles can see their mass increase by large amounts. This
is particularly the case in galaxy formation simulations, where some processes
such as enrichment from stellar evolution (see \S~\ref{ssec:eagle:feedback}) can
increase some particle masses by large, sometimes unwanted, factors. To mitigate
this problem, the \swift code can optionally be run with a mechanism to split
particles that reach a specific mass. We note that this is a mere mitigation
tool and should not be confused for a more comprehensive multi-resolution
algorithm where particle would adapt their masses dynamically in different
regions of the simulation volume and/or based on refinement criteria.

When a particle reaches a user-defined mass $m_{\rm thresh}$, we split the
particle into two equal mass particles. The two particles are exact copies of
each other but they are displaced in a random direction by a distance
$0.2h$. All the relevant particle-carried properties are also halved in this
process. One of the two particles then receives a new unique
identifier\footnote{Depending on how the IDs are distributed in the initial
conditions, we either generate a new random ID or append one to the maximal ID
already present in the simulation.}. To keep track of the particles' history, we
record the number of splits a particle has undergone over its lifetime and the
ID of the original progenitor of the particle present in the initial
conditions. Combined with a binary tree of all the splits, also stored in the
particle, this leads to fully traceable, unique, identifier for every particle
in the simulation volume.

\subsection{The nIFTy cluster}
\label{ssec:sph:nifty}

\begin{figure}
  \includegraphics[]{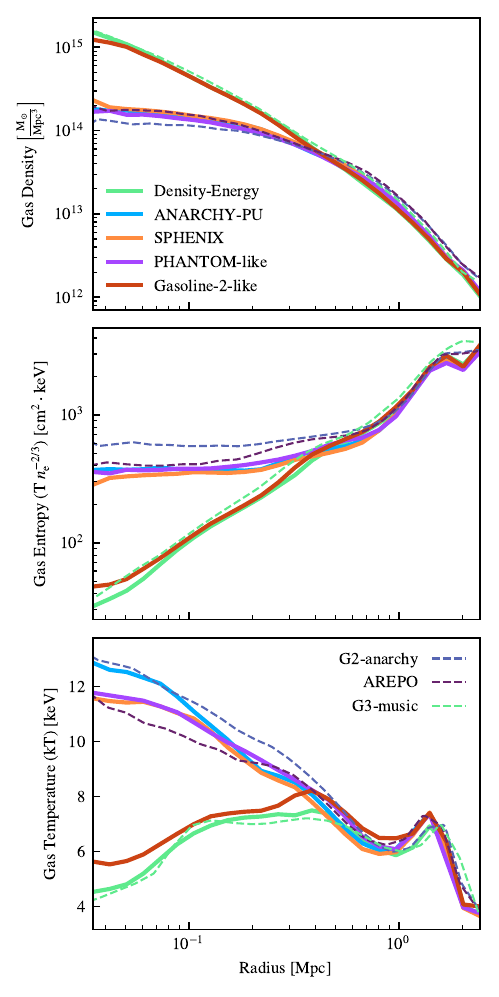}
  \vspace{-0.4cm}
  \caption{\emph{Top panel}: The gas density profile of the nIFTy cluster when
  simulated with five models within \swift (thick solid lines of various colours),
  and three external codes (dashed thin lines), shown at redshift $z=0$.
  \emph{Middle panel}: Gas entropy profile of the cluster (as extracted from the
  temperature and electron density profiles). \emph{Bottom panel}: Gas temperature
  profile of the cluster with the same models.}
  \label{fig:sph:nifty}
\end{figure}

In Fig. \ref{fig:sph:nifty}, we demonstrate the performance of a selection of
the hydrodynamics solvers within \swift on the (non-radiative) nIFTy cluster
\citep{Sembolini2016} benchmark. The initial conditions used to perform this
test are available for download as part of the \swift package in \libraryname{hdf5}
format. All necessary data, like the parameter file required to run the test, is
also provided in the repository as a ready-to-go example.

In the figure, we demonstrate the performance of five models from \swift
(Density--Energy (\S \ref{ssec:sph:basic}) in green, \anarchy-PU (\S
\ref{ssec:sph:anarchy}) in blue, \sphenix (\S \ref{ssec:sph:sphenix}) in orange,
\phantomSPH (\S \ref{ssec:sph:phantom}) in purple, and \gasoline-2 (\S
\ref{ssec:sph:gasoline}) in red)\footnote{ We remind the reader that all solvers
are independent re-implementations within \swift rather than using their
original codes, and all use the same neighbour-finding and time-step limiting
procedures.}. All simulations use the same Wendland-C2 kernel and
$\eta=1.2$. For comparison purposes, we display the results on this problem from
the \gadget-2 flavour of \anarchy (based upon Pressure-Entropy;
\sphflavour{G2-anarchy} in dashed blue), the \codename{Arepo} code and moving mesh-based
solver (dashed purple), and a more standard SPH flavour implemented in \gadget-3
(\sphflavour{G3-music}). These additional curves were extracted from the original
\citet{Sembolini2016} work.

Outside of radius $R > 0.5$ Mpc, all models show very similar
behaviour. Internally to this radius, however, two classes of hydrodynamics
model are revealed: those that form a flat entropy profile (i.e. the entropy
tends towards very low values within the centre, driven by high densities and
low temperatures), or a declining entropy profile (entropy flattens to a level
of $k_{\rm B} T n_{\rm e}^{-2/3} \approx 10^{2.5}$~cm$^2$~keV, driven by a low
central density and high temperature). There has been much debate over the
specific reasons for this difference between solvers. Here, we see that we form
a flat profile with the \gasoline-2-like (GDF) and Density-Energy models within
\swift, and the \codename{G3-music} code. These models have relatively low levels
of diffusion or conduction (or none at all, in the case of Density--Energy and
\codename{G3-music}). For instance, within our \gasoline-2-like implementation, we
choose the standard value of the conduction parameter $C=0.03$, consistent with
the original implementation.  Using a similar model \citet{Wadsley2008}
demonstrated that the formation of flat or declining entropy profiles was
sensitive to the exact choice of this parameter (only forming flat profiles for
$0.1 < C < 1.0$), and it is likely that this is the case within our \swift
implementation too, though any such tuning and parameter exploration is out of
the scope of this technical paper.

\section{Gravity solver}
\label{sec:gravity}

We now turn our attention towards the equations solved in \swift to account for
self-gravity \citep[see][for reviews]{Dehnen2011, Angulo2022}.  We start by
introducing the gravity softening kernels (\S \ref{ssec:potential_softening}),
then move on to summarise the Fast-Multipole-Method at the core of the algorithm
(\S \ref{ssec:fmm_summary}), and describe how it is implemented in our
task-based framework (\S \ref{ssec:gravity:walk}). We then present our choice of
opening angle (\S \ref{ssec:gravity:mac}) and the coupling of the method to a
traditional Particle-Mesh algorithm (\S \ref{ssec:gravity:mesh_summary}).  We
finish by showing a selection of test results (\S \ref{ssec:gravity:convergence}) before
discussing how massive neutrinos are treated
(\S \ref{ssec:gravity:nu}).

\subsection{Gravitational softening}
\label{ssec:potential_softening}

To avoid artificial two-body relaxation and avoid singularities when particles
get too close, the Dirac $\delta$-distribution of the density field
corresponding to each particle is convolved with a softening kernel of a given
fixed, but possibly time-varying, scale-length $H$. Beyond $H$, a purely
Newtonian regime is recovered.

Instead of the commonly used spline kernel of \cite{Monaghan1985} we use a C2
kernel \citep{Wendland1995}, which leads to an expression for the force that is
cheaper to compute whilst yielding a very similar overall shape. We modify the
density field generated by a point-like particle $\tilde\delta(\mathbf{r}) =
\rho(|\mathbf{r}|) = W(|\mathbf{r}|, 3\epsilon_{\rm Plummer})$, where
\begin{align}
W(r,H) =& \frac{21}{2\pi H^3} \times \nonumber \\
&\left\lbrace\begin{array}{rcl}
4u^5 - 15u^4 + 20u^3 - 10u^2 + 1 & \mbox{if} & u < 1,\\
0 & \mbox{if} & u \geq 1,
\end{array}
\right.
\end{align}
with $u = r/H$, and $\epsilon_{\rm Plummer}$ is a free parameter linked to the
resolution of the simulation \citep[e.g.][]{Power2003, Ludlow2019}. The
potential $\varphi(r,H)$ corresponding to this density distribution reads

\begin{align}
\varphi(r,H) = 
\left\lbrace\begin{array}{rcl}
f\left(\frac{r}{H}\right) \times H^{-1} & \mbox{if} & r < H,\\
r^{-1} & \mbox{if} & r \geq H,
\end{array}
\right.
\label{eq:fmm:potential}
\end{align}
with $f(u) \equiv -3u^7 + 15u^6 - 28u^5 + 21u^4 - 7u^2 + 3$. These choices lead
to a potential at $|\mathbf{x}| = 0$ that is equal to the central potential of a
\citet{Plummer1911} sphere (i.e. $\varphi(r=0) = 1/\epsilon_{\rm
  Plummer}$)\footnote{Note the factor of $3$ in the definition of
$\rho(|\mathbf{x}|)$ differs from the factor $2.8$ used for the cubic spline
kernel, as a consequence of the change of the functional form of $W$.}. From
this expression the softened gravitational force can be easily obtained:

\begin{align}
\mathbf{\nabla}\varphi(r,H) = \mathbf{r} \cdot
\left\lbrace\begin{array}{rcl}
g(\frac{r}{H}) \times H^{-3} & \mbox{if} & r < H,\\
r^{-3} & \mbox{if} & r \geq H,
\end{array}
\right.
\label{eq:fmm:force}
\end{align}
with $g(u) \equiv f'(u)/u = -21u^5+90u^4-140u^3+84u^2-14$. This last expression
has the advantage of not containing any divisions or branching (besides the
always necessary check for $r<H$), making it faster to evaluate than the
softened force derived from the \cite{Monaghan1985} spline kernel\footnote{A
Plummer softening would also be branch-free but would have undesirable
consequences on the dynamics \citep[see e.g.][]{Dehnen2001}.}. It is hence well
suited to target modern hardware, for instance to exploit SIMD instructions. In
particular, the use of a C2 kernel here allows most of the commonly used
compilers to automatically generate vectorised code, which is not the case when
using a spline-based kernel with branches. On the realistic scenario used as a
convergence test of \S \ref{ssec:gravity:convergence}, we get a speed-up of 2.5x
when using \techjargon{AVX2} vectorisation over the regularly optimised
code\footnote{Note that switching off all optimisation levels slows down the
code by a factor 3.6x compared to the non-vectorised baseline.}. The same code
using a spline kernel forfeits that speed-up and is even slightly slower due to
the extra operations even in the non-vectorised case.

\begin{figure}
\includegraphics[width=\columnwidth]{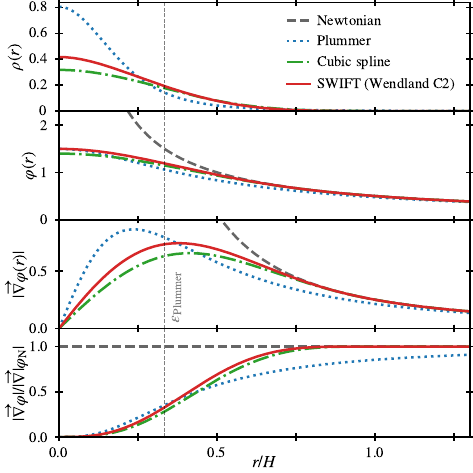}
\vspace{-0.3cm}
\caption{The density, potential, force, and force ratio to the Newtonian case
  generated by a point unit mass in our softened gravitational scheme. We use
  distances in units of the kernel cut-off $H$ to normalise the figures. A
  Plummer-equivalent sphere is shown for comparison. The spline kernel of
  \citet{Monaghan1985} is depicted for comparison but note that it has not been
  normalised to match the Plummer-sphere potential at $r=0$ (as is done in
  simulations) but rather normalised to the Newtonian potential at $r=H$ to
  better highlight the differences in shapes.}
\label{fig:fmm:softening}
\end{figure}

The softened density profile, with its corresponding potential and resulting
forces\footnote{For more details about how these are constructed see section 2
of~\cite{Price2007}.}  are shown in Fig. \ref{fig:fmm:softening}. For comparison
purposes, we also implemented the more traditional spline-kernel softening in
\swift. For a recent discussion of the impact of different softening kernel
shapes see section 8 of \cite{Hopkins2023}.

\subsection{Evaluating the forces using the Fast Multipole Method}
\label{ssec:fmm_summary}

The algorithmically challenging aspect of the \nbody problem is to generate the
potential and associated forces received by \emph{each} particle in the system
from \emph{every other} particle in the system. Mathematically, this means
evaluating

\begin{equation}
  \phi(\mathbf{x}_a) = \sum_{b \neq a} G_{\rm{N}} m_b\varphi(\mathbf{x}_a -
  \mathbf{x}_b)\qquad \forall~a\in N
  \label{eq:fmm:n_body}
\end{equation}
efficiently for large numbers of particles $N$ (with $G_{\rm{N}}$ the
gravitational constant). In the case of collisionless dynamics, the particles
are a mere Monte--Carlo sampling of the underlying coarse-grained phase-space
distribution \citep[e.g.][]{Dehnen2011}, which justifies the use of approximate
methods to evaluate eq.~\ref{eq:fmm:n_body}. The \emph{Fast Multipole Method}
\citep[FMM][]{Greengard1987, Cheng1999} is an $\mathcal{O}(N)$ approximation of
eq.~\ref{eq:fmm:n_body}, popularised in astronomy and adapted specifically for
gravity solvers by \cite{Dehnen2000, Dehnen2002} (see also \cite{Warren1995} for
related ideas). The FMM works by expanding the potential in a Taylor series
around \emph{both} $\mathbf{x}_a$ and $\mathbf{x}_b$ and grouping similar terms
arising from nearby particles to compute long-distance interactions between
well-separated groups only once. In other words, we consider groups of particles
with a large enough separation that the forces between them can be approximated
well enough by just the forces between their centres of mass. Higher-order
expressions, as used in \swift and other FMM codes, then not only approximate
these groups as interacting point masses, but also take into account their
shape, i.e. use the next order terms such as inertia tensors and beyond. A more
rigorous derivation is given below.

The convergence of FMM and its applicability to a large range of gravity
problems have been explored extensively \citep[see e.g.][]{Dehnen2002,
  Dehnen2014, pkdGrav3, Abacus, Springel2021}. For comparison, a
\cite{Barnes1986} tree-code, used in other modern codes such as \codename{2Hot}
\citep{2HOT} and \gadget-4 \citep[][in its default operating
  mode]{Springel2021}, only expands the potential around the sources
$\mathbf{x}_b$. The formal complexity of such a method is
$\mathcal{O}(N\log{N})$.

\subsubsection{Double expansion of the potential}

\begin{figure}
\includegraphics[width=\columnwidth]{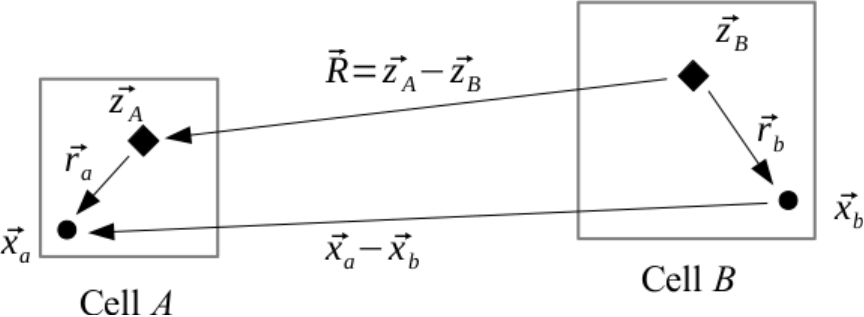}
\vspace{-0.3cm}
\caption{The basics of the Fast Multipole Method: The potential generated by a
  particle at position $\mathbf{x}_b$ on a particle at location $\mathbf{x}_a$
  is replaced by a double Taylor expansion of the potential around the distance
  vector $\mathbf{R}$ linking the two centres of mass ($\mathbf{z}_A$ and
  $\mathbf{z}_B$) of cell $A$ and $B$. The expansion converges towards the exact
  expression provided $|\mathbf{R}|>|\mathbf{r}_a + \mathbf{r}_b|$. In contrast,
  in a traditional \citet{Barnes1986} tree-code, all the particles in the cell
  $A$ receive direct contributions from $\mathbf{z}_B$ without involving the
  centre of expansion $\mathbf{z}_A$ in $A$.}
\label{fig:fmm:cells}
\end{figure}

In this section, we use the compact multi-index notation of \cite{Dehnen2014}
(repeated in appendix \ref{sec:multi_index_notation} for completeness) to
simplify expressions and ease comparisons with other published work. In what
follows $\mathbf{k}$, $\mathbf{m}$, and $\mathbf{n}$ denote the multi-indices
and $\mathbf{r}$, $\mathbf{R}$, $\mathbf{x}$, $\mathbf{y}$, and $\mathbf{z}$ are
vectors, whilst $a$ and $b$ denote particle indices. Note that no assumptions
are made on the specific functional form of the potential $\varphi$.

For a single pair of particles $a$ and $b$ located in respective cells $A$ and
$B$ with centres of mass $\mathbf{z}_A$ and $\mathbf{z}_B$, as shown in
Fig.~\ref{fig:fmm:cells}, the potential generated by $b$ at the location of $a$
can be written as
\begin{align}
  \varphi(\mathbf{x}_a - \mathbf{x}_b)
  &= \varphi\left(\mathbf{x}_a - \mathbf{z}_A - \mathbf{x}_b +
  \mathbf{z}_B + \mathbf{z}_A - \mathbf{z}_B\right)  \nonumber \\
  &= \varphi\left(\mathbf{r}_a - \mathbf{r}_b + \mathbf{R}\right)
  \nonumber \\
  &= \sum_\mathbf{k} \frac{1}{\mathbf{k}!} \left(\mathbf{r}_a -
  \mathbf{r}_b\right)^{\mathbf{k}} \nabla^{\mathbf{k}}\varphi(\mathbf{R})
  \nonumber \\
  &= \sum_\mathbf{k} \frac{1}{\mathbf{k}!} \sum_{\mathbf{n} <
    \mathbf{k}} \binom{\mathbf{k}}{\mathbf{n}} \mathbf{r}_a^{\mathbf{n}}
  \left(-\mathbf{r}_b\right)^{\mathbf{k} - \mathbf{n}}
  \nabla^{\mathbf{k}}\varphi(\mathbf{R})\nonumber \\
  &= \sum_\mathbf{n} \frac{1}{\mathbf{n}!} \mathbf{r}_a^{\mathbf{n}}
  \sum_\mathbf{m} \frac{1}{\mathbf{m}!}
  \left(-\mathbf{r}_b\right)^\mathbf{m} \nabla^{\mathbf{n}+\mathbf{m}} \varphi(\mathbf{R}),
  \label{eq:fmm:expansion}
\end{align}
where the Taylor expansion of $\varphi$ around $\mathbf{R} \equiv \mathbf{z}_A -
\mathbf{z}_B$ was used on the third line, $\mathbf{r}_a \equiv \mathbf{x}_a -
\mathbf{z}_A$, $\mathbf{r}_b \equiv \mathbf{x}_b - \mathbf{z}_B$ is defined
throughout, and $\mathbf{m} \equiv \mathbf{k}-\mathbf{n}$ is defined for the
last line. Expanding the series only up to order $p$, we get
\begin{equation}
  \varphi(\mathbf{x}_a - \mathbf{x}_b) \approx \sum_{\mathbf{n}}^{p}
  \frac{1}{\mathbf{n}!} \mathbf{r}_a^{\mathbf{n}} \sum_{\mathbf{m}}^{p
    -|\mathbf{n}|} 
  \frac{1}{\mathbf{m}!} \left(-\mathbf{r}_b\right)^\mathbf{m}
  \nabla^{\mathbf{n}+\mathbf{m}} \varphi(\mathbf{R}),
  \label{eq:fmm:fmm_one_part}
\end{equation}
with the approximation converging towards the correct value provided
$|\mathbf{R}|>|\mathbf{r}_a + \mathbf{r}_b|$ as $p\rightarrow\infty$. If we now
consider all the particles within $B$ and combine their contributions to the
potential at location $\mathbf{x}_a$ in cell $A$, we get
\begin{align}
  \phi_{BA}(\mathbf{x}_a) &= \sum_{b\in B}G_{\rm{N}} m_b\varphi(\mathbf{x}_a -
  \mathbf{x}_b)  \label{eq:fmm:fmm_one_cell}  \\
  &\approx G_{\rm{N}}\sum_{\mathbf{n}}^{p}
  \frac{1}{\mathbf{n}!} \mathbf{r}_a^{\mathbf{n}} \sum_{\mathbf{m}}
    ^{p -|\mathbf{n}|}
  \frac{1}{\mathbf{m}!} \sum_{b\in B} m_b\left(-\mathbf{r}_b\right)^\mathbf{m}
  \nabla^{\mathbf{n}+\mathbf{m}} \varphi(\mathbf{R}) \nonumber. 
\end{align}
This last equation forms the basis of the FMM. The algorithm decomposes
eq.~\ref{eq:fmm:n_body} into three separated sums, evaluated at different
stages.

\subsubsection{The FMM algorithm}

As a first step, multipoles are constructed from the innermost sum. For each
cell, we compute up to order $p$ all the necessary multi-poles (i.e. all terms
$\mathsf{M}$ whose norm of the multi-index $\mathbf{m} \leq p$)
\begin{equation}
  \mathsf{M}_{\mathbf{m}}(\mathbf{z}_B) = \frac{1}{\mathbf{m}!}
  \sum_{b\in B} m_b\left(-\mathbf{r}_b\right)^\mathbf{m} =
  \sum_{b\in B} m_b \mathsf{X}_{\mathbf{m}}(-\mathbf{r}_b),
  \label{eq:fmm:P2M} 
\end{equation}
where we re-used the tensors $\mathsf{X}_{\mathbf{m}}(\mathbf{r}_b) \equiv
\frac{1}{\mathbf{m}!} \mathbf{r}_b^{\mathbf{m}}$ to simplify the notation.  This
is the first kernel of the method, commonly labelled as \textsc{P2M} (particle
to multipole). In a second step, we compute the second kernel, \textsc{M2L}
(multipole to local expansion), which corresponds to the interaction of a cell
with another one:
\begin{equation}
  \mathsf{F}_{\mathbf{n}}(\mathbf{z}_A) = G_{\rm{N}}\sum_{\mathbf{m}}^{p -|\mathbf{n}|}
  \mathsf{M}_{\mathbf{m}}(\mathbf{z}_B)
  \mathsf{D}_{\mathbf{n}+\mathbf{m}}(\mathbf{R}), \label{eq:fmm:M2L} 
\end{equation}
where $\mathsf{D}_{\mathbf{n}+\mathbf{m}}(\mathbf{R}) \equiv
\nabla^{\mathbf{n}+\mathbf{m}} \varphi(\mathbf{R})$ is an order $n+m$ derivative
of the potential. This is the computationally expensive step of the FMM
algorithm, as the number of operations in a naive implementation using Cartesian
coordinates scales as $\mathcal{O}(p^6)$. More advanced techniques
\citep[e.g.][]{Dehnen2014} can bring the cost down to $\mathcal{O}(p^3)$, albeit
at a considerable algebraic cost. In the case of collisionless dynamics, accuracy down to
machine precision for the forces is not required, and low values of $p$ are thus
sufficient, which maintains a reasonable computational cost for the M2L kernel
(even in the Cartesian form).

Finally, the potential is propagated from the local expansion centre back to the
particles (L2P kernel) using
\begin{equation}
  \phi_{BA}(\mathbf{x}_a) = \sum_{\mathbf{n}}^{p}
  \frac{1}{\mathbf{n}!} \mathbf{r}_a^{\mathbf{n}}
  \mathsf{F}_{\mathbf{n}}(\mathbf{z}_A) = \sum_{\mathbf{n}}^{p}
  \mathsf{X}_{\mathbf{n}}(\mathbf{r}_a)
  \mathsf{F}_{\mathbf{n}}(\mathbf{z}_A). \label{eq:fmm:L2P}
\end{equation}
This expression is purely local, and can be efficiently implemented in a loop
that updates all the particles in cell $A$.\\

In summary, the potential generated by a cell $B$ on the particles in cell $A$
is obtained by the successive application of the P2M, M2L and L2P kernels. The
P2M and L2P kernels need only be applied once per particle, whilst one M2L
calculation must be performed for each pair of cells. \\

The forces applied to the particles are obtained by the same procedure, now
using an extra order in the Taylor expansion. For instance, for the acceleration
along the $x$ axis, we have:
\begin{equation}
  a_x(\mathbf{x}_a) = \sum_{\mathbf{n}}^{p-1}
  \mathsf{X}_{\mathbf{n}}(\mathbf{r}_a)
  \mathsf{F}_{\mathbf{n}+\left(1,0,0\right)}(\mathbf{z}_A). \label{eq:fmm:L2P_force} 
\end{equation}
Higher-order terms, such as tidal tensors, can be constructed using the same
logic. Note that only the last step in the process, the L2P kernel, needs to be
modified for the accelerations or tidal tensors. The first two steps of the FMM,
and in particular the expensive M2L phase, remain identical.

In practice, the multipoles can be constructed recursively from the leaves of
the tree to the root, and the local expansions from the root to the leaves by
shifting the $\mathsf{M}$ and $\mathsf{F}$ tensors and adding their
contributions to their parent or child cell's tensors respectively. This can be
done during the tree construction phase, for instance. Similarly, the local
expansion tensors ($\mathsf{F}$) can be propagated downwards using the opposite
expressions.

While constructing the multipoles $\mathsf{M}$, we also collect the centre of
mass velocity of the particles in the cells. This allows us to drift the
multipoles forward in time. This is only first-order accurate, but is sufficient
in most circumstances, especially since once the particles have moved too much a
full reconstruction of the tree (and hence of the multipoles) is triggered. Here, we follow the same logic as employed in many codes
\citep[e.g. \gadget][]{Springel2005} and force a tree reconstruction once a
fixed cumulative fraction (typically 1\%) of the particles have received an
update to their forces.

One final useful expression that enters some of the interactions between
tree-leaves is the P2M kernel. This directly applies the potential due to a
multipole expansion in cell B to a particle in cell A without using the
expansion of the potential $\mathsf{F}$ at the centre of mass of cell A. This
kernel is obtained by setting $\mathbf{r}_a$ to zero in
eq.~\ref{eq:fmm:expansion}, re-defining $\mathbf{R}\equiv\mathbf{x}_{\rm a} -
\mathbf{z}_{\rm B}$, and constructing the same $\mathsf{M}$ and $\mathsf{D}$
tensors as for the other kernels:
\begin{align}
  \phi_{Ba}(\mathbf{x}_a) &= G\sum_{\mathbf{m}}^p \mathsf{M}_{\mathbf{m}} \mathsf{D}_{\mathbf{m}}(\mathbf{R}),\\
  a_x(\mathbf{x}_a) &= G\sum_{\mathbf{m}}^p \mathsf{M}_{\mathbf{m}} \mathsf{D}_{\mathbf{m}+\left(1,0,0\right)}(\mathbf{R}).
  \label{eq:fmm:M2P}
\end{align}
The P2M kernel acts identically to traditional \cite{Barnes1986} tree-codes, 
which use solely that kernel to obtain the forces from the multipoles 
(or often just monopoles, i.e. setting $p=0$ throughout) to the particles.

With all the kernels defined, we can construct a tree walk by recursively
applying the M2L operation in a similar fashion to the double tree-walk
introduced by \cite{Dehnen2000}.

\subsubsection{Implementation choices}

All the kernels (eqs.~\ref{eq:fmm:P2M}-\ref{eq:fmm:M2P}) are rather
straightforward to evaluate as they are only made of additions and
multiplications (provided $\mathsf{D}$ can be evaluated quickly), which are
extremely efficient instructions on modern architectures.
However, the fully expanded sums can lead to rather large, and prone to typos,
expressions. To avoid any mishaps, we use a \techjargon{python} script to
generate the \techjargon{C} code in which all the sums are unrolled, ensuring
they are correct by construction. This script is distributed as part of the code
repository. In \swift, FMM kernels are implemented up to order $p=5$, more than
accurate enough for our purposes (see \S\,\ref{ssec:gravity:convergence}), but
this could be extended to higher order easily. At order $p=5$, this implies
storing $56$ numbers per cell for each $\textsf{M}$ and $\textsf{F}$ plus three
numbers for the location of the centre of mass. Our default choice is to use
multipoles up to order $p=4$; higher or lower implementations can be chosen at
compile time. For leaf-cells with large numbers of particles, as in \swift, this
is a small memory overhead. One further small improvement consists in choosing
$\mathbf{z}_A$ to be the centre of mass of cell $A$ rather than its geometrical
centre. The first order multipoles
($\mathsf{M}_{100},\mathsf{M}_{010},\mathsf{M}_{001}$) then vanish by
construction. This allows us to simplify some of the expressions and helps
reduce, albeit by a small fraction, the memory footprint of the tree structure.

\subsection{The tree walk and task-parallel implementation}
\label{ssec:gravity:walk}

The three main kernels of the FMM methods (eq. \ref{eq:fmm:P2M},
\ref{eq:fmm:M2L}, and \ref{eq:fmm:L2P}) are evaluated in different sections of
the code. The construction of the multipoles is done during the tree building
phase. This is performed outside of the task-based section of the code. As there
is no need to handle dependencies or conflicts during the construction, we use a
simple parallelisation over the threads for this phase. As is done in other
codes, this is achieved by recursively accumulating information
from the tree leaves to the root level. ~\\

Once the tree and associated multipoles have been constructed, the remaining
work to be performed is laid out. In a similar fashion to the hydrodynamics case
(\S\,\ref{ssec:design:parallel_strategy}), all the calculations (M2L kernels and
direct leaf-leaf interactions) can, in principle, be listed. The only difference
lies in the definition of which cells need to interact using which kernel. This
is based on the distance between the cells and information gathered from the
multipoles (see \S\,\ref{ssec:gravity:mac} for the exact expression). In the
case of a calculation using multiple nodes, the multipole information of
neighbouring cells located on another node is exchanged after the tree
construction (see \S \ref{ssec:mpi}). Whilst in the SPH case, the cells were
constructed such that only direct neighbours need to be considered, one may,
here, need to consider longer-range pairs of cells.

\begin{figure}
\includegraphics[width=0.95\columnwidth]{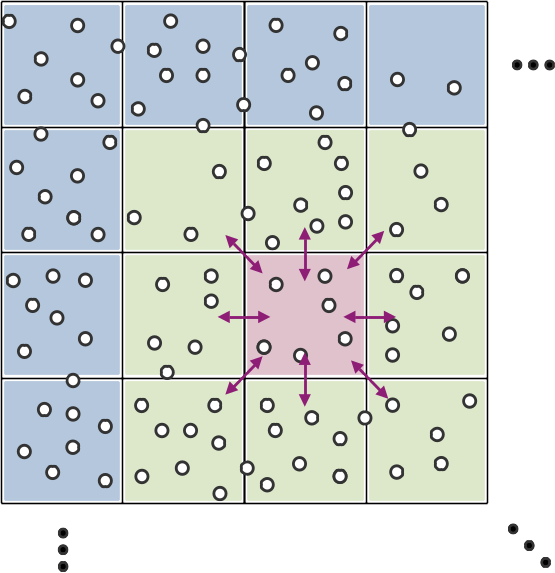}
\vspace{-0.1cm}
\caption{The basic decomposition of the FMM tree-walk into tasks for a set of
  particles in their cells, shown in 2D for clarity. The operations involving
  the red cell are as follows: (1) one \emph{self} task computing the gravity
  kernels within the cell itself, (2) eight \emph{pair} tasks computing the
  kernels for each pair of the red-green pairs of cells (the arrows), and
  (3) a single long-range task computing the M2L kernel contribution of all the
  blue cells to the red cell. In a realistic example, there will be many more
  blue cells beyond what is depicted here, but all their contributions to the
  cell of interest's potential will be handled by a single task looping over all
  of them. The green cells are too close, based on the criterion of
  \S\,\ref{ssec:gravity:mac} to use a multipole-multipole (M2L) interaction;
  their interactions with the red cell are hence treated as individual tasks as
  they contain a substantial amount of calculation to perform. In some cases,
  the distance criterion may be such that cells slightly further away also need
  to be treated by the pair tasks rather than just the directly neighbouring
  layer. This depends on the exact particle configuration and on the user's
  opening angle choices.}
\label{fig:fmm:gravity_tasks}
\end{figure}

In practice, we start from the top-level grid of cells and identify all the
pairs of cells that cannot interact via the M2L kernel. We then construct a
\emph{pair} task for each of them. Each cell also gets a \emph{self} task which
will take care of all the operations inside itself. Finally, for each cell, we
create a \emph{long-range} task, which will take care of all the interactions
involving this cell and any cell far enough that the M2L kernel can be directly
used. This third task is generally very cheap to evaluate as it involves only
the evaluation of eq. \ref{eq:fmm:M2L}. This is illustrated on
Fig. \ref{fig:fmm:gravity_tasks} for a simple case. ~\\

In most cases, the number of operations to perform within a single \emph{self}
or \emph{pair} task is large. These cells are also very likely to be split into
smaller cells in the tree. The tasks will hence attempt to recurse down the tree
and perform the operations at the level that is most suitable. To this end, they
use a double tree-walk logic akin to the one introduced by
\citet{Dehnen2002}. At each level, we verify whether the children cells are far
enough from each other based on the opening angle criterion
(\S\,\ref{ssec:gravity:mac}). If that is the case, then the M2L kernel is
used. If not, then we move further down the tree and follow the same logic at
the next level. The algorithm terminates when reaching a leaf cell. At this
point, we either apply the M2P kernel, if allowed by the criterion, or default
to the basic direct summation (P2P kernel) calculation.

Finally, the L2P kernel is applied on a cell-by-cell basis from the root to the
leaves of the tree using a per-cell task. These tasks are only allowed to run
once all of the self, pair, and long-range gravity tasks described above have
run on the cell of interest. This is achieved using the dependency mechanism of
the task scheduling library.

As the gravity calculation updates different particle fields (or even different
particles) from the SPH tasks, we do not impose any dependency between the
gravity and hydrodynamics operations. Both sets of tasks can run at the same
time on the same cells and particles. This differs from other codes where an
ordering is imposed. Our choice allows for better load-balancing since we do not
need to wait for all the gravity operations (say) to complete before the
hydrodynamics ones.

\subsection{The multipole acceptance criterion}
\label{ssec:gravity:mac}

The main remaining question is to decide when two cells are far enough from each
others that the truncated Taylor expansion used as approximation for the
potential (eq. \ref{eq:fmm:expansion}) is accurate enough. The criterion used to
make that decision is called the \emph{multipole acceptance criterion}
(MAC).\\

\noindent We know that eq. \ref{eq:fmm:expansion} converges towards the correct
answer as $p$ increases provided that $1>|\mathbf{r}_a + \mathbf{r}_b| /
|\mathbf{R}|$. This is hence the most basic (and always necessary) MAC that can
be designed. If this ratio is lower, the accuracy (at a fixed expansion order)
is improved and it is hence common practice to define a critical \emph{opening
angle} $\theta_{\rm cr}$ and allow the use of the multipole approximation
between two cells of size $\rho_{\rm A}$ and $\rho_{\rm B}$ if

\begin{equation}
  \theta_{\rm cr} > \frac{\rho{\rm _A} + \rho_{\rm B}} {|\mathbf{R}|}.
  \label{eq:fmm:angle}
\end{equation}
This lets users have a second handle on the accuracy on the gravity calculation
besides the much more involved change in the expansion order $p$ of the FMM
method. Typical values for the opening angle are in the range $[0.3, 0.7]$, with
the cost of the simulation growing as $\theta_{\rm cr}$ decreases. Note that
this MAC reduces to the original \cite{Barnes1986} criterion when individual
particles are considered (i.e. $\rho_{\rm A} = 0$).

This method has the drawback of using a uniform criterion across the entire
simulation volume and time evolution, which means that the chosen value of
$\theta_{\rm cr}$ could be too small in some regions (leading to too many
operations for the expected accuracy) and too large in some other ones (leading
to a lower level of accuracy than expected). \swift instead uses a more adaptive
criterion to decide when the multipole approximation can be used. This is based
on the error analysis of FMM by \cite{Dehnen2014} and is summarised below for
completeness\footnote{See also \cite{Springel2001} for similar ideas in the
regular tree case, based on the detailed error analysis of the tree code by
\cite{Salmon1994}.}. The key idea is to exploit the additional information about
the distribution of particles that is encoded in the higher-order multipole
terms.

We start by defining the scalar quantity $P_{\rm A,n}$, the \emph{power} of the
multipole of order $n$ of the particles in cell $A$, via
\begin{equation}
  P_{\rm A,n}^2 = \sum_{|\mathbf{m}|=n}
  \frac{\mathbf{m}!}{|\mathbf{m}|!}\mathsf{M}_{A,\mathbf{m}}^2,
\end{equation}
where the sum runs over all multipole terms of order $n$ in the
cell\footnote{Note that $P_{0} \equiv \mathsf{M}_{(0,0,0)}$ is just the mass of
the cell and since \swift uses the centre of mass as the centre of expansion of
the multipoles, $P_{1} = 0$.}. This quantity is a simple upper bound for the
amplitude of the multipole ($\mathsf{M}_{A, \mathbf{m}} <
P_{\rm{A},|\mathbf{m}|}/|\mathbf{m}|!$) and can hence be used to estimate the
importance of the terms of a given order in the Taylor series of the
potential. Following \cite{Dehnen2014} we then consider a sink cell $A$ and a
source cell $B$ (Fig. \ref{fig:fmm:cells}) for which we evaluate at order $p$
the scalar
\begin{equation}
  E_{\rm BA,p} = \frac{1}{M_{\rm B}|\mathbf{R}|^p} \sum_{n=0}^p \binom{p}{n} P_{\rm B,n}
  \rho_{\rm A}^{p-n},
  \label{eq:fmm:e_ab}
\end{equation}
with $M_{\rm B} \equiv \mathsf{M}_{{\rm B},(0,0,0)}$, the sum of the mass of the
particles in cell $B$. Note that since $P_{\rm B,n} \leq M_{\rm B} \rho_{\rm
  B}^n$, we have $E_{\rm BA, p} \leq \left((\rho_{\rm A} + \rho_{\rm
  B})/|\mathbf{R}|\right)^p$, where the right-hand side is the expression used
in the basic opening angle condition (eq.\,\ref{eq:fmm:angle}). We finally scale the
$E_{\rm BA,p}$'s by the relative size of the two cells to define the error
estimator $\tilde{E}_{\rm BA,p}$:
\begin{equation}
  \tilde{E}_{\rm BA,p} = 8\frac{\max(\rho_{\rm A}, \rho_{\rm B})}{\rho_{\rm A} + \rho_{\rm B}}E_{\rm BA,p}.
  \label{eq:fmm:e_ab_tilde}
\end{equation}
As shown by \cite{Dehnen2014}, these quantities are excellent estimators of the
error made in computing the accelerations between two cells using the M2L and
M2P kernels at a given order. We can hence use this property to design a new MAC
by demanding that the estimated acceleration error is no larger than a certain
fraction of the smallest acceleration in the sink cell $A$. This means we can
use the FMM approximation to obtain the accelerations in cell $A$ due to the
particles in cell $B$ if
\begin{equation}
  \tilde{E}_{\rm BA,p} \frac{M_{\rm B}}{|\mathbf{R}|^2} < \epsilon_{\rm FMM} \min_{a\in
    A}\left(|\mathbf{a}_a|\right) \quad \rm{and} \quad \frac{\rho_{\rm A} +
    \rho_{\rm B}} {|\mathbf{R}|} < 1,
  \label{eq:fmm:mac}  
\end{equation}
where $\mathbf{a}_a$ is the acceleration of the particles in cell $A$ and
$\epsilon_{\rm FMM}$ is a tolerance parameter. Since this is self-referencing
(i.e. we need the accelerations to decide how to compute the accelerations), we
need to use a an estimator of $|\mathbf{a}_a|$. In \swift, we follow the
strategy commonly used in other software packages and use the acceleration of
the previous time-step\footnote{On the first time-step of a simulation this
value has not been computed yet. We hence run a fake ``zeroth'' time-step with
the simpler MAC (eq. \ref{eq:fmm:angle}), which is good enough to obtain
approximations of the accelerations.}. The minimal norm of the acceleration in a
given cell can be computed at the same time as the P2M kernels which are
obtained in the tree construction phase. The second condition in
eq.\,\ref{eq:fmm:mac} is necessary to ensure the convergence of the Taylor
expansion.

One important difference between this criterion and the purely geometric one
(eq. \ref{eq:fmm:angle}) is that it is not symmetric in $A \leftrightarrow B$
(i.e. $E_{\rm AB,p} \neq E_{\rm BA,p}$). This implies that there are cases where
a multipole in cell $A$ can be used to compute the field tensors in cell $B$ but
the multipole in $B$ cannot be used to compute the $\mathsf{F}$ values of cell
$A$ and vice versa. This affects the tree walk by breaking the symmetry and
potentially leading to cells of different sizes interacting. That is handled
smoothly by the tasking mechanism which naturally adapts to the amount of work
required. Note that an alternative approach would be to force the symmetry by
allowing the multipoles to interact at a given level only if the criterion is
satisfied in both directions. We additionally remark that this breaking of the
symmetry formally leads to a breaking of the momentum-conserving property of the
FMM method. We, however, do not regard this as an important issue as the
momentum conservation is already broken by the use of per-particle time-step
sizes.

\subsection{Coupling the FMM to a mesh for periodic long-range forces}
\label{ssec:gravity:mesh_summary}

To account for periodic boundary conditions in the gravity solver, the two main
techniques present in the literature are: (1) apply an \cite{Ewald1921}-type
correction to every interaction \citep[e.g.][]{Hernquist1989, Klessen1997,
  Springel2001, Springel2005, Hubber2011, pkdGrav3, Abacus, Springel2021}; and
(2) split the potential in two (or more) components with one of them solved for
in Fourier space and thus accounting for the periodicity \citep[e.g.][]{Xu1995,
  Bagla2002, Springel2005, HACC, Springel2021}. We implement the latter of these
two options in \swift and follow the same formalism as presented by
\citet{Bagla2003}, adapted for FMM.

We start by truncating the potential and forces computed via the FMM using a
smooth function that drops quickly to zero at some scale $r_{\rm s}$ set by the
size of the gravity mesh. The Newtonian potential in eq. \ref{eq:fmm:n_body} is
effectively replaced by
\begin{equation}
  \phi_{\rm s}(r) = \frac{1}{r} \cdot \chi\left(r, r_{\rm s}\right) \equiv
  \frac{1}{r} \cdot {\rm erfc}\left(\frac{1}{2}\frac{r}{r_{\rm s}}\right),
  \label{eq:gravity:phi_s}
\end{equation}
where the subscript $s$ indicates that this is the short-range part of the
potential. As $\chi(r, r_{\rm s})$ rapidly drops to negligible values, the
potential and forces need only be computed via the tree walk for distances up to
$r_{\rm cut} = \beta r_{\rm s}$; interactions at larger distances are considered
to contribute exactly zero to the potential. Following \cite{Springel2005}, we
use $\beta = 4.5$ as our default\footnote{At this distance, the suppression is
almost three orders of magnitude already, as $\chi(4.5r_{\rm s}, r_{\rm s}) <
1.5\times10^{-3}$.}. This maximal distance for tree interaction means that the
long-range task (the one taking care of all the blue cells in
Fig.~\ref{fig:fmm:gravity_tasks}) only needs to iterate over the cells up to a
distance $\beta r_s$. This reduces further the amount of work to be performed
for the long-range operations by the tree.

The long-range part of the potential ($\phi_{\rm l}(r) = \frac{1}{r}\times {\rm
  erf}\left(\frac{1}{2}\frac{r}{r_{\rm s}}\right)$) is solved using a
traditional particle-mesh \citep[PM, see][]{Hockney1988} method. We assign all
the particles onto a regular grid of $N_{\rm mesh}^3$ cells using a
cloud-in-cell (CIC) algorithm. The mesh also sets the cut-off size $r_{\rm
  s}\equiv\alpha L/N_{\rm mesh}$, where $\alpha$ is a dimensionless order-unity
factor and $L$ is the size-length of the simulation volume. We use $\alpha=1.25$
as our default parameter value. In a second phase, we apply a Fourier transform
to this density field using the Fast-Fourier-Transform (FFT) algorithm
implemented in the \libraryname{fftw} library \citep{fftw}.

With the potential in Fourier space, Poisson's equation is solved by multiplying
each cell's value by the transform of the long-range potential
\begin{equation}
  \hat\phi_{\rm l}(k) = -\frac{4 \pi G_{\rm N}}{|\mathbf{k}|^2} \cdot
  \exp\left(-|\mathbf{k}|^2r_{\rm s}^2\right).
  \label{eq:gravity:phi_l}
\end{equation}
We then deconvolve the CIC kernel twice (once for the assignment, once for the
potential interpolation) and apply an inverse (fast) Fourier transform to
recover the potential in real space on the mesh. Finally, the particles'
individual potential and forces are obtained by interpolating from the mesh
using the CIC method.

The functional form of eq. \ref{eq:gravity:phi_s} might, at first, appear
sub-optimal. The error function is notoriously expensive to evaluate
numerically. In our formulation, we must evaluate it for every pair of
interactions (P2P or M2L) at every step. On the other hand, 
eq.~\ref{eq:gravity:phi_l} needs to be evaluated only $N_{\rm mesh}^3$ times at
every global step (see below). Typically, $N_{\rm mesh}\sim N^{1/3}$ but each of
the $N$ particles will perform many P2P kernel calls every single step. Using a
simpler form for $\chi$ in real space with a more expensive one to evaluate
correction in $k$-space may hence seem like an improvement. We experimented with
sigmoid-like options such as
\begin{equation}
  \chi(r,r_{\rm s}) = \left[2 - 2\sigma\left(\frac{2r}{r_{\rm s}}\right)\right],
  \qquad \sigma(w) \equiv \frac{e^w}{1 + e^w}
\end{equation}
but found little benefit overall. The solution we adopted instead is to stick
with eq.~\ref{eq:gravity:phi_s} and use an approximation to $\rm{erfc}$
sufficient for our needs. Specifically, we used eq.~7.1.26 of \cite{AS64}. Over
the range of interest, ($r \leq 4.5r_{\rm s}$), this approximation has a
relative error of less than $10^{-4}$ and the error tends to $0$ as
$r\rightarrow 0$. An alternative would be to store exact values in a table and
interpolate between entries, but that approach has the disadvantage of requiring non-local
memory accesses to this table shared between threads. Comparing simulations run
with an exact $\rm{erfc}$ to simulations using the approximation above, we find
no differences in the results.

Time integration of the forces arising from the long-range gravitational
potential is performed using a long time step and the symplectic algorithm for
sub-cycling of \cite{Duncan1998}. We split the Hamiltonian in long and short
timescales, corresponding to the long- and short-range gravity forces. The
short-range Hamiltonian also contains the hydrodynamics forces. The time-steps
then follow a sequence of kick \& drift operators for the short-range forces
embedded in-between two long-range kick operators \citep[See also][]{Quinn1997,
  Springel2005, Springel2021}.

As the mesh forces involve all particles and require all compute cores to
perform the FFT together, we decided to implement the PM calculation (i.e. the
CIC density interpolation, the calculation of the potential via Fourier space,
and the interpolation of the accelerations back onto the particles) outside of
the tasking system. In large calculations, the PM steps are rare (i.e. the
long-range, global, time-step size is long compared to the smallest individual
particle short-range time-step sizes). These steps are also where all particles
will have to update their short-range forces, which will trigger a full tree
rebuild. Having the PM calculation then perform a global operation outside of
the tasking framework whilst locking all the threads is hence not an issue. To
speed up operations, the PM calculation also uses parallel operations. The
assignment of the particles onto the density grid is performed using a simple
threading mechanism on each compute node. The Fourier transforms themselves are
then performed using the \techjargon{MPI}+~threads version of the
\libraryname{fftw} library. All nodes and cores participate in the
calculation. Once the potential grid has been obtained, the assignment of
accelerations to the particles is done using the same basic per-node threading
mechanism used for the construction of the density.

\subsection{Convergence tests}
\label{ssec:gravity:convergence}

The fast multipole method has been thoroughly tested both in the context of
collisional dynamics and for collisionless applications \citep[see
  e.g.][]{Dehnen2014,Springel2021}.  Many tests of simple scenarios, including
cells with uniform particle distributions or isolated halos with different
profiles can be found in the literature.  As the behaviour of the method is well
established and since our implementation does not differ from other reference
codes besides the parallelisation aspects, we do not repeat such a detailed
study here. We report having successfully tested the FMM implementation in
\swift on a wide range of cases, most of which are distributed as part of the
examples in the code. We thus verified that the code converges towards the
correct solution and presents the correct behaviour when the free parameters
(e.g. the MAC or the gravity mesh parameters) are varied.  We report here on one
such experiment with potential relevance to end users. \\

\begin{figure*}
\includegraphics[width=1.\columnwidth]{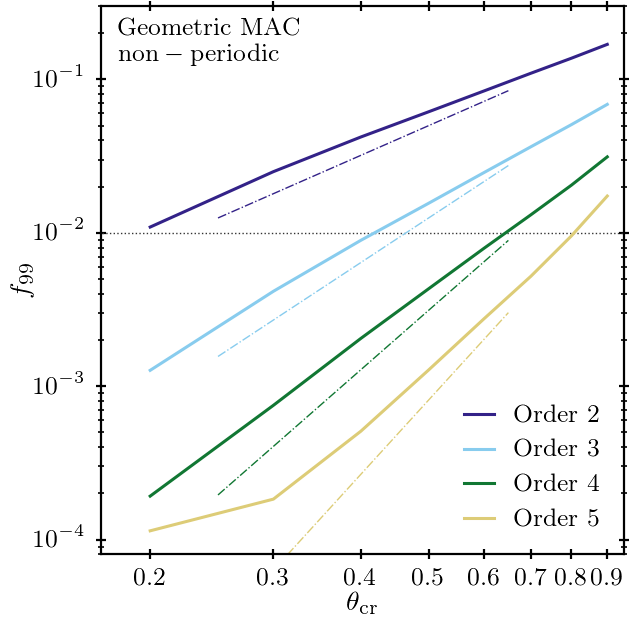}
~\,~\,~
\includegraphics[width=1.\columnwidth]{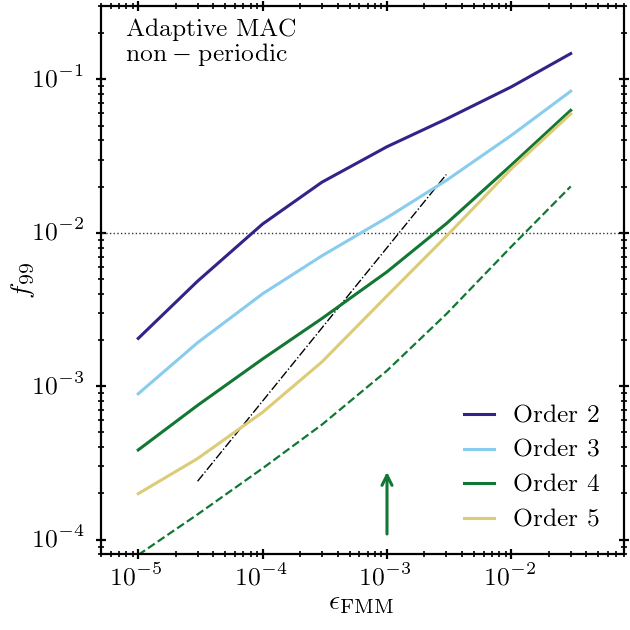}
\vspace{-0.1cm}
\caption{Accuracy of the gravity calculation (solid lines) for the two multipole
  acceptance criteria (MAC) on a low-redshift ($z=0.1$) $2\times376^3$particles,
  $25~{\rm Mpc}$ cosmological hydrodynamical simulation extracted from the
  \eagle suite. For 1 in every 100 particles, we calculated the exact forces
  using direct summation for comparison with the FMM-obtained prediction. We
  switch off periodic boundary conditions, and hence the gravity mesh, for this
  test. The \nth{99} percentile of the relative force error distribution is
  plotted against the geometric MAC, the classic tree opening angle, on the
  left, and against the adaptive MAC parameter on the right. Various multipole
  calculation orders $p$ are shown using different colours. Theoretical
  predictions for the convergence rates ($f_{99}\propto\theta^p$ for the
  geometric and $f_{99}\propto\epsilon_{\rm FMM}$ for the adaptive case at all
  orders) are shown using thin dot-dashed lines in the background (only one line
  for the adaptive case as the predictions is independant of $p$). The horizontal
  dotted line indicates where $99$ percent of the particles achieve a relative
  accuracy of better than $1$ percent, a commonly adopted accuracy target. Our
  default MAC choice, indicated by an arrow on the right panel, corresponds to a
  \nth{99} percentile of the relative error of $5\times10^{-3}$ for our standard
  setup using the \nth{4} order FMM implementation. We additionally show the
  \nth{90} percentile of the error ($f_{90}$) for the order four adaptive MAC
  case using a dashed line. The \swift implementation converges at a lower rate
  than theoretical expectations in the adaptive case. In the geometric case, the
  deviation from the theoretically expected power-law behaviour for $\theta_{\rm
    cr}<0.3$ and $p=5$ is due to truncation errors in single precision.}
\label{fig:gravity:validation_np}
\end{figure*}

\noindent Our test setup is a snapshot from a cosmological simulation of the
\eagle \citep{Schaye2015} suite. We take the $z=0.1$ snapshot from their
$(25~{\rm Mpc})^3$ volume. This setup comprises $2\times376^3 \approx10^7$
particles with a very high degree of clustering and is hence directly relevant
to all galaxy formation applications of the code. The combination of haloes and
voids present in the test allows us to test \swift's accuracy in a variety of
regimes.  We randomly select $1$ percent of the particles for which the exact
forces are computed using a direct summation algorithm. An \cite{Ewald1921}
correction is applied to take into account the periodicity of the volume. We
then run \swift and compute the forces via the FMM-PM code described above. We
finally compute the relative force error for our sample of particles and
evaluate the \nth{99} percentile ($f_{99}$) of the error distribution. We chose
to show the \nth{99} percentile error over lower ones as it provides better
guidance for users for their accuracy requirements by taking into account
outliers. We show this error percentile as a function of the opening angle
parameters in Fig.~\ref{fig:gravity:validation_np} for the case where periodic
boundary conditions have been switched off. In this test, only the FMM part of
the code is thus exercised. The left panel corresponds to the case of a purely
geometric MAC (eq.~\ref{eq:fmm:angle}) and the right panel to the case of the
adaptive MAC (eq.~\ref{eq:fmm:mac}). On both panels, we show different orders of
the method using different line colours. The dotted line is used to indicate the
$1\%$-error level. We find that, as expected, the forces converge towards the
correct, direct-summation-based, solution when the accuracy parameters are
tightened. Similarly, when using the geometric MAC the relationship between
$f_{99}$ and $\theta_{\rm cr}$ is found to be a power law whose slope steepens
for higher values of $p$ as predicted by theoretical arguments
\citep[e.g.][]{Dehnen2014, Springel2021}. These expectations are displayed on
the figure using thin dash-dotted lines. In the geometric case, the expected
behaviour is recovered.  The deviation from a power law at $\theta_{\rm cr}<0.3$
for $p=5$ is taking place in the regime where the results start to be affected
by single precision floating-point truncation. We have verified that when
switching to double precision the power-law behaviour continues for smaller
values of $\theta_{\rm cr}$, demonstrating that our implementation of the FMM
algorithm matches theoretical expectations. In practice, this truncation error
takes place much below the regime used in production runs. In the adaptive MAC
case, the theoretical expectation is for the scheme to converge as $f_{99}
\propto \epsilon_{\rm FMM}$ for all orders $p$. This is shown as a thin black
dash-dotted line on the figure. The current \swift implementation converges at a
rate below these theoretical predictions. Our recommended default value for the
adaptive MAC parameter is shown as a green arrow on the right panel. Using our
default setup where we construct multipoles to fourth order, $99$ percent of the
particles have a relative error of less than $5\times10^{-3}$ for their force
calculation. For comparison with the often used in the literature \nth{90}
percentile of the error \citep[e.g.][]{Springel2021}, we additionally show it
using a dashed line on the right panel for our default \nth{4}-order FMM setup.

\begin{figure*}
\includegraphics[width=1.\columnwidth]{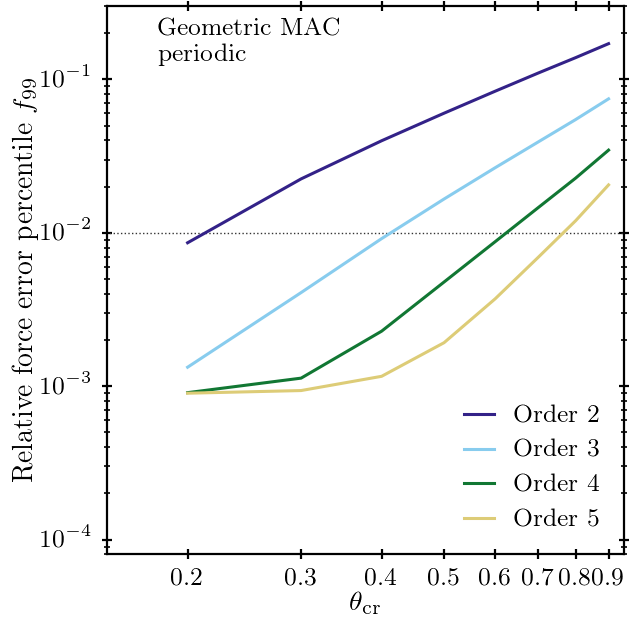}
~\,~\,~
\includegraphics[width=1.\columnwidth]{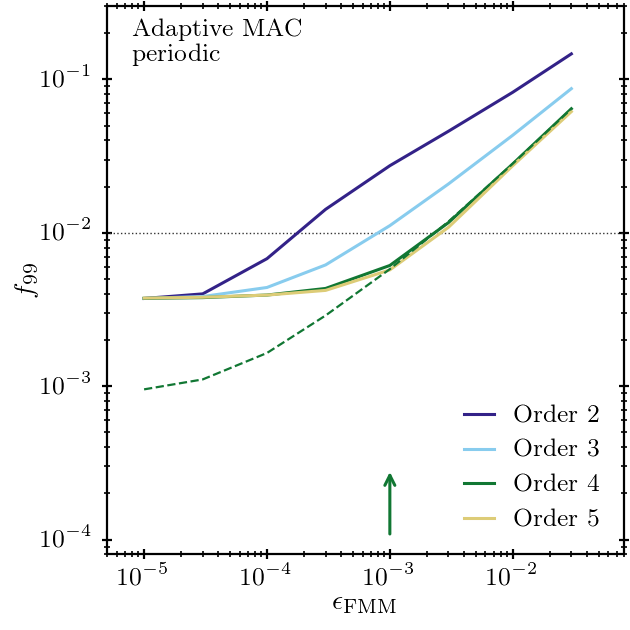}
\vspace{-0.1cm}
\caption{The same as Fig.~\ref{fig:gravity:validation_np}, but now considering
  periodic boundary conditions.  A gravity mesh of size $N_{\rm mesh}=512$ with
  $a_{\rm smooth}=1.25$ was used.  The \nth{99} percentile of the relative error
  rapidly reaches a plateau set by the accuracy of the force calculations
  computed by the PM part of the algorithm. The dashed line on the right panel
  corresponds to the order four scheme but using $a_{\rm smooth} = 3$,
  illustrating the effect of the mesh parameters on the calculation's
  accuracy. For our default setup (green arrow), the scheme reaches a relative
  force accuracy of better than $6\times10^{-3}$ for $99$ percent of the
  particles, a level only reached with very small opening angle values in the
  geometric case. }
\label{fig:gravity:validation_p}
\end{figure*}

We repeat the same exercise but with periodic boundaries switched on and display
the results in Fig.~\ref{fig:gravity:validation_p}. The FMM part of the
algorithm is unchanged, we only additionally add the PM part using a grid of
$512^3$ cells and a smoothing factor of $a_{\rm smooth} = 1.25$ (our default
value). In this case, the force error reaches a plateau for low values of the
opening angle $\theta_{\rm cr}$ or adaptive MAC parameter $\epsilon_{\rm FMM}$.
This is where the algorithm reaches the accuracy limit of the PM part of the
method. This is illustrated on the right panel by the dashed line which
corresponds to the same run but with $a_{\rm smooth} = 3$. In our default setup
(\nth{4} order FMM, $\epsilon_{\rm FMM}=10^{-3}$, $a_{\rm smooth} = 1.25$)
indicated by the green arrow, $99$ percent of the particles have a relative
force accuracy of better than $6\times10^{-3}$.

\subsection{Treatment of massive neutrinos}
\label{ssec:gravity:nu}

Accurately modelling neutrinos is of great interest for large-scale structure
simulations, due to their outsized effect on matter clustering (see
\citealt{Lesgourgues2006} for a review). We implemented two schemes for the
treatment of neutrino effects in \swift: one based on the linear response method
\citep{AliHaimoud2013} and another based on the $\delta f$ method
\citep{Elbers2020}. In terms of the total matter power spectrum they produce,
the two schemes are in good agreement.\\

\noindent The linear response method is a grid-based approach that accounts for
the presence of neutrino perturbations by applying a linear correction factor in
Fourier space to the long-range gravitational potential:
\begin{align}
\hat{\phi}_l(\mathbf{k}) = \hat{\phi}_{l,\text{cb}}(\mathbf{k}) \cdot \left[1 + \frac{f_\nu}{f_\text{cb}}\frac{\delta^\text{lin}_{\nu}(k)}{\delta^\text{lin}_\text{cb}(k)}\right],
\end{align}
where $\hat{\phi}_{l,\text{cb}}$ is the long-range gravitational potential
computed from the cold dark matter and baryon particles
(\S\,\ref{ssec:gravity:mesh_summary}).  The correction factor depends on the
ratio of linear theory transfer functions ($\delta$) for neutrinos and cold dark
matter plus baryons, as well as their relative mass fractions ($f$). \\

\noindent The second scheme, based on the $\delta f$ method, actively solves for
the neutrino perturbations. It is a hybrid approach that combines a
particle-based Monte Carlo sampling of the neutrino phase-space distribution
with an analytical background solution. The aim is to solve for the nonlinear
gravitational evolution of the neutrinos, while suppressing the shot noise that
plagues traditional particle implementations. In this method, the nonlinear
phase-space density $f$ of neutrinos is decomposed as
\begin{align}
f(\mathbf{x},\mathbf{p},t) = \bar{f}(p,t) + \delta f(\mathbf{x},\mathbf{p},t),
\end{align}
where $\bar{f}(p,t)=\left(1+\exp(p / k_{\rm B} T_\nu)\right)^{-1}$ is the
background Fermi--Dirac distribution (expressed in terms of the neutrino
temperature $T_\nu$) and $\delta f$ is a possibly non-linear perturbation. In
contrast to traditional, pure particle, implementations, only $\delta f$ is
estimated from the particles hence reducing the shot noise. To achieve this
decomposition, the contribution of neutrino particles to the mass density is
statistically weighted. The weight of particle $i$ is given by
\begin{align}
w_i = \frac{\delta f_i}{f_i} = \frac{f_i-\bar{f}_i}{f_i}, \label{eq:deltaf_weights}
\end{align}
where $f_i$ is the phase-space density at its location. Weights express the
deviation from the background, they can be positive or negative, and are ideally
small. The reduction in shot noise is proportional to $\left\langle
w^2\right\rangle$ for the neutrino power spectrum. The weights must be updated
on the fly, which involves a single loop over neutrino particles. We make use of
the fact that $\bar{f}_i$ depends only on the current particle momentum, while
the value of $f_i$ is conserved. To avoid storing $f_i$, \swift uses the
particle ID as a deterministic pseudo-random seed to sample the initial
Fermi--Dirac momentum. The value of $f_i$ is then recomputed when needed. As a
result, the memory footprint of neutrinos is identical to that of cold dark
matter particles. The neutrino particles then enter the gravity calculation
identically to all the other species but see their mass multiplied by their
weight.

The possibility of negatively weighted particles requires some attention. In
exceptional circumstances, which nevertheless occur for simulations involving
billions of particles and thousands of steps, the centre of mass of a group of
neutrinos can lie far beyond the geometric perimeter of the particles. Since
\swift uses a multipole expansion around the centre of mass, this possibility
causes a breakdown of the multipole expansion in eq.~\ref{eq:fmm:fmm_one_part},
when truncated at finite $p$. Although the multipole expansion could, in
principle, be performed around another point \citep{Elbers2020}, we instead
additionally implemented a version of the $\delta f$ method that only applies
the weights in the long-range PM gravity calculation. This choice ensures that
the spurious back-reaction of neutrino shot noise, which is most prominent on
large scales and therefore feeds through the long-range force, is eliminated,
while the possibility of neutrinos affecting smaller scales through short-range
forces is not excluded. An added benefit is that PM steps are rare for large
calculations, such that the computational overhead of the $\delta f$ step is
minimal.

In addition, the $\delta f$ weights are always used to reduce the noise in
on-the-fly power spectra and are provided in snapshots for use in post
processing.

A final point concerns the relativistic nature of neutrino particles at high
redshift. To ensure that neutrino velocities do not exceed the speed of light
and to recover the correct free streaming lengths, we apply the relativistic
correction factor $c/\sqrt{c^2+(v/a)^2}$ to neutrino drifts, where $v$ is the
internal velocity variable described in Section \ref{ssec:ccordinates} and $a$
is the scale factor. Relativistic corrections to the acceleration can be
neglected in the time frame typical for cosmological simulations
\citep{Elbers2022b}.

\section{Cosmological integration}
\label{sec:cosmo}

\subsection{Background evolution}
\label{ssec:flrw}

In \swift we assume a standard FLRW metric for the evolution of the background
density of the Universe and use the Friedmann equations to describe the
evolution of the scale-factor $a(t)$.  We scale $a$ such that its present-day
value is $a_0 \equiv a(t=t_{\rm now}) = 1$. We also define redshift $z \equiv
1/a - 1$ and the Hubble parameter
\begin{equation}
H(t) \equiv \frac{\dot{a}(t)}{a(t)},
\end{equation}
with its present-day value denoted as $H_0 \equiv H(t=t_{\rm now})$. Following
usual conventions we write $H_0 = 100
h~\rm{km}\cdot\rm{s}^{-1}\cdot\rm{Mpc}^{-1}$ and use $h$ as the input parameter
for the Hubble constant.

To allow for general expansion histories we use the full Friedmann equations
and write
\begin{align}
H(a) &\equiv H_0 E(a), \\ E(a) &\equiv\sqrt{\Omega_{\rm m} a^{-3} + \Omega_{\rm r}
  a^{-4} + \Omega_{\rm k} a^{-2} + \Omega_\Lambda \exp\left(3\tilde{w}(a)\right)},
  \label{eq:comso:Ea} \\
\tilde{w}(a) &= (a-1)w_a - (1+w_0 + w_a)\log\left(a\right),
\label{eq:friedmann}
\end{align}
where we followed \cite{Linder2003} to parameterise the evolution of the
dark-energy equation of state\footnote{Note that $\tilde{w}(z)\equiv \int_0^z
\frac{1+w(z')}{1+z'}{\rm d}z'$, which leads to the analytic expression we use.}
as:
\begin{equation}
w(a) \equiv w_0 + w_a~(1-a).
\end{equation}
The cosmological model is hence fully defined by specifying the dimensionless
constants $\Omega_{\rm m}$, $\Omega_{\rm r}$, $\Omega_{\rm k}$,
$\Omega_\Lambda$, $h$, $w_0$, and $w_a$ as well as the starting redshift (or
scale-factor of the simulation) $a_{\rm start}$ and final time $a_{\rm
  end}$. \\ At any scale-factor $a_{\rm age}$, the time $t_{\rm age}$ since the
Big Bang (age of the Universe) is computed as \citep[e.g.][]{Wright2006}:
\begin{equation}
  t_{\rm age} = \int_{0}^{a_{\rm age}} {\rm d}t = \int_{0}^{a_{\rm age}}
  \frac{{\rm d}a}{a H(a)} = \frac{1}{H_0} \int_{0}^{a_{\rm age}}
  \frac{{\rm d} a}{a E(a)}. \label{eq:flrw:age}
\end{equation}
For a general set of cosmological parameters, this integral can only be
evaluated numerically, which is too slow to be evaluated accurately during a
run. At the start of the simulation we tabulate this integral for $10^4$ values
of $a_{\rm age}$ equally spaced between $\log(a_{\rm start})$ and $\log(a_{\rm
  end})$. The values are obtained via adaptive quadrature using the 61-points
Gauss--Konrod rule implemented in the {\sc gsl} library \citep{GSL} with a
relative error limit of $\epsilon=10^{-10}$. The value for a specific $a$ (over
the course of a simulation run) is then obtained by linear interpolation of the
table.

\subsection{Addition of neutrinos}
\label{ssec:cosmo:nu}

Massive neutrinos behave like radiation at early times, but become
non-relativistic around $a^{-1}\approx1890 (m_\nu/1\text{ eV})$. This changes
the Hubble rate $E(a)$ and therefore most integrated quantities described in the
previous section. We optionally include this effect by specifying the number of
massive neutrino species $N_\nu$ and their non-zero neutrino masses $m_{\nu,i}$
in eV $(m_{\nu,i}\neq 0, i=1,\dots,N_\nu)$. Multiple species with the same mass
can be included efficiently by specifying mass degeneracies $g_i$. In addition,
the present-day neutrino temperature $T_{\nu,0}$ must also be set\footnote{To
match the neutrino density from an accurate calculation of decoupling
\citep{Mangano2005}, one can use the value
$T_{\nu,0}/T_{\mathrm{CMB},0}=0.71599$ \citep{Lesgourgues2011}.} as well as an
effective number of ultra-relativistic (massless) species
$N_\mathrm{ur}$. Together with the present-day CMB temperature
$T_{\mathrm{CMB},0}$, these parameters are used to compute the photon density
$\Omega_\gamma$, the ultra-relativistic species density $\Omega_\mathrm{ur}$,
and the massive neutrino density $\Omega_\nu(a)$, replacing the total radiation
density parameter $\Omega_{\rm r}$.  In our conventions, the massive neutrino
contribution at $a=1$ is \emph{not} included in the present-day matter density
$\Omega_{\rm{m}}=\Omega_{\rm{cdm}}+\Omega_{\rm{b}}$. The radiation term
appearing in eq. \ref{eq:comso:Ea} is simply replaced by
\begin{align}
    \Omega_{\rm r} a^{-4} &= \left[\Omega_\gamma + \Omega_\mathrm{ur} + \Omega_\nu(a)\right] a^{-4}.
\end{align}
In this expression, the constant $\Omega_\gamma$ describes the CMB density and is given by
\begin{align}
    \Omega_\gamma &= \frac{\pi^2}{15}\frac{(k_{\rm B} T_\text{CMB,0})^4}{(\hbar c)^3}\frac{1}{\rho_{\rm crit}c^2},
\end{align}
while the ultra-relativistic neutrino density is given by
\begin{align}
    \Omega_\mathrm{ur} = \frac{7}{8}\left(\frac{4}{11}\right)^{4/3} N_\mathrm{ur}\,\Omega_\gamma.
\end{align}
Note that we assume instantaneous decoupling for the ultra-relativistic
species. The time-dependent massive neutrino density parameter is
\citep{Zennaro2016}:
\begin{align}
    \Omega_\nu(a) = \Omega_\gamma \sum_{i=1}^{N_\nu}\frac{15}{\pi^4}g_i\left(\frac{T_{\nu,0}}{T_\text{CMB}}\right)^4 \mathcal{F}\left(\frac{a m_{\nu,i}}{k_{\rm B} T_{\nu,0}}\right), \label{eq:nudensity}
\end{align}
where the function $\mathcal{F}$ is given by the momentum integral
\begin{align}
    \mathcal{F}(y) = \int_0^{\infty} \frac{x^2\sqrt{x^2+y^2}}{1+e^{x}}\mathrm{d}x.
\end{align}
As $\Omega_\nu(a)$ is needed to compute other cosmological integrals, this
function should be calculated with sufficient accuracy. At the start of the
simulation, values of eq. \ref{eq:nudensity} are tabulated on a piece-wise
linear grid of $2\times3\times10^4$ values of $a$ spaced between
$\log(a_{\nu,\text{begin}})$, $\log(a_{\nu,\text{mid}})$, and
$\log(a_{\nu,\text{end}}) = \log(1)=0$. The value of $a_{\nu,\text{begin}}$ is
automatically chosen such that the neutrinos are still relativistic at the start
of the table. The value of $\log(a_{\nu,\text{mid}})$ is chosen just before the
start of the simulation. The integrals $\mathcal{F}(y)$ are evaluated using the
61-points Gauss--Konrod rule implemented in the {\sc gsl} library with a
relative error limit of $\epsilon=10^{-13}$. Tabulated values are then linearly
interpolated whenever $E(a)$ is computed.

Besides affecting the background evolution, neutrinos also play a role at the
perturbation level. These effects can be included in \swift using the linear
response method of \cite{AliHaimoud2013} or the particle-based $\delta f$ method
of \cite{Elbers2020}, as described in \S\,\ref{ssec:gravity:nu}.

\subsection{Choice of co-moving coordinates}
\label{ssec:ccordinates}

Note that, unlike many other solvers, we do not express quantities with ``little
h'' ($h$) included\footnote{See e.g. \cite{Croton2013} for a rational.}; for
instance units of length are expressed in units of $\rm{Mpc}$ and not
${\rm{Mpc}}/h$. As a consequence, the time integration operators (see below)
also include an $h$-factor via the explicit appearance of the Hubble constant.\\

\noindent In physical coordinates, the Lagrangian for a particle $i$ in an
energy-based flavour of SPH with gravity reads

\begin{equation}
  \Lag =
  \frac{1}{2} m_i \dot{\mathbf{r}}_i^2 -
  m_iu_i -
  m_i \phi_i.
\end{equation}

\noindent Introducing the comoving positions $\mathbf{r}'$ such that $\mathbf{r}
= a(t) \mathbf{r}'$, we get

\begin{equation}
  \Lag = \frac{1}{2} m_i \left(a\dot{\mathbf{r}}_i'
  + \dot{a}\mathbf{r}_i' \right)^2 - m_i\frac{u_i'}{a^{3(\gamma-1)}} - m_i \phi,
\end{equation}
where the comoving internal energy $u'=ua^{3(\gamma-1)}$ is \emph{chosen} such
that the equation of state for the gas and thermodynamic relations between
quantities have the same form (i.e. are scale-factor free) in the primed frame
as well. Together with the definition of comoving densities $\rho' \equiv
a^3(t)\rho$, this implies

\begin{equation}
  P' = a^{3\gamma}P,\quad A'=A, \quad c'=a^{3(\gamma-1)/2}c,
\end{equation}
for the pressure, entropy, and sound-speed respectively.  Following
\cite{Peebles1980} (chapter 7), we introduce the gauge transformation $\Lag
\rightarrow \Lag + \frac{d}{dt}\Psi$ with $\Psi \equiv
\frac{1}{2}a\dot{a}\mathbf{r}_i^2$ and obtain

\begin{align}
  \Lag &= \frac{1}{2}m_ia^2 \dot{\mathbf{r}}_i'^2 -
  \ m_i\frac{u_i'}{a^{3(\gamma-1)}}
  -\frac{\phi'}{a},\\
  \phi' &= a\phi + \frac{1}{2}a^2\ddot{a}\mathbf{r}_i'^2,\nonumber
\end{align}
and call $\phi'$ the peculiar potential.  Finally, we introduce the velocities
used internally by the code:

\begin{equation}
  \mathbf{v}' \equiv a^2\dot{\mathbf{r}'},
\end{equation}
allowing us to simplify the first term in the Lagrangian.  Note that these
velocities \emph{do not} have a direct physical interpretation. We caution that
they are not the peculiar velocities ($\mathbf{v}_{\rm p} \equiv
a\dot{\mathbf{r}'} = \frac{1}{a}\mathbf{v}'$), nor the Hubble flow
($\mathbf{v}_{\rm H} \equiv \dot{a}\mathbf{r}'$), nor the total velocities
($\mathbf{v}_{\rm tot} \equiv \mathbf{v}_{\rm p} + \mathbf{v}_{\rm H} =
\dot{a}\mathbf{r}' + \frac{1}{a}\mathbf{v}'$) and also differ from the
convention used in outputs produced by \gadget \citep{Springel2005,
  Springel2021} and other related simulation codes ($\mathbf{v}_{\rm out,Gadget}
= \sqrt{a} \dot{\mathbf{r}'}$)\footnote{One inconvenience of our choice of
generalised coordinates is that our velocities $\mathbf{v}'$ and sound-speed
$c'$ do not have the same dependencies on the scale-factor. The signal velocity
entering the time-step calculation will hence read $v_{\rm sig} =
a\dot{\mathbf{r}'} + c = \frac{1}{a} \left( |\mathbf{v}'| + a^{(5 -
  3\gamma)/2}c'\right)$.}.

\subsubsection{SPH equations}

Using the SPH definition of density, $\hat{\rho}_i' =
\sum_jm_jW(\mathbf{r}_{j}'-\mathbf{r}_{i}',h_i') = \sum_jm_jW_{ij}'(h_i')$, we
follow \cite{Price2012} and apply the Euler-Lagrange equations to write

\begin{alignat}{3}
  \dot{\mathbf{r}}_i'&= \frac{1}{a^2} \mathbf{v}_i'&  \label{eq:cosmo_eom_r}, \\
  \dot{\mathbf{v}}_i' &= -\sum_j m_j &&\left[\frac{1}{a^{3(\gamma-1)}}f_i'P_i'\hat{\rho}_i'^{-2}\mathbf{\nabla}_i'W_{ij}'(h_i)\right. \nonumber\\
  &   && + \left. \frac{1}{a^{3(\gamma-1)}}f_j'P_j'\hat{\rho}_j'^{-2}\mathbf{\nabla}_i'W_{ij}'(h_j)\right. \nonumber\\
  &   && + \left. \frac{1}{a}\mathbf{\nabla}_i'\phi'\right], \label{eq:cosmo_eom_v}
\end{alignat}
with
\begin{equation}
    f_i' = \left[1 + \frac{h_i'}{3\rho_i'}\frac{\partial
      \rho_i'}{\partial h_i'}\right]^{-1}, \qquad \mathbf{\nabla}_i'
  \equiv \frac{\partial}{\partial \mathbf{r}_{i}'}. \nonumber
\end{equation}
These correspond to the equations of motion for density-entropy SPH
\citep[e.g. eq. 14 of][]{Hopkins2013} with cosmological and gravitational
terms. Similarly, the equation of motion describing the evolution of $u'$ is
expressed as:

\begin{equation}
  \dot{u}_i' = \frac{1}{a^2}\frac{P_i'}{\hat{\rho}_i'^2} f_i'\sum_jm_j\left(\mathbf{v}_i' -
    \mathbf{v}_j'\right)\cdot\mathbf{\nabla}_i'W_{ij}'(h_i).
  \label{eq:cosmo_eom_u}
\end{equation}
In all these cases, the scale-factors appearing in the equations are later
absorbed in the time-integration operators such that the RHS of the equations of
motions is identical for the primed quantities to the ones obtained in the
non-cosmological case for the physical quantities. Additional terms in the SPH
equations of motion (e.g. viscosity switches) often rely on the velocity
divergence and curl. We do not give a full derivation here but the co-moving
version of all these terms can easily be constructed following the same
procedure we employed here.

\subsection{Time-integration operators}
\label{ssec:operators}
For the choice of cosmological coordinates made in \swift, the normal
\emph{kick} and \emph{drift} operators get modified to account for the expansion
of the Universe. The rest of the leapfrog algorithm is identical to the
non-comoving case. The derivation of these operators from the system's
Lagrangian is given in appendix A of \cite{Quinn1997} for the collisionless
case. We do not repeat that derivation here but, for completeness, give the
expressions we use as well as the ones used for the hydrodynamics.  The drift
operator gets modified such that $\Delta t$ for a time-step running from a
scale-factor $a_{n}$ to $a_{n+1}$ becomes

\begin{equation}
  \Delta t_{\rm drift} \equiv \int_{a_n}^{a_{n+1}} \frac{{\rm d}t}{a^2} = \frac{1}{H_0} \int_{a_n}^{a_{n+1}} \frac{{\rm d}a}{a^3E(a)},
\end{equation}
with $E(a)$ given by eq.~\ref{eq:friedmann} and the $a^{-2}$ chosen to absorb
the one appearing in eq.~\ref{eq:cosmo_eom_r}.  Similarly, the
time-step-entering kick operator for collisionless acceleration reads
\begin{equation}
  \Delta t_{\rm kick,g} \equiv \int_{a_n}^{a_{n+1}} \frac{{\rm d}t}{a} = \frac{1}{H_0} \int_{a_n}^{a_{n+1}} \frac{{\rm d}a}{a^2E(a)}.
\end{equation}
However, for the case of gas dynamics, given our choice of coordinates, the kick
operator has a second variant that reads
\begin{equation}
  \Delta t_{\rm kick,h} \equiv \int_{a_n}^{a_{n+1}} \frac{{\rm d}t}{a^{3(\gamma-1)}} = \frac{1}{H_0} \int_{a_n}^{a_{n+1}} \frac{{\rm d}a}{a^{3\gamma - 2}E(a)}.
\end{equation}
Accelerations arising from hydrodynamic forces (\nth{1} and \nth{2} term in
eq.~\ref{eq:cosmo_eom_v}) are integrated forward in time using $\Delta t_{\rm
  kick,h}$, whilst the accelerations given by the gravity forces (\nth{3} term
in eq.~\ref{eq:cosmo_eom_v}) use $\Delta t_{\rm kick,g}$. The internal energy
(eq. \ref{eq:cosmo_eom_u}) is integrated forward in time using $\Delta t_{\rm
  kick,u} = \Delta t_{\rm drift}$.
 

Following the same method as for the age of the Universe (\S \ref{ssec:flrw}),
these three non-trivial integrals are evaluated numerically at the start of the
simulation for a series $10^4$ values of $a$ placed at regular intervals between
$\log \left( a_{\rm begin}\right)$ and $\log \left(a_{\rm end}\right)$. The
values for a specific pair of scale-factors $a_n$ and $a_{n+1}$ are then
obtained by interpolating that table linearly.

\subsection{Validation}
\label{ssec:cosmology:validation}

To assess the level of accuracy of \swift, it is important to compare results
with other codes. This lets us assess the level of systematic differences and
uncertainties left in the code. This is especially important for the studies of
non-linear structure formation, as there is no possibility to use an exact
solution to compare against. One such benchmark was proposed by
\cite{Schneider2016} in the context of the preparation for the \simulationname{Euclid}
survey. Their goal was to assess whether cosmological codes can converge towards
the same solution, within the targeted $1$ percent accuracy of the survey. They
focused on the matter density power spectrum as their observable and used three
different \nbody codes for their study. Importantly, their work utilised three
codes using three different algorithms to solve for the gravity forces:
\codename{Ramses} \citep[][multi-grid technique]{Ramses}, \codename{Pkdgrav3}
\citep[][FMM tree algorithm]{pkdGrav3}, and \codename{Gadget-3} \citep[][tree-PM
  technique]{Springel2005}. The setup evolves a cosmological simulation in a
$(500~{\rm{Mpc}}/h)^3$ volume from $z=49$ to $z=0$, assuming a $\Lambda$CDM
cosmology, sampled using $2048^3$ particles. The setup only considers
gravitational interactions and comoving time integration. The same setup was
later adopted by \cite{Garrison2019} to compare their \codename{Abacus} code and
by \cite{Springel2021} for the \codename{Gadget-4} code\footnote{We thank Lehman
Garrison and Volker Springel for graciously providing their data and analysis
tools.}. It is a testimony to the advances of the field in general and to the
increase in available computing power that a run akin to the
then-record-breaking \simulationname{Millennium} simulation \citep{Millennium} is nowadays
used as a mere benchmarking exercise.\\

\begin{figure}
\centering
\includegraphics[width=0.95\columnwidth]{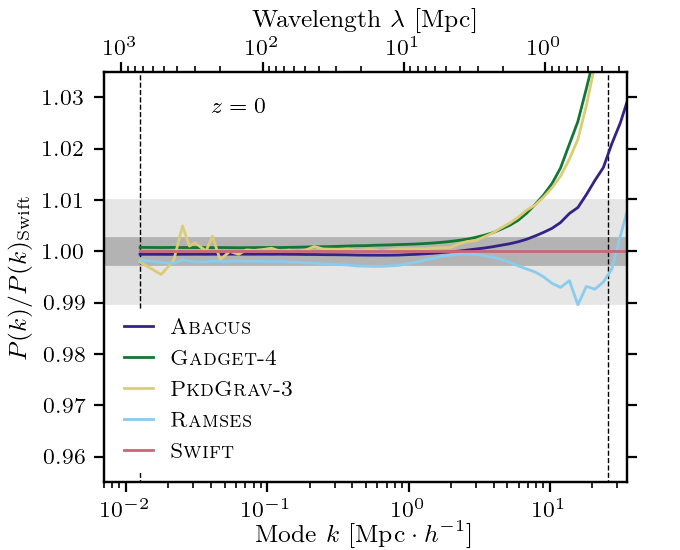}
\vspace{-0.1cm}
\caption{ Comparison of the matter power-spectra as a function of scale for four
  different $n$-body codes (see text) relative to the \swift prediction on the
  test problem introduced by \citet{Schneider2016}. The simulation evolves
  $2048^3$ dark matter particles in a $(500~{\rm{Mpc}}/h)^3$ volume run from
  $z=49$ to $z=0$ assuming a $\Lambda$CDM cosmology. All power spectra were
  measured using the same tool (see text). The dark- and light-shaded regions
  correspond to $\pm 0.25\%$ and $\pm 1\%$ level agreement between codes.  The
  fundamental mode (left) and the Nyquist frequency (right) are indicated using
  vertical dashed lines. Over the range of interest for modern cosmological
  applications, all codes agree to within $1\%$.}
\label{fig:cosmology:validation}
\end{figure}

We ran \swift on the same initial conditions and analysed the results as
described below. The exact configuration used for the \swift run is released as
part of the code package, namely: a $2048^3$ gravity mesh for the PM code, the
adaptive MAC with $\epsilon_{\rm FMM} = 10^{-3}$, and a Plummer-equivalent
softening length $\epsilon = 10/h~{\rm kpc}$. The top-left panel of
Fig.~\ref{fig:publicity} shows the projection of the matter density field in a
$10~{\rm Mpc}/h$ slice rendered using the \swiftsimio tool
\citep{swiftsimio}. To ease the comparison to published results, and eliminate
any possible discrepancy coming from binning choices or exact definitions, we
used the power-spectrum measurement tool embedded in the \codename{Gadget}-4
code on our output to allow for a direct comparison with the data presented by
\cite{Springel2021} (who had also reanalysed the other runs with their tool).
We show our results alongside the published measurements from other codes in
Fig.~\ref{fig:cosmology:validation}, each presented as ratios to the \swift
prediction. The shaded regions correspond to $\pm 0.25$ percent and $\pm 1$
percent differences with respect to our results. Over the range of wavelengths
of interest for this problem, the \swift results are in excellent agreement with
the other codes. This agreement extends from the linear regime to the non-linear
regime ($k \gtrsim 0.1 \rm{Mpc} / h$). This confirms \swift's ability to make
solid predictions for modern cosmological applications.~\\

\noindent Note also that a similar exercise was independently presented by
\cite{Grove2021} in the context of the \simulationname{DESI} survey code comparison effort, for
which \swift, \codename{Abacus}, and \codename{Gadget-2} were compared. Comparing
outputs at $z=1$ and $z=2$, they obtained results in excellent agreement with
the ones presented here.

\section{Input \& Output strategy}
\label{sec:io}

We now turn our attention towards the input and output strategy
used by the \swift code.

\subsection{Initial Conditions}
\label{ssec:io:ics}

To ease the use of the code and given the large number of legacy initial
conditions (ICs) existing in the community using this format, we adopt the same
file format for input as the ``mode 3'' option of the \gadget-2 code
\citep{Springel2005}, i.e. the mode based on the \libraryname{hdf5} library
\citep{hdf5}.  \swift is fully compatible with any valid \gadget-2 set of
initial conditions, but we also provide additional optional features. Firstly,
we allow for different units to be used internally and in the ICs. \swift would
then perform a conversion upon start-up to the internal units. This can be
convenient when a certain set of ICs uses a range of values problematic when
represented in single-precision. Secondly, for cosmological runs, \swift can
also apply the necessary $h$-factor and $a$-factor corrections (see
\S\,\ref{ssec:ccordinates}) to convert to the system of co-moving coordinates
adopted internally.  A departure from the strict \gadget-2 format is that \swift
only allows for the data to be distributed over a single file; we do, however,
provide scripts to transform such distributed input files to our format.

Some tools also exist to directly generate \swift ICs with all the optional
features added. The \swiftsimio
\footnote{\url{https://github.com/SWIFTSIM/swiftsimio}} python package
\citep{swiftsimio} can be used to generate simple setups. The
\seagen \footnote{\url{https://github.com/jkeger/seagen}} \citep{Kegerreis2019}
and \woma\footnote{\url{https://github.com/srbonilla/WoMa}} \citep{RuizBonilla2021}
packages are designed to generate spherical or spinning planetary bodies in
equilibrium for collision problems (See sec. \ref{ssec:planetary}). For
cosmological simulations, the public version of the state-of-the-art ICs code
\monofonic
\citep{Michaux2021,Hahn2021} has been extended to be able to produce files that
are directly compatible with the format expected by \swift. In particular,
information about the adopted cosmological parameters, phases, and all the
information required to re-generate the ICs are added to the files, read by
\swift, and propagated to the snapshots. This allows for runs to be reproduced
based solely on the information given in the \swift outputs.

\subsection{Snapshots}
\label{ssec:io:snaps}

For the same convenience reasons as for the ICs, we also adopt an output file
format designed as a fully-compatible extension to the \gadget-2
\citep{Springel2005} ``mode 3'' format based on the \libraryname{hdf5} library
\citep{hdf5}.  We extend the format by creating new particle groups for the
species not existing in the original \gadget-2 code. We also add to the
snapshots a full copy of the parameters used to perform the simulation,
information about the version of the code, details of the cosmological models,
and information about the ICs. Another noteworthy extension is the extensive use
of units metadata in the snapshots. We attach full units information to every
field in the snapshots. That information includes human-friendly and
machine-readable conversion factors to the cgs system, as well as the conversion
factor needed to move between the co-moving and physical frame (See
sec. \ref{ssec:ccordinates}).  These metadata can be read by python packages
such as \swiftsimio \citep{swiftsimio} to then propagate this information
through the simulation analysis. This mechanism is based on the \libraryname{unyt}
\citep{unyt} library. The particles are stored in the snapshots in order of the
domain cells they belong to (See \S\,\ref{ssec:parallel:cells}).  Efficiently
retrieving the particles located in a small sub-region of the computational
domain is hence possible; for instance extracting the particles in the region
around a single halo only. In large simulations, this is much more efficient
than reading all the randomly ordered particles and then masking out the ones
that do not fall in the region of interest. Metadata to ease such reading
patterns are added to the snapshots. That information is picked up by tools such
as \swiftsimio to aid analysis of these massive simulations. The commonly used
visualisation package \libraryname{yt}\footnote{\url{https://yt-project.org/}} 
\citep{yt} has also been extended to directly read in \swift snapshots, including
the relevant meta-data.

The snapshots can either be written into one single file, with all nodes writing
collectively to the same dataset in parallel, or by splitting the data such that
each node writes a file with its local subset of particles. That second option
is preferable when using file systems that are not able to handle parallel
writes to a single file efficiently. When writing such a distributed snapshot,
an additional meta-snapshot is written; it contains all the information of a
regular single-file snapshot, but uses \libraryname{hdf5}'s virtual dataset
infrastructure to present the data distributed over many files as a single
contiguous array. The links between files are handled in the background by the
library. These meta-snapshots can then be read as if they were standard
snapshots, for instance via tools like
\libraryname{Gadgetviewer}\footnote{\url{https://github.com/jchelly/gadgetviewer/}}.
\swift can also optionally apply lossless compression to the snapshots (via
\libraryname{hdf5}'s own \libraryname{gzip} filter) as well as a per-field lossy
compression where the number of bits in the mantissa of the numbers can be
reduced to save disk space. This option is particularly interesting when
considering particle fields where the 23 bits of relative precision
(i.e. $\approx 7$ decimal digits) of a standard \techjargon{float} type are more
than sufficient for standard analysis\footnote{Classic examples are the
temperature field or the particles' metallicity.}. Similar filters can be
applied to \techjargon{double}-precision variables. Finally, \swift implements an
option to down-sample the particles of a given type in the snapshots by writing
only a fraction of the particles chosen at random.

As an example of i/o performance in a realistic scenario, the snapshots for the
recent flagship \flamingo run \citep{Schaye2022} were written in 200
seconds. They contain $2.65\times10^{11}$ particles of different types spread
over $960$ files totalling $39$ terabytes of data. This corresponds to a writing
speed of $200~\rm{GB}/\rm{s}$.  As this test only used 65\% of the systems'
nodes, this compares favourably to the raw capability ($350~\rm{GB}/\rm{s}$) of
the full cluster. Compressing the data using both lossy and lossless filters
reduces the snapshot size to $11$ terabytes but the writing time increases to
$1260$ seconds. This corresponds to a sustained writing speed of
$9~\rm{GB}/\rm{s}$; the difference is due to the compression algorithm embedded
within the \libraryname{hdf5} library. Additionally, by making use of the library's
parallel writing capability, we can repeat the uncompressed test but with all
nodes writing to a single file. In this configuration, we require $463$ seconds,
effectively achieving a sustained parallel writing speed of $86~\rm{GB}/\rm{s}$.

Snapshots can be written at regular intervals in time or change in scale-factor.
Alternatively, the user can provide a list of outputs in order to specify output
times more precisely. This list can be accompanied by a list of fields (or of
entire particle types) the user does not want to be written to a snapshot. This
allows for the production of reduced snapshots at high-frequency; for instance
to finely track black holes. Any of the structure finders (\S\,\ref{sec:stf})
can be run prior to the data being written to disk to include halo membership
information of the particles in the outputs.

\subsection{Check-pointing mechanism}
\label{ssec:io:restart}

When running simulations at large computing centres, limits on the length of a
given compute job are often imposed. Many simulations will need to run for
longer than these limits and a mechanism to cleanly stop and resume a simulation
is thus needed. This \emph{check-pointing} mechanism can also be used to store
backups of the simulation's progress in case one needs to recover from a
software or hardware failure. Such a mechanism is different from the writing of
science-ready snapshots as all the information currently in the memory needs to
be saved; not just the interesting fields carried by the particles. These
outputs are thus typically much larger than the snapshots and are of the same
size as the memory used for the run.

In \swift, we choose to write one file per \techjargon{MPI} rank. No
pre-processing of any kind is done during writing. Each of the code's modules
writes its current memory state one after the other. This includes the raw
particle arrays, the cells, the tasks, and the content of the extensions (see
\S~\ref{sec:extensions}) among many other objects. At the start each module's
writing job we include a small header with some information about the size of
the data written. This allows us to verify that the data was read in properly
when resuming a simulation. As these are simple, unformatted, large, and
distributed writing operations, the code typically achieves close to the maximal
writing speed of the system. For the same \flamingo run mentioned above, the
whole procedure took $260~\rm{s}$ for 64 TB of data in $960$ files. This
corresponds to a raw writing speed of $250~\rm{GB/s}$. As the check-pointing is
fast, it is convenient to write files at regular intervals (e.g. every few
hours) to serve as a backup.

When restarting a simulation from a check-point file, the opposite operation is
performed. Each rank reads one file and restores the content of the memory. At
this point, the simulation is in exactly the same state as it was when the files
were written. The regular operations can thus resume as if no stoppage and
restarting operation had ever occurred.

As is the case in many software packages, our implementation is augmented with a
few practical options such as the ability to stop an on-going run or to ask the
simulation to run for a set wall-clock time before writing a check-point file
and stopping itself.

\subsection{Line-Of-Sight outputs}
\label{ssec:los}

In addition to full-box snapshots, \swift can also produce so-called
\emph{line-of-sight} outputs. Randomly-positioned rays (typically perpendicular
to a face) are cast through the simulation volume and all gas particles whose
volumes are crossed by the infinitely thin rays are stored in a list. We then
write all the properties of these particles for each ray to a snapshot with a
format similar to the one described above but much reduced in volume. These
outputs can then be used to produce spectra via tools such as
\codename{SpecWizard} \citep{SpecWizard1,SpecWizard2}. Thanks to their small data
footprints, these line-of-sight snapshots are typically produced at high time
frequencies over the course of a run. This type of output is particularly
interesting for simulations of the IGM and Lyman-$\alpha$ forest (See
\S\,\ref{ssec:qla}).

\subsection{Lightcone outputs}
\label{ssec:io:lightcones}

To bring the cosmological simulation outputs closer to observation mock
catalogs, \swift implements two separate mechanisms to record information as
particles cross the past light cone of a selection of observers placed in the
simulation box. The first mechanism writes the particles to disk as they reach a
distance from the observer corresponding to the light-travel distance of the
look-back time to the outputs. The second mechanism accumulates particle
information in redshift shells onto pixels to directly construct maps as the
simulation runs. See the Appendix of \cite{Schaye2022} for a detailed use case
of both these mechanisms.

\subsubsection{Particle data}

The position of each observer, the redshift range over which light-cone particle
output will be generated, and the opening angle of the cone are specified at run
time. At each time-step we compute the earliest and latest times that any
particles could be integrated forward to and the corresponding co-moving
distances. This defines a shell around each observer in which particles might
cross the past light cone as a result of drift operations carried out during
this time-step. An additional boundary layer is added to the inside of the shell
to account for particles that move during the time-step and assuming that they
have sub-luminal speeds.

For simulations employing periodic boundary conditions, we must additionally
output any periodic copy of a particle which crosses the observer's light
cone. We therefore generate a list of all periodic copies of the simulation
volume that overlap the shell around the observer. Then, whenever a particle is
moved, we check every periodic copy for a possible overlap with any of the
shells. If so, the particle's position is interpolated to the exact redshift at
which it crossed the lightcone and the particle is added to a writing
buffer. When the buffer reaches a pre-defined size, we write out the particles
including all their properties to disk.

To optimise the whole process, we take advantage of the way that \swift
internally arranges the particles in a cubic grid of cells
(\S\,\ref{ssec:parallel:cells}). We can use this structure to identify which
tree cells overlap with the current lightcone shells. This allows us to reduce
the number of periodic replications to check for every particle. Only the
particles in the cells previously identified need to undergo this process.

In most cases, the raw data generated by the particle lightcone requires some
post-processing; for instance to reorganise the particles inside the files in
terms of angular coordinates on the sky and redshift. 

\subsubsection{\healpix maps}

Light-cone particle outputs as well as the internal memory requirement rapidly
grow in size as the upper redshift limit is increased, especially if many box
replications occur, and can become impractical to store. \swift therefore also
contains a scheme to store spherical maps of arbitrary quantities on the light
cone with user specified opening angle, angular resolution, and redshift bins.

To this end, the observer's past light cone is split into a set of concentric
spherical shells in co-moving distance. For each shell we create one full-sky
\healpix \citep{Gorski2005} map for each quantity to be recorded. Whenever a
particle is found to have entered one of these shells, we accumulate the
particles' contributions to the \healpix maps for that shell. Typical examples
are the construction of mass or luminosity maps. Particles can also, optionally,
be smoothed onto the maps using an SPH kernel.

As the maps do not overlap in redshift, it is not necessary to store all of the
shells simultaneously in memory. Each map is only allocated and initialised when
the simulation first reaches the time corresponding to the outer edge of the
shell. It is then written to disk and its memory freed once all the particles
have been integrated to times past that corresponding to the light travel time
to the inner edge of the shell. In practice, the code will hence only need to
have a maximum of two maps in memory at any point in time.

\subsection{On-the-fly Power Spectra}
\label{ssec:io:ps}

Finally, \swift can compute a variety of auto- and cross- power
spectra at user-specified intervals. These include the mass density in
different particle species (and combinations thereof) as well as the
electron pressure. For the neutrino density, we also implement the
option to randomly select one half of the particles only or the
other. This helps reduce the shot-noise by computing a cross-spectrum
between the two halves.

The calculation is performed on a regular grid (usually of size $256^3$ and
hence allowing for the Fourier transform to be performed on a single node). 
Foldings \citep{Jenkins1998} are used to extend the range probed to smaller
scales with a typical folding factor of $4$ between iterations. Different window
functions from nearest-grid-point, to CIC, to triangular-shaped-clouds can be
used and are compensated for self-consistently \citep[see
e.g.][]{Colombi2009}. This could easily be extended to higher-order schemes and
to more particle properties.

\subsection{Continuous non-blocking adaptive output strategy}
\label{ssec:io:csds}

In \swift we also include a novel output strategy called the \emph{Continuous
Simulation Data Stream} (CSDS), described by \cite{Hausammann2022}. The key
principles are summarised here \citep[for related ideas, see][]{Faber2010,
  Rein2017}.

In classic output strategies (\S\,\ref{ssec:io:snaps}), the simulation is
stopped at fixed time intervals and the current state of the system is written
to disk, similar to the frames of a movie. This is an expensive operation where
all the compute nodes suddenly stop processing the physics and instead put an
enormous stress on the communication network and file-system. During these
operations, the state of the system is not advanced, leading to an overall loss
in performance as the whole simulation has to wait until the i/o operations have
completed. Furthermore, in simulations with deep time-step hierarchies, only few
particles are active on most steps, with most particles just drifting
forward. In a cosmological context, a large fraction of the particles have
fairly simple trajectories, barely departing from \nth{1}- or \nth{2}-order
perturbation theory tracks. Only the small fraction of particles deep inside
haloes follow complex trajectories. For the first group of particles,
simulations typically have more snapshots than necessary to trace them, whilst
for the second group, even one thousand snapshots (say) over a Hubble time may
not be sufficient to accurately re-create their trajectory. It is hence natural
to consider a more adaptive approach.

The CSDS departs from the snapshot idea by instead creating a database of
updates. At the start of a simulation an entry is written for each particle. We
then start the simulation and progress the particles along. In its simplest
form, the CSDS then adds an entry for a particle to the database every few
($\sim10$) particle updates. As the writing is done on a particle-by-particle
basis, it can easily be embedded in the tasking system. Writing is no longer a
global operation where the whole simulation stops; rather updates are made
continuously. By writing an update every few particle steps, the trajectory of
each particle is, by construction, well-sampled, irrespective of whether it is
in a very active region (e.g. haloes) or not (e.g. in voids).  With this
mechanism, particles outside of structures can have as little as two entries
(start time and end time of the simulation) whilst some particles will have
thousands of entries. Since the time-step size of a particle is designed to
correctly evolve a particle, relying on this information to decide when to write
a database entry guarantees that the particles' evolution can later be
faithfully recreated.  Each entry for a particle contains a pointer to the
previous entry such that particles can easily be followed in time.

An improved version of this approach would be to write a database entry every
time a particle field has changed by some pre-defined fraction
$\varepsilon$. This is an important philosophical change; instead of creating
frames at fixed intervals, we can demand that the evolution of \emph{any}
quantity be reconstructed to some accuracy from the output and get the CSDS to
create the individual particle entries at the required times. The somewhat
arbitrary choice of time interval between snapshot is hence replaced by an
objective accuracy threshold.

This database of particle updates allows for many new simulation analysis
options. The trajectory and evolution of any particle can be reconstructed to
the desired accuracy; that is we have all the information for a high
time-resolution tracking of all the objects in a run. The first use is to
produce classic snapshots at any position in time. We simply interpolate all the
particle entries to that fixed time. But, one can also demand to construct
slices in space-time, i.e. a light-cone from the output. New possibilities
arising from this new output format will undoubtedly appear in the future. Tools
to perform the basic operations described here are part of the CSDS package
linked to \swift. The tools, and most of the analysis performed thus far, are
currently focused on dark-matter simulations, but we expect to extend this to
more complex scenarios in the future.

\section{Structure finding}
\label{sec:stf}

\subsection{Friends-Of-Friends group finder}
\label{ssec:stf:fof}

The classic algorithm to identify structures in simulations is
\emph{Friends-Of-Friends} \citep[FOF, see e.g.][]{Davis1985}. Particles are
linked together if they are within a fixed distance (linking length) of each
other. Chains of links form groups, which in a cosmological context are
identified as haloes. For a linking length of $0.2$ of the mean inter-particle
separation, the haloes found are close (by mass) to the virialised structures
identified by more sophisticated methods. The FOF method falls into the wider
class of \emph{Union-Find} algorithms \citep{UnionFind} and very efficient
implementations have been proposed over the last decade for a variety of
computing architectures \citep[e.g.][]{Creasey2018}.

The implementation in \swift is fully described by \cite{Willis2020}. In brief,
the algorithm operates on a list of disjoint sets. The \emph{Union} operation
merges two sets and the \emph{Find} operation identifies the set a given element
resides in. Initially, each set contains a single element (one particle), which
plays the role of the set identifier. The algorithm then searches for any two
pairs of particles within range of each other. When such a pair is identified,
the Find operation is used to identify which set they belong to. The Union
operation is then performed to merge the sets if the particles do not already
belong to the same one. To speed-up the pair-finding process, we use the same
base principles as the ones discussed in \S\,\ref{sec:design}. More precisely,
by using the linking length as the search radius, we can construct a series of
nested grids down to that scale. The search for links between particles can then
be split between interactions within cells and between pairs of neighbouring
cells. The tasking infrastructure can then be used to distribute the work over
the various threads and nodes. When running a simulation over multiple compute
nodes, the group search is first performed locally, then fragments of groups are
merged together across domains in a second phase. This is however very different
from other particle-particle interactions like the ones used for
e.g. hydrodynamics, where the interactions are performed simultaneously,
i.e. strictly within a single phase. Additional optimisations are described by
\cite{Willis2020}, alongside scaling results demonstrating excellent strong and
weak scaling of the implementation.

Structures identified via \swift's internal FOF can either be used to seed black
holes (see \S\,\ref{ssec:eagle:agn}) or be written as a halo or group catalogue
output. Additionally, the FOF code can be run as stand-alone software to
post-process an existing snapshot and produce the corresponding group catalogue.

\subsection{Coupling to \velociraptor}
\label{ssec:stf:vr}

Many algorithms have been proposed to identify bound structures and
sub-structures inside FOF objects \citep[for a review, see][]{Knebe2013}. Many
of them can be run on simulation snapshots in a post-processing phase. However,
that is often inefficient as it involves substantial i/o work. In some cases, it
can also be beneficial to have access to some of the (sub-)halo membership
information of a particle inside the simulation itself. For these reasons, the
\swift code contains an interface to couple with the \velociraptor code
\citep{Elahi2011, Elahi2019}. \velociraptor uses phase-space information to
identify structures using a 6D FOF algorithm. An initial 3D FOF is performed to
identify haloes, however, this process may artificially join haloes together via
a single particle, which is known as a {\it particle bridge}. These haloes are
split apart by running a 6D FOF to identify particle bridges based upon their
velocity dispersion. Large mergers are then identified in an iterative search
for dense phase-space cores. Gravitationally unbound particles can optionally be
removed from the identified structures. Such a substructure algorithm has the
advantage over pure configuration-space algorithms of being able to identify
sub-haloes deep within a host halo, where the density (or potential) contrasts
relative to the background are small.

Over the course of a \swift run, the \velociraptor code can be invoked to
identify haloes and sub-haloes.  To this end, the public version of the
structure finder was modified to be used as a library. At user-specified
intervals (typically at the same time as snapshots), \swift will create a copy
of the particle information and format it to be passed to \velociraptor. This
process leads to some duplication of data but the overheads are small as only a
small subset of the full particle-carried information is required to perform the
phase-space finding. This is particularly the case for simulations which employ
a full galaxy-formation model, where particles carry many additional tracers
irrelevant to this process.

When the structure identification is completed, the list of structures and the
particle membership information is passed back from the library to \swift. This
information can then either be added to snapshots or be acted upon if any of the
sub-grid models so require.

\begin{figure}
\includegraphics[width=\columnwidth]{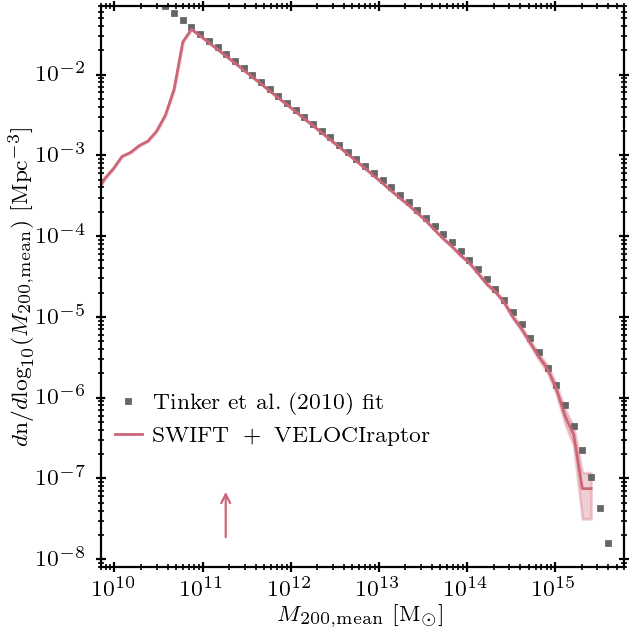}
\caption{The halo mass function, computed using \velociraptor as the structure
finder, extracted from the benchmark cosmological simulation of \citet{Schneider2016} 
run with \swift (See \S\,\ref{ssec:cosmology:validation}) and compared with the 
fitting function of  \citet{Tinker2010}. The shaded region depicts the $1-\sigma$ 
Poisson errors on the counts, while the arrow indicates the mass corresponding to 
$100$ particles. 
}
\label{fig:stf:hmf}
\end{figure}

As an example, we ran \swift with \velociraptor halo finding on the benchmark
simulation of \cite{Schneider2016} introduced in
\S\,\ref{ssec:cosmology:validation}. The resulting halo mass function is shown
on Fig.~\ref{fig:stf:hmf} alongside the reference fitting function of
\citet{Tinker2010} for the same cosmology. Our results are in excellent
agreement with the predictions from the literature.

\section{Extensions}
\label{sec:extensions}

Besides the coupled hydrodynamics and gravity solver, the \swift code also
contains a series of extensions. These include complete galaxy formation models,
AGN models, multi-material planetary models, and a series of external
potentials. These features are briefly summarised over the next pages.

\subsection{The {\bfseries \scshape Swift-Eagle} galaxy formation model}
\label{ssec:eagle}

An implementation of an evolution of the sub-grid models used for the \eagle
project \citep{Schaye2015, Crain2015} is part of the \swift code. The model is
broadly similar to the original \gadget-based implementation but was improved in
several areas. Some of these changes also arose from the change of SPH flavour
from a pressure-based formulation \citep[see][for the version used in \eagle]
{Schaller2015} to the \sphenix energy-based flavour tailored specifically for
galaxy formation simulations (\S\,\ref{ssec:sph:sphenix}).  We summarise here
the main components of the model. All the parameters presented below have values
that can be adjusted for specific simulation campaigns and are stored in
parameter files that \swift reads in upon startup. The example parameter files
provided in the \swift repository contain the parameter values for this model
that were obtained via the calibration procedure of
\citet{Borrow2022calibration}.

\subsubsection{Radiative cooling and heating}
\label{ssec:eagle:cooling}

The radiative cooling and heating rates are pre-calculated on an
element-by-element basis given the element abundance of each particle. The gas
mass fractions of H, He, C, N, O, Ne, Mg, Si, and Fe are explicitly tracked in
the code and directly affected by metal enrichment, while the abundance of S and
Ca is assumed to scale with the abundance of Si using solar abundance
ratios. \swift can use the tabulated cooling rates from
\cite{Wiersma2009cooling} (W09) for optically thin gas from the original \eagle
runs, as well as the various public tables from \cite{Ploeckinger2020}
(PS20). Compared to W09, the PS20 tables are computed with a more recent version
of Cloudy: \techjargon{c07} \citep{Ferland1998} in W09 and \techjargon{c17}
\citep{Ferland2017} in PS20, use an updated version of the UV and X-ray
background (\citet{HM2001} in W09 and a background based on \citet{FG2020} in
PS20) and include physical processes relevant for optically thick gas, such as
cosmic rays, dust, molecules, self shielding, and an interstellar radiation
field.

\subsubsection{Entropy floor and star formation}
\label{ssec:eagle:sf}

In typical \eagle-like simulations, the resolution of the model is not
sufficient to resolve the cold dense phase of the ISM, its fragmentation, and
the star formation that ensues. We hence implement an entropy floor following
\cite{Schaye2008}, which is typically set with a normalisation of 8000~K at a
density of $n_{\rm{H}} = 0.1~\rm{cm}^{-3}$ with a slope expressed by the
equation of state for pressure as $P \propto \rho^{4/3}$.

The star formation model uses the pressure-law model of \cite{Schaye2008} which
relates the star formation rates to the surface density of gas.  Particles are
made eligible for star formation based on two different models. The first one
follows \eagle and uses a metallicity-dependent density threshold based on the
results of \cite{Schaye2004}. The second model exploits the
\cite{Ploeckinger2020} tables. By assuming pressure equilibrium, we find the
density and temperatures on the thermal equilibrium curve for the particles
limited by the entropy floor. A combination of density and temperature threshold
is then used with these sub-grid quantities (typically $n_{\rm H} > 10~{\rm
  cm}^{-3}$ and $T < 1000~{\rm K}$). In practice, both models lead to broadly
similar results.

Once a gas particle has passed the threshold for star formation, we compute its
star formation rate based on two different models. We either assume a
\cite{Schmidt1959} law with a fixed efficiency per free-fall time, or use the
pressure-law of \cite{Schaye2008}, which is designed to reproduce the
\cite{Kennicutt1998} relation. Based on the particle masses and computed star
formation rate, random numbers are then drawn to decide whether the particles
will indeed be converted into a star particle or not. The star particles 
formed in this manner inherit the metal content and unique ID of their parent gas particle.

\subsubsection{Stellar enrichment \& feedback}
\label{ssec:eagle:feedback}

Stellar enrichment is implemented for the SNIa, core-collapse, and AGB channels
using the age- and metal-dependant yields compilation of
\cite{Wiersma2009enrichment}. The light emitted by the stars in various filters,
based on the model of \cite{Trayford2015}, is written to the snapshots. Stellar
feedback is implemented using a stochastic thermal form \citep{DV2012} with
various options to choose which neighbour in a star particle's kernel to heat
\citep{Chaikin2022}.  The energy per supernova injection can either be kept
fixed or be modulated by the local metallicity or density
\citep{Crain2015}. Additionally, \swift includes the modified version of the
stochastic kinetic feedback model of \cite{Chaikin2023} that was used in the
\flamingo simulations \citep{Schaye2022,Kugel2023}. The SNe can either inject
their energy after a fixed delay or can stochastically sample the stars'
lifetimes. The energy injection from SNIa is done by heating all the particles
in the stars' SPH kernel during each enrichment step.

\subsubsection{Black holes \& AGN feedback}
\label{ssec:eagle:agn}

Black hole (BH) particles are created by converting the densest gas particle in
FOF-identified haloes (see \S\,\ref{ssec:stf:fof}) that do not yet contain a BH
and are above a user-defined mass threshold.  BHs grow by accreting mass from
their neighbourhood, using a \cite{Bondi1952} model, possibly augmented by
density-dependent boosting terms \citep{Booth2009} or angular-momentum terms
\citep{Rosas2015}.  BH particles can swallow neighbouring gas particles when
they have accreted enough mass or can ``nibble'' small amounts of mass from them
\citep[see][]{Bahe2022}.  Feedback from AGN is implemented using a stochastic
thermal heating mechanism where energy is first stored into a reservoir until a
pre-defined number of particles can be heated to a set temperature
\citep{Booth2009}.  Finally, the various modes of repositioning BHs presented in
\cite{Bahe2022} are available as part of the \eagle model in \swift.

\subsubsection{Results}
\label{ssec:eagle:results}

\begin{figure}
\includegraphics[width=\columnwidth]{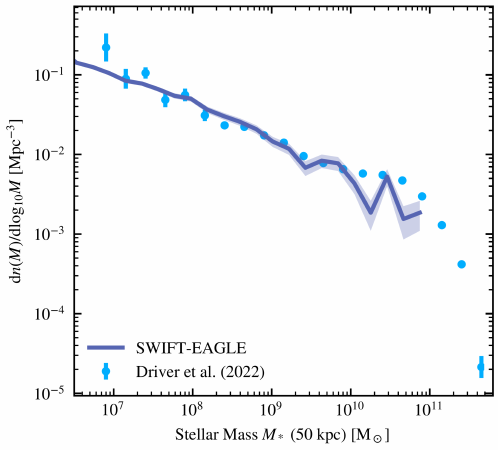}
\vspace{-0.6cm}
\caption{The galaxy stellar mass function, computed using \velociraptor as the
  structure finder and measured in $50~\rm{kpc}$ spherical apertures, extracted
  from a $(25~\rm{Mpc})^3$ volume run with \swift-\eagle model and compared to
  the \citet{Driver2022} data inferred from the \simulationname{GAMA} survey. The shaded region on
  the simulation corresponds to Poisson error counts in each
  $0.2~\rm{dex}$ mass bin.  }
\label{fig:eagle:gsmf}
\end{figure}

The model and the calibration of its free parameters are fully described by
\cite{Borrow2022calibration}, alongside a comprehensive set of results. For
completeness, we show here the $z=0$ galaxy stellar mass function measured in
$50~\rm{kpc}$ spherical apertures \citep[see appendix of][]{Graaff2022} from a
$(25~\rm{Mpc})^3$ simulation with $2\times376^3$ particles in
Fig.~\ref{fig:eagle:gsmf}. The baryon particle mass in this simulation is
$m_{\rm gas}=1.81\times10^6~\rm{M}_\odot$, the resolution of the \eagle
simulations and the resolution at which the model was calibrated. For
comparison, we show the \cite{Driver2022} estimates of the mass function
obtained from the \simulationname{GAMA} survey. Over the range where the masses are resolved and
the galaxies are not too rare to feature in such a small volume, the
\swift-\eagle model produces is in good agreement with the data. That same model
was used by \cite{Altamura2022} for their studies of groups and clusters; a map
of the gas temperature weighted by its velocity dispersion extracted from one of
their simulated clusters is displayed on panel (b) of Fig.~\ref{fig:publicity}.

We note that the exact parameters and initial conditions for this simulation are
provided as part of the code release.

\subsection{{\bfseries \scshape Gear}-like galaxy formation model}
\label{ssec:gear}

The \gear physical model implemented in \swift is based on the model initially
implemented in the \gear code \citep{Revaz2012, Revaz2016,Revaz2018}, a fully
parallel chemo-dynamical Tree/SPH code based on \gadget-2
\citep{Springel2005}. While \gear can be used to simulate Milky Way-like
galaxies \citep{Kim2016,Roca-Fabrega2021} its physical model has been mainly
calibrated to reproduce Local Group dwarf galaxies
\citep{Revaz2018,Harvey2018,Hausammann2019,Sanati2020} and ultra-faint dwarfs
\citep{Sanati2022}.  We review hereafter the main features of the model; more
details about the \swift implementation can be found in \citet{LoicThesis}. An
example galaxy from the \emph{Agora}-suite \citep{Kim2016} run using
\swift-\gear is displayed in panel (c) of Fig.~\ref{fig:publicity}.

\subsubsection{Gas radiative cooling and heating}

Radiative gas cooling and heating is computed using the \libraryname{Grackle} library
\citep{Smith2017}. In addition to primordial gas cooling, it includes
metal-lines cooling, obtained by interpolating tables, and scaled according to
the gas metallicity.  \libraryname{Grackle} also includes UV-background radiation
heating based on the prediction from \citet{Haardt2012}.  Hydrogen
self-shielding against the ionising radiation is incorporated. Two shielding
options can be used: (1) the UV-background heating for gas densities above
$n_{\rm{H}} = 0.007$ $\mathrm{cm}^{-3}$ \citep{Aubert2010}, and (2) the
semi-analytic prescriptions of \citet{Rahmati2013} directly included in the
\libraryname{Grackle} cooling tables.

\subsubsection{Pressure floor}

To prevent gas from artificially fragmenting at high density and low
temperature, i.e. when the Jeans length is not resolved
\citep{Truelove1997,Bate1997,Owen1997}, the gas' normal adiabatic equation of
state is supplemented by a non-thermal pressure term.  This additional term,
interpreted as the non-thermal pressure of the unresolved ISM turbulence,
artificially increases the Jeans length to make it comparable to the gas
resolution \citep{Robertson2008,Schaye2008}.  The \gear model uses the following
pressure floor, a modified version of the formulation proposed by
\citet{hopkins2011}:
\begin{equation}
P_{{\rm Jeans}} = \frac{\rho}{\gamma} \left(  \frac{4}{\pi} G h^2 \rho N_{\rm{Jeans}}^{2/3} - \sigma^2  \right),
\label{eq:gear:Pjeans}
\end{equation}
where $G$ is the universal gravitational constant and, $\gamma$ the adiabatic
index of the gas fixed to $5/3$.  $h$, $\rho$, and $\sigma$ are respectively the
SPH smoothing length, density, and velocity dispersion of the gas particle.  The
parameter $N_{\rm{Jeans}}$ (usually set to 10) is the ratio between the SPH mass
resolution and the Jeans mass.

\subsubsection{Star formation and pressure floor}

Star formation is modelled using a modified version of the stochastic
prescription proposed by \citet{Katz1992} and \citet{Katz1996} that reproduces
the \citet{Schmidt1959} law.  In the \gear model star formation proceeds only in
dense and cold gas phases where the physics is unresolved, i.e.  where the
artificial Jeans pressure dominates.  Inverting eq.~\ref{eq:gear:Pjeans}, the
temperature and resolution-dependent density threshold that delimits the
resolved and unresolved gas phases is defined:
\begin{equation}
\rho_{\rm{SFR},i} = \frac{\pi}{4} G^{-1} N_{\rm{Jeans}}^{-2/3} h_i^{-2} \left( \gamma \frac{ k_{\rm{B}}}{\mu m_{\rm H}} T  + \sigma_i^2  \right).
\label{eq:gear:rho_sfr}
\end{equation}
Above this limit, the gas particles are eligible to form stars.  It is possible
to supplement this threshold with a constant density threshold, which prevents
the stars from forming in cold and low-density gas regions, or by a temperature
threshold, which prevents stars from forming in hot phases.  Finally, only
particles with a negative divergence of the velocity are eligible to form stars.

Once a particle of mass $m_{\rm{g}}$ is eligible, it will have a probability
$p_{\star}$ to form a stellar particle of mass $m_{\star}$ during a time
interval $\Delta t$ \citep{Springel2003}:
\begin{equation}
p_{\star} = \frac{m_{\rm{g}}}{m_{\star}}\left[ 1-\exp\left(  -\frac{c_\star}{t_{\rm{g}}}\Delta t  \right) \right],
\label{eq:gear:pstar}
\end{equation}
where $c_\star$ is a free parameter and $t_{\rm{g}}$ the local free fall time.
Each gas particle can form a maximal number $N_\star$ of stellar particles over
the whole simulation. $N_\star$ is a free parameter set by default to 4.

The \gear model can use a critical metallicity $[\rm{Fe/H}]_{\rm c}$ parameter
to differentiate stellar populations.  Below $[\rm{Fe/H}]_{\rm c}$, a stellar
particle will represent a Pop III (metal-free) population and above the critical
metallicity, it will be considered a Pop II star. Both populations are
characterised by different initial mass functions (IMF), stellar yields, stellar
lifetimes, and energies of supernova explosions. All this information is
provided to \swift by tables that can be generated by the
\textsc{PyChem}\footnote{\url{http://lastro.epfl.ch/projects/PyChem}} utility.

\subsubsection{Stellar feedback, chemical evolution and metal mixing}

At each time step following the creation of a stellar particle, the IMF and
stellar lifetimes-dependent number of exploding supernova (core collapse and
Type Ia) is computed.  This number that can be less than one and is turned into
an integer number using a stochastic procedure called the random discrete IMF
sampling (RIMFS) scheme in which the IMF is considered as a probability
distribution \citep{Revaz2016}.  Once a supernova explodes, its energy and
synthesised elements are injected into the surrounding gas particles using
weights provided by the SPH kernel.  A parameter $\epsilon_{\rm{SN}}$ may be
used to decide the effective energy that will impact the ISM, implicitly
assuming that the remainder will be radiated away.

To avoid instantaneous radiation of the injected energy, the delayed cooling
method, which consists in disabling gas cooling for a short period of time of
about $5\,\rm{Myr}$ \citep{Stinson2006}, is used.

The released chemical elements are further mixed in the ISM using either the
smooth metallicity scheme
\citep{Okamoto2005,Tornatore2007,Wiersma2009enrichment} or explicitly solving a
diffusion equation using the method proposed by \citet{Greif2009}.

\subsection{Spin-driven AGN jet feedback}
\label{ssec:spin_jets}

This model for AGN feedback is fully described by \cite{Husko2022b} and
\cite{Husko2023}.  We summarise here its main features. This sub-grid model only
contains a prescription for AGN and can be used in combination with the
\eagle-like model described above for the rest of the galaxy formation
processes.

In this model for AGN feedback, additional sub-grid physics related to accretion
disks is included, allowing the evolution of spin (angular momentum) for each
black hole in the simulation. This in turn means that one can use the
spin-dependent radiative efficiency, instead of using a constant value
(e.g. $10$ percent) for the thermal feedback channel employed in the fiducial
model. More significantly, tracking black hole spins also allows for the
inclusion of an additional mode of AGN feedback in the form of kinetic jets. The
hydrodynamic aspects of the jets and their interaction with the CGM were tested
by \cite{Husko2022a}.  These jets are included in a self-consistent way by using
realistic jet efficiencies (that depend strongly on spin), and by accounting for
the jet-induced spindown of black holes. In the standard version of the model,
at high accretion rates it is assumed that thermal feedback corresponds to
radiation from sub-grid thin, radiatively-efficient accretion discs
\citep{Shakura1973}. At low accretion rates, jets are launched from unresolved,
thick, advection-dominated accretion disks \citep{Narayan1994}. In more
complicated flavours of the model, jets are also launched at high accretion
rates and radiation (thermal feedback) at low accretion rates, as well as strong
jets and thermal feedback from slim discs at super-Eddington accretion rates --
all of which is motivated by either observational findings or simulations.

These modifications to the AGN feedback may lead to more realistic populations
of galaxies, although they probably have a stronger impact on the properties of
the CGM/ICM. Although the model comes with the price of a more complicated
feedback prescription (which involves some number of free parameters), it also
opens an avenue for further observational comparisons between simulations and
observations. The model yields predictions such as the spin--mass relation for
black holes or the AGN radio luminosity function. These relations can be used to
constrain or discriminate between versions of the model.

\subsection{Quick-Lyman-alpha implementation}
\label{ssec:qla}

Besides galaxy formation models, another popular application of cosmological
hydrodynamical simulations is the study of the inter-galactic medium (IGM) via
the Lyman-$\alpha$ forest. So-called ``Quick-Lyman-alpha'' codes have been
developed \citep[e.g.][]{qla, Regan2007} to simulate the relevant physics. As
the focus of such simulations is largely on the low-density regions of the
cosmic web, a very simplified network of sub-grid model can be employed. In
particular, for basic applications at least, the chemistry and cooling can be
limited to only take into account the primordial elements. Similarly, any
high-density gas can be turned into dark matter particles as soon as the gas
reaches a certain over-density (typically $\Delta=1000$). In such a case, no
computing time is wasted on evolving the interior of haloes, which allows for a
much shallower time-step hierarchy than in a full galaxy formation model and
thus much shorter run times.

We implement such a model in \swift. The ``star formation'' is designed as
described above: any gas particle reaching an over-density larger than a certain
threshold is turned into a dark matter particle. The cooling makes use of the
table interpolation originally designed for the \swift-\eagle model
(\S\,\ref{ssec:eagle}). Either the W09 or the P20 tables can be used.  Of
particular interest for Quick-Lyman-alpha applications, these are based on two
different models of the evolution of the UV background: \cite{HM2001} and
\cite{FG2020} respectively. A simulation using the W09 tables would be similar
to the ones performed by \cite{Garzilli2019}.

\subsection{Material extensions and planetary applications}
\label{ssec:planetary}

\swift also includes features that can be used to model systems with more
complicated and/or multiple equations of state (EoS), and to better deal with
density discontinuities.  They are organised under a nominal `planetary' label,
given their initial application to giant impacts \citep{Kegerreis2019}.  These
extensions can be applied either onto a `\sphflavour{Minimal}'-like solver, with the
inclusion of the \citet{Balsara1995} viscosity switch, or in combination with
the other, more sophisticated SPH modifications described below.

\subsubsection{Equations of state}

Many applications of SPH involve materials for which an ideal gas is not
appropriate, and may also require multiple different materials.  Included in
\swift are a wide variety of EoS, which use either direct formulae
\citep[e.g.][]{Tillotson1962} or interpolation of tabulated data
\citep[e.g.][]{Stewart+2020,Chabrier+2021} to compute the required thermodynamic
variables.  Each individual SPH particle is assigned a material ID that
determines the EoS it will use.  By default, no special treatment is applied
when particles of different EoS are neighbours: the smoothed densities are
estimated as before, and the pressure, sound speed, and other thermodynamic
variables are then computed by each particle using its own EoS.

Currently implemented are EoS for several types of rocks, metals, ices, and
gases. Custom user-provided EoS can also be used.  Some materials can, for
example, yield much more dramatic changes in the pressure for moderate changes
in density than an ideal gas, and can also account for multiple phase states.
In practice, in spite of the comparative complexity of some of these EoS,
invoking them does not have a significant effect on the simulation run speed, because they
are called only by individual particles instead of scaling over multiple
neighbours.

Some input EoS may include a tension regime, where the pressure is negative for
a cold, low-density material.  This is usually undesired behaviour in a typical
SPH simulation and/or implies an unphysical representation of the material in this state
as a fluid, and can lead to particles accelerating towards each other and
overlapping in space.  As such, by default, a minimum pressure of zero for these
EoS is applied.

\subsubsection{Special treatment for initial conditions}

Prior to running a simulation, it is a common practice to first perform a
`settling' run to relax the initial configuration of particles.  This is
particularly pertinent to planetary and similar applications, where the
attempted placement of particles to model a spherical or spinning body will
inevitably lead to imperfect initial SPH densities
\citep{Kegerreis2019,RuizBonilla2021}.  If the applied EoS includes specific
entropies, then \swift can explicitly enforce the settling to be adiabatic,
which may be a convenient way to maintain an entropy profile while the particles
relax towards equilibrium.

\subsubsection{Improvements for mixing and discontinuities}

Standard SPH formulations assume a continuous density field, so can struggle to
model contact discontinuities and to resolve mixing across them
\citep[e.g.][]{Price2008}.  However, density discontinuities appear frequently
in nature.  For example, in a planetary context, sharp density jumps might
appear both between a core and mantle of different materials, and at the outer
vacuum boundary.  Smoothing particles' densities over these desired
discontinuities can lead to large, spurious pressure jumps, especially with
complex EoS.

We have developed two approaches to alleviate these issues in \swift, briefly
summarised here, in addition to the significant benefits of using more SPH
particles for higher resolutions than were previously feasible.  First, a simple
statistic can be used to identify particles near to material and/or density
discontinuities and to modify their estimated densities to mitigate the
artificial forces and suppressed mixing \citep{RuizBonilla2022}.  This method is
most effective when combined with the geometric density-average force (GDF)
equations of motion \citep{Wadsley2017}.

Second, a more advanced scheme in which density discontinuities are addressed by
directly reducing the effects of established sources of SPH error
\citep{Sandnes2024}. This combines a range of novel methods with recent SPH
developments, such as gradient estimates based on linear-order reproducing
kernels \citep{Frontiere2017}. The treatment of mixing in simulations with
either one or multiple equations of state is significantly improved both in
standard hydrodynamics tests such as Kelvin–Helmholtz instabilities and in
planetary applications \citep{Sandnes2024}.

Each of these modifications may be switched on and off in \swift in isolation.
Further improvements are also in active development -- including the
implementation of additional features such as material strength models.

\subsection{External potentials}
\label{ssec:extensions:potentials}

Several external potentials intended for use in idealised simulations are
implemented in \swift. The simplest external potentials include an unsoftened
point mass, a softened point mass (i.e. a \citet{Plummer1911} sphere), an
isothermal sphere, a \citet{navarro1997} (NFW) halo, and a constant
gravitational field.

Besides these traditional options, \swift includes two \citet{hernquist1990}
profiles that are matched to a NFW potential. The matching can be performed in
one of two ways: (1) we demand that the mass within $R_{200, {\rm cr}}$ is
$M_{200,{\rm cr}}$
\footnote{$M_{200,\rm{cr}}$ is the mass within the radius $R_{200, {\rm cr}}$, at which the average internal density
$\langle \rho_{} \rangle = 200 ~\rho_{\rm crit}$, and $\rho_{\rm crit}$ is the
critical density of the Universe.} 
for the \citet{hernquist1990} profile, i.e.
$M_{\rm Hern}(R_{\rm match})= M_{\rm NFW}(R_{200, {\rm cr}})$ at some specific matching
radius. (2) We demand that the density profile in the centre is equivalent i.e. $\rho_{\rm
Hern} (r) = \rho_{\rm NFW}(r)$ for $r\ll R_{200, {\rm cr}}/c$, where $c$ is the NFW
concentration of the halo.

The first of these profiles follows \citet{springel2005b} and uses $M_{\rm
  Hern}(r\rightarrow\infty)= M_{\rm NFW}(R_{200, {\rm cr}})= M_{200, {\rm cr}}$
and $\rho_{\rm Hern} (r) = \rho_{\rm NFW}(r)$. Using this they can derive a
matched scale factor with the assumption that $a/R_{200, {\rm cr}}\ll 1$ of the
halo given by $a=\sqrt{b} R_{200, {\rm cr}} $ where
\begin{align}
    b=\frac{2}{c^2}\left(\ln\left(1+c \right) - \frac{c}{1+c} \right)
\end{align}
The second profile follows \cite{Nobels2023}, who match $M_{\rm
  Hern}(R_{200, {\rm cr}})= M_{\rm NFW}(R_{200, {\rm cr}})$, $\rho_{\rm Hern}
(r) = \rho_{\rm NFW}(r)$ and do not assume a $a/R_{200, {\rm cr}}\ll 1$. This
gives a different \citet{hernquist1990} scale length and $M_{\rm Hern}(R_{200,
  {\rm cr}})$, producing a better match with the NFW profile. Both approaches
are similar for haloes with large concentration parameters.

In order to reduce errors in the integration of orbits, each of the
spherically-symmetric potentials optionally imposes a minimum time-step to each
particle \citep[see e.g.][]{nobels2022}. We compute the distance from the centre
$r$ of each particle and the corresponding circular velocity $V_{\rm
  circ}(r)$. We then impose a minimum time-step of $\Delta t_{\rm pot} =
\varepsilon_{\rm pot} \frac{r}{V_{\rm circ}(r)}$, where $\varepsilon_{\rm pot}$
is a free parameter typically defaulting to $\varepsilon_{\rm pot}=0.01$
(i.e. 100 time-steps per orbit).

\section{Implementation details \& Parallelisation}
\label{sec:details}

In this Section, we present some of the important implementation details,
especially surrounding the multi-node parallelism, and discuss the results of a
scaling test on a realistic problem testing the entirety of the code modules.

\subsection{Details of the cells \& tasking system}
\label{ssec:parallel:cells}

The basic decomposition of the computational domain in meaningfully-sized cells
was introduced in \S\,\ref{ssec:design:hydro_first}. We present some more
technical details here. \\

\noindent In all the calculations we perform, we start by laying a Cartesian
grid on top of the domain. This defines the most basic level in the cell
hierarchy and is referred to as the top-level grid\footnote{Note that this grid
is not related to the one used for the periodic gravity calculation
(\S\,\ref{ssec:gravity:mesh_summary}). It is, however, the base grid used to
retrieve particles efficiently in small sections of the snapshots
(\S\,\ref{ssec:io:snaps}).}. The size of this grid varies from about 8 cells on
a side for small simple test runs to 64 elements for large calculations. In most
cases, there will be many thousands or millions of particles per cell. We then
use a standard oct-tree construction method to recursively split the cells into
8 children cells until we reach a number of particles per cell smaller than a
set limit, typically $400$. This leads to a relatively shallow tree when
compared to other codes which create tree nodes (cells) down to a single
particle, and implies a much smaller memory footprint for the tree itself than
for other codes.  As discussed in \S\,\ref{ssec:design:hydro_first}, \swift can
perform interactions between cells of different size.

Once the tree has been fully constructed, we sort the particles into their
cells. By using a depth-first ordering, we can guarantee that the particles
occupy a contiguous section of memory for all the cells in the tree and at any
level. This greatly helps streamline operations on single or pairs of cells as
all the particles will simply be located between two known addresses in memory;
no speculative walk will be necessary to find all the particles we need for a
set of interactions. This sorting of particles can be relatively expensive on
the very first step as we inherit whatever order the particles were listed in
the initial conditions. However, in the subsequent constructions, this will be
much cheaper because the particles only move by small amounts with respect to
their cells in between constructions. This is also thanks to the relatively
shallow tree we build, which permits for comparatively large cell sizes.  For
this reason, we use a \emph{parallel merge sort} here to sort the particles in
their cells as it is an efficient way to sort almost-sorted lists, which is the
case in all but the first step. Recall also that we do not need to sort the
particles very finely, thanks to the high number of them we accept in tree
leaves. Whilst this operation is technically a sort, we refer to it as binning
of the particles in what follows to avoid confusion with the sorting of
particles on the interaction axis used by the pseudo-Verlet algorithm.

With the tree constructed and the particles all in their cell hierarchies, we
have all the information required to decide which cells will need to interact
for SPH (based on the cells' maximum smoothing lengths) and for gravity (based
on the multipoles). All the quantities required for this decision making were
gathered while binning the particles. We start by constructing the tasks on the
top-level grid only, as described in \S\,\ref{ssec:design:parallel_strategy} and
\S\,\ref{ssec:gravity:walk} for SPH and gravity respectively. In most
non-trivial cases, however, this will lead to tasks with very large numbers of
particles and hence a large amount of work to perform. If there are only a few
expensive tasks, then the scheduler will not be able to load-balance the work
optimally as its options are limited. We ideally want significantly more tasks
to be enqueued and waiting for execution than there are compute cores. It is
hence key to fine-grain the problem further. To achieve this, we attempt to
split the tasks into smaller units. For instance, a task acting on a single cell
might be split into eight tasks, each acting on its eight children cells
independently. For some tasks, in particular when there are no particle-particle
interactions involved, this is trivially done (e.g. time integration or for a
cooling sub-grid model) but other tasks may lead to more complex scenarios. An
SPH task for instance cannot be split into smaller tasks if the smoothing length
of the particles is larger than the size of the children cells. In most
non-pathological cases, however, the tasks can be moved down the tree by several
levels, thus multiplying their overall number many times over and ultimately
satisfying our request to have many more tasks than computing units. In cases
where more than one loop over the neighbours are needed, only the tasks
corresponding to the first loop are moved down the tree levels by assessing
whether refinement criteria are met. The tasks corresponding to the subsequent
interaction loops however are created by duplicating the already existing tasks
of the first loop. As an example, the SPH \emph{force} loop is created by
copying all the tasks needed for the \emph{density} loop and relabelling
them. Similarly, all the sub-grid feedback or black hole-related loops are
created in this fashion. This approach has the advantage of keeping the
task-creation code as simple as possible. While duplicating the loops, we also
set dependencies between tasks to impose the logical order of operations between
them (see Fig.~\ref{fig:design:tasks}).

With the tasks created, the code is ready to perform many time-steps. That is,
we can re-use the infrastructure created above until the geometrical conditions
are violated by particle movement. For SPH, these conditions would be too large
a change in smoothing length or a particle moving too far out of its cell
meaning that the assumption that all the neighbours are in the same cell or any
directly adjacent one is broken. For gravity, this would be too large a particle
movement, leading to it being impossible to recompute multipoles without
changing the cell geometry. Our shallow tree with large leaves has the advantage
of remaining valid for many steps. We also note that other criteria (such as a
global mesh gravity step or a certain number of particle updates leading to a
tree rebuild) do, in practice, trigger a tree and tasks construction more often
than these.

At the start of each step, we perform a quick tree walk starting, in parallel,
in each of the many top-level cells. In this walk, we simply identify which
cells contain active particles (i.e. particles which need to be integrated
forward in time on this step) and activate the corresponding tasks. This
operation is very rapid (much less than $1$ percent of the total runtime in
production runs) and can easily be parallelised given the large number of cells
present in a run. Once all the tasks have been activated, they are handed over
to the \libraryname{QuickSched} engine which will launch them when ready.

As described by \citet{Gonnet2016}, the tasks whose dependencies are all
satisfied (i.e. for which all the tasks taking place earlier in the graph have
already run) are placed in queues. We typically use one of these queues per
thread and assign the tasks to the queues (and hence threads) either randomly or
based on their physical location in the compute domain. The threads then run
through their queues and attempt to fetch a task. When doing so they have to
verify that the tasks they get are not conflicting with another, already-running
operation. To this end, a mechanism of per-cell locks and semaphores is used. If a
thread cannot acquire the lock on a given cell, it abandons this task and
attempts to fetch the next one in the queue. If it can acquire a task, it will
run the physics operations and upon completion will unlock all the dependencies
associated with this task, hence enabling the next tasks to be placed in the
queues. We highlight once more that the physics operations themselves are taking
place inside a single thread and that no other thread can access the same data
at the same time. This places the physics and maths operations taking place in a
very safe space, allowing users with only limited programming experience to
easily modify or extend the physics contained inside the tasks. No intimate
knowledge of parallel programming or even of task-based parallelism is needed to
alter the content of a task. If a thread reaches the end of its queue, it starts
again from the beginning until there are no more tasks it can process. When that
happens, the thread will attempt to steal work from the other threads' queues, a
unique feature, at the time this project started, of the \libraryname{QuickSched}
library.  Once all tasks in all queues have been processed, the time-step has
been completed and the threads are paused until the start of the next step.

\subsection{Multi-node strategy}
\label{ssec:mpi}

\noindent The top-level grid described in the previous section serves as the
base decomposition unit of the simulated domain. When decomposing the problem
into multiple domains, which would be necessary to run a simulation over multiple
compute nodes, we assign a certain number of these cells to each of them. The
tree construction algorithm is then run in parallel in each domain for each
cell. The only addition is the possible exchange of particles which have left
their domain entirely. They are sent to their new region and placed in the
appropriate cells.

With the tree fully constructed, we send the sections of the trees (the cell
geometry information and multipoles, not the particles) that border a domain to
the nodes on the other side of the divide. Each compute node has henceforth full
knowledge of its own trees and of any of the directly adjacent ones. With that
information in hand, each node will be able to construct all of its tasks, as
described above. It will do so for all the purely local cells as well as for the
pair tasks operating on one local cell and one foreign cell. The compute node on
the other side of the divide will create the exact same task as it bases its
decision-making on exactly the same information. The only remaining operation is
the creation of send and receive tasks for each task pair overlapping with a
domain edge. By adding the appropriate dependencies, we create a task graph
similar to the one depicted in Fig.~\ref{fig:design:tasks_mpi}.

With this logic, any task spanning a pair of cells that belong to the same
partition needs only to be evaluated on that rank/partition, whilst tasks
spanning more than one partition need to be evaluated on both ranks/partitions.
This is done in the shallow tree walk that performs the task activation at the 
start of a step. A minor
optimisation can be used in the cases where only one of the two cells in a pair
task contains active particles. In that situation, we can skip the sending and
receiving of data to the node hosting the inactive cell since it will not be
using it for any local updates.

All the tasks are put in queues in exactly the same way as in the single-node
case. The only difference applies to the communication tasks. These are treated
slightly differently. As soon as their dependencies are satisfied, the data is
sent \emph{asynchronously}. Similarly, as soon as the receiving node is ready,
it will post a call to an asynchronous receive operation. Note that these
communication tasks are treated like any other task; in particular, any of the
threads can act on them and thus perform the inter-node communications. We then
use the conflict mechanism of the queues to ask the \MPI communication library
whether the data has effectively been sent or received, respectively. Once that
has happened, we simply unlock the corresponding tasks' dependencies and the
received data can safely be used from that point onward. This allows us to
effectively hide all the communications in the background and perform local work
while the data move. We also note that once the data have arrived, nothing
distinguishes them from data that were always on that node. This means that the
physics operations in tasks can be agnostic of which data they work on. There is
no need for special treatment when dealing with remote data; once more helping
developers of physics modules to focus on the equations they implement rather
than on the technicalities of distributed parallelism.

\subsection{Domain decomposition}
\label{ssec:dd}

When running a large simulation over \MPI using many ranks, an important question
is how to share the workload across all the ranks and their host compute
nodes. This is important, beyond the obvious reasons like limited memory and CPU
cores per node, as the progression of a simulation with synchronisation points
is determined by the slowest part.

The simulation workload consists of not just particles and their memory, but
also the associated computation, which can vary depending on the types of
particles, the current state and environment of the particles, as well as 
the costs of inter-node communication. All these elements play their part.

\begin{figure}
\includegraphics[width=\columnwidth]{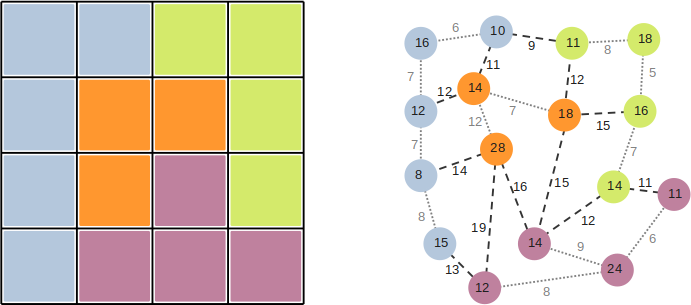}
\vspace{-0.3cm}
\caption{The representation of the top-level cells as a graph to be split over
  domains. The cells of the grid (on the left) correspond to the vertices of the
  graph (on the right), while the tasks spanning two cells constitute its edges
  (dashed and dotted lines). For simplicity, we consider here a $4\times4$
  non-periodic grid in 2D and only show the pair tasks for cells that share an
  edge. Each vertex and graph edge has a weight associated with it, shown here
  as the numbers on each vertex and edge. The weights correspond to the cost of
  the task execution. If a pair operation is taking place over the network
  (shown here using dashed lines), its cost will be increased since
  communications will have to take place and the task will be executed on both
  of the involved ranks. The domain decomposition algorithm splits the graph so
  that the work (vertices and edges) is as evenly distributed as possible among
  all computing ranks (the four colours), minimising the total cost by creating
  as few communications as possible. In the case shown here, this corresponds to
  the domain decomposition presented on the left. Note in particular that the
  number of cells assigned to each domain may not necessarily be the same.}
\label{fig:details:graph}
\end{figure}

A representation of the workload and communication can be constructed by
considering the hyper-graph of all top-level cells, where graph vertices
represent cells and the edges represent the connections to the nearest
neighbours (so each vertex has up to 26 edges). In this graph the vertices
represent the computation done by the cell's tasks and the edges represent only
the computation done in pair-interaction tasks. This follows since pair
interactions are the only ones that could involve non-local data, so the
computation in tasks spanning an edge should be related to the communication
needed. Now, any partition of this graph represents a partition of the
computation and communication, i.e. the graph nodes belonging to each partition
will belong to an \MPI rank, and the data belonging to each cell resides on the
rank to which it was assigned. Such a decomposition is shown in
Fig.~\ref{fig:details:graph} for a simple toy example.

The weighting of the vertices and edges now needs to reflect the actual work and
time expected to be used for communication. Initially, the only knowledge we
have of the necessary weights is the association of particles and cells, so we
only have vertex weights. However, when a simulation is running, every task is
timed to CPU tick accuracy and thus has a direct wall-clock measurement to
reflect the computation. This will never be perfect, as other effects like
interruptions from other processes will add time, but should be good enough.
Note that it also naturally accounts for unknowns, like CPU speed and compiler
optimisations, that a non-timed system would need to know about for all the
different task types. So, once all the tasks of a simulation have run, we then
know how long they take and can then use these real-world weights in the graph.

Decomposing such graphs is a standard problem in computer science and multiple
packages exist in the literature. We chose to use \metis and \parmetis
\citep{metis}.

Using this simple weights scheme is sufficient, as shown in the next
section. Note also that we are not demanding a perfect partition of the
graph. In typical simulations, the workload evolves with time (which task times
naturally take into account), and it is hence counterproductive to spend a large
amount of time identifying the perfect partition. We prefer to use a partition
that is good enough but quick to obtain. For realistic simulations, we find that
we can maintain the imbalance between compute domains to less than 10 percent
\citep[see also][and Fig.~\ref{fig:details:weak_scaling} below]{Schaller2016}. 
We caution that this approach does not
explicitly consider any geometric constraints, nor does it attempt to distribute
the data uniformly. The only criterion is the relative computational cost of
each domain, for which the task decomposition provides a convenient model. We
are therefore partitioning the computation, as opposed to just the data. There
could, in principle, be cases where the work-based decomposition leads to
problematic data distributions leading to the code running out of memory on a
given compute node. We have so far never encountered such a situation in
practice.

In addition to this default mechanism, \swift also offers other domain
decomposition algorithms. The first one just attempts to split the data evenly
between the compute nodes, so maintains the initial state. This is similar to
what other simulation packages do, though here it is based on the top-level
cells. This is also used as a backup mechanism in case the work-based
decomposition leads to too much data imbalance. Finally, a mode where the
top-level grid is simply split into regular chunks is also implemented. This is
never recommended but the code will default to this if the \metis library is not
available.

\subsection{Scaling results \& code performance}

The scaling performance of the \swift code on various test problems has been
reported in different publications thus far. We give a quick overview here and
complement it with a test exploiting the full cosmological simulation engine in
a realistic scenario.

\noindent In their original \swift feasibility study, \cite{Schaller2016}
analysed the original SPH-only code's performance on cosmological test
boxes. They reported a strong-scaling efficiency of 60 percent when scaling a
problem from $512$ cores to $131\,072$ cores of a BlueGene system. This
demonstrated the viability of the task-based approach combined with a
graph-based domain decomposition mechanism and set the foundation for the
current version of the code.

In their analysis, \cite{Borrow2018} took low-redshift cosmological simulations
from the \eagle suite and ran strong- and weak-scaling tests of the code. They
focused on the scaling of the SPH operations by running only the hydrodynamics
tasks. However, by using late-time cosmological boxes, they analysed the
performance of the code with a realistic density (and hence time-step)
distribution. They demonstrated the importance of running the drift operation
only on the region of the volumes that directly contribute to the calculation.

Finally, \cite{Rogers2022} analysed the performance of \swift in the context of
future exa-scale developments with engineering-type SPH applications in mind. To
this end, they ran a fixed time-step, fairly uniform, test volume with more than
$5.5\times10^{11}$ gas particles and demonstrated excellent weak-scaling
performance up to the size of their test cluster ($\approx 50\,000$ cores).

\begin{figure}
\includegraphics[width=\columnwidth]{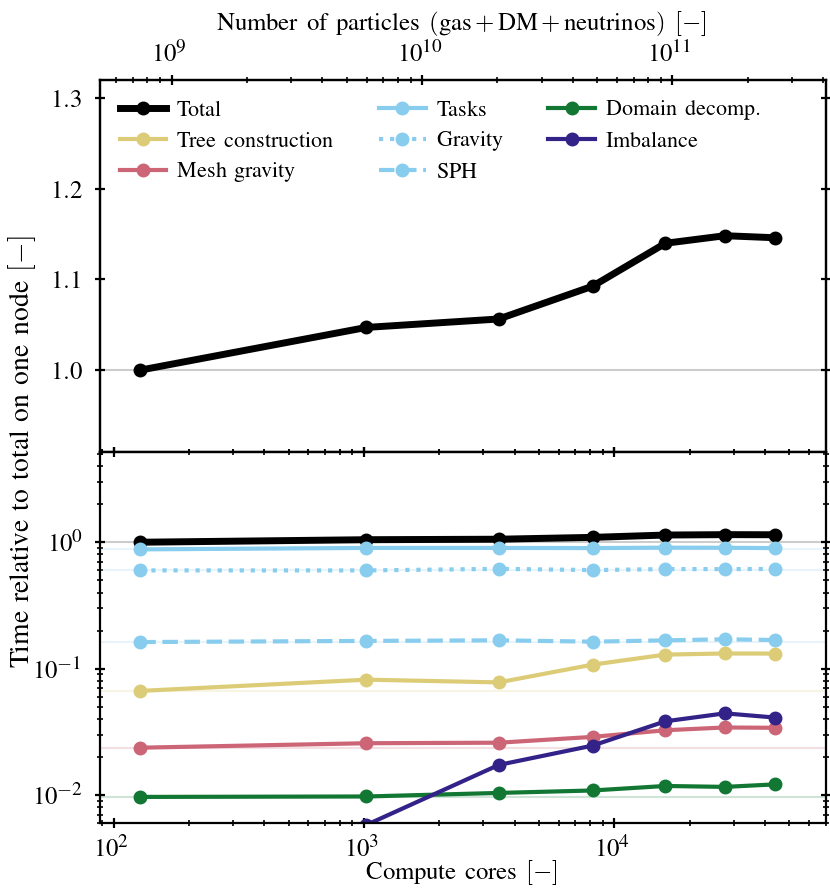}
\vspace{-0.3cm}
\caption{Weak-scaling performance of the \swift code on a representative
  cosmological simulation test problem. We use a $400^3~{\rm Mpc}^3$ volume
  extracted from the \flamingo series with $720^3$ baryon, $720^3$ dark matter,
  and $144^3$ neutrino particles at $z=1$. That base unit is then replicated
  periodically in all three directions; the top-level grid, as well as the
  gravity mesh, are also scaled alongside the replications. The number of
  compute nodes is grown proportionally, starting from a single node ($128$
  cores) for the base volume. The top axis indicates the total number of
  particles used in each of the tests. When scaling the problem by a factor
  $7^3=343$, the total runtime (black line) increases by only $15$ percent, as
  shown on the top panel (note the linear y-axis). The bottom panel shows the
  breakdown of the total time in different categories (note the log y-axis). The
  time spent in the tasks (aka. actually solving physics equations, blue line)
  is remarkably constant as the problem size increases. The task time can be
  further subdivided in gravity (the FMM part) and SPH operations (dotted and
  dashed lines); all other tasks, including the sub-grid operations, correspond
  to a negligible fraction of the runtime. The ``mesh gravity'' category
  corresponds to all the operations performed by the PM-part of the
  algorithm. The loss of performance is dominated by the lack of scalability of
  some operations within the tree construction (yellow) as well as by the
  accumulation of residual imbalance between nodes (purple). The domain
  decomposition itself (green) only requires a negligible amount of time.}
\label{fig:details:weak_scaling}
\end{figure}

To complement these earlier tests, we present here a scaling test exploiting all
the main physics modules, including a galaxy formation model. To be as
representative as possible, we use a $z=1$ setup such that the density structure
and hence time-step hierarchy is well developed. We use a $400^3~{\rm Mpc}^3$
volume with $720^3$ baryon, $720^3$ dark matter, and $144^3$ neutrino particles
extracted from the \flamingo \citep{Schaye2022} suite and run it for 1024
time-steps. The sub-grid model is broadly similar to the one described in
\S\,\ref{ssec:eagle} but with parameters calibrated to match observational
datasets at a lower resolution than \eagle did \citep[for details,
  see][]{Kugel2023}.  We use this volume as a base unit and run it on a single
node (128 cores) of the \computername{cosma-8} system\footnote{The
\computername{cosma-8} system is run by DiRAC (\url{www.dirac.ac.uk}) and hosted
by the University of Durham, UK. The system is made of 360 compute nodes with 1
TB RAM and dual 64-core AMD EPYC 7H12 at 2.6 GHz (4 \techjargon{NUMA} regions /
CPU) with \techjargon{AVX2} vector capability. The interconnect is Mellanox HDR,
200GBit/s, with a non-blocking fat-tree topology. The machine has a theoretical
1.9 PF peak performance and achieved 1.3 PF on the standard HPL benchmark.}.  We
use 4 \MPI ranks per node even when running on a single node to include the \MPI
overheads also in the smallest run. The 4 \MPI ranks are distributed over the
various \techjargon{NUMA} regions of the node. We then scale up the problem by
replicating the box periodically along the three axes and increasing the number
of nodes proportionally. We also use 8 top-level cells per unit volume and an
FFT gravity mesh of size $512^3$. Both are scaled up when increasing the problem
size. We increase the problem size by a factor $7^3=343$, which corresponds to
the largest setup we can fit on the system.  The results of this test are shown
in Fig. \ref{fig:details:weak_scaling}, where we plot the time to solution in
units of the time taken on one node. Perfect weak-scaling hence corresponds to
horizontal lines. When the problem size is increased by a factor $343$, the
performance loss is only $15$ percent.  We also decompose the time spent in the
main code sections. The tasks (i.e. physics operations, blue line) dominate the
run time and display an excellent scaling performance. Decomposing the task work
into the gravity and SPH parts, we see that gravity is the dominant component,
validating the hydrodynamics-first approach of the overall code design. All
other operations, including all of the sub-grid model tasks, are a negligible
contribution to the total. The loss of performance when scaling up comes from
the tree construction (orange) and from the overall imbalance between the
different nodes (purple) due to an imperfect domain decomposition leading to
slightly non-uniform work-load between the nodes despite the problem being
theoretically identical. As discussed in \S\,\ref{ssec:dd}, we can maintain the
node-to-node imbalance below 10 percent. We also report that the time spent
deciding how to distribute the domains and performing the corresponding exchange
of particles (green line) is a negligible fraction of the total runtime.

Finally, we note that the right-most points in
Fig. \ref{fig:details:weak_scaling} correspond to a test as large as the largest
cosmological hydrodynamical simulation (by particle number) ever run to $z=0$
\citep[the flagship $2\times5040^3$ \flamingo volume of][]{Schaye2022},
demonstrating \swift's capability to tackle the largest problems of interest to
the community.\\

\noindent We started the presentation of the design decisions that lead to the
architecture of \swift in \S\,\ref{sec:design} by a brief discussion of the
performance of the previous generation of cosmological hydrodynamical
simulations and in particular of the \eagle suite. To demonstrate improvements
we could have repeated the flagship simulation of \cite{Schaye2015} with \swift
using our updated SPH implementation and the \eagle-like model of
\S\,\ref{ssec:eagle}. Even with \swift's enhanced performance, this would still
be a large commitment of resources for a benchmarking exercise, so we decided to
instead compare the time taken by the codes on a smaller simulation volume using
the same model. The $(25~{\rm Mpc})^3$ volume run with $2\times376^3$ particles
presented in \S\,\ref{ssec:eagle:results} took 159 hours using 28 compute cores
of the \computername{cosma-7} system\footnote{The \computername{cosma-7} system
is run by DiRAC (\url{www.dirac.ac.uk}) and hosted by the University of Durham,
UK. The system is made of 448 compute nodes with 512 GB RAM and dual 14-core
Intel Xeon Gold 5120 CPU at 2.2 GHz (1 \techjargon{NUMA} region / CPU) with
\techjargon{AVX512} vector capability. The interconnect is Mellanox EDR,
100GBit/s, using a fat tree topology with a 2:1 blocking configuration.}; this
corresponds to a total of $4452$ CPU core hours. The \gadget-based run, using
the same initial conditions, from the original \eagle suite took $32900$ CPU
core hours, meaning that our software is $>7\times$ faster on that
problem. Recall however, that the flavours of SPH and the implementation of the
sub-grid models are different from the original \eagle code making a more
detailed comparison difficult.

We also note that this \swift-based \eagle-like run only required $92~\rm{GB}$
of memory meaning that it would easily fit in the memory of a single compute
node of most modern facilities. By contrast, the \gadget-based \eagle run
required $345~\rm{GB}$ of memory; a factor of nearly 4x more.

\subsection{Random number generator}

Many extensions of the base solvers, in particular sub-grid models for galaxy
formation, make use of (pseudo-)random numbers in their algorithms. Examples of
this are stochastic star formation models or feedback processes (see
\S~\ref{ssec:eagle:sf} and \S~\ref{ssec:eagle:feedback} for such models in
\swift). Simulation packages can generate random numbers in various ways, often
based on direct calls to a generator such as the base one part of
\techjargon{UNIX} or the more advanced ones in \libraryname{GSL} \citep{GSL}. To
speed things up or to make the sequence independent of the number of
\libraryname{MPI} nodes, these calls can then be bundled into tables and
regenerated every so often. The particles and physics modules then access these
tables to retrieve a random number. This approach can lead to different issues
of reproducibility between runs if the particles or modules are not calling the
generator in the same order. These issues can arise due to task ordering
choices\footnote{Note that in \libraryname{MPI} codes, the same
order-of-operations-issue can also occur if rounding choices change the
time-step size of a particle, thus altering the sequence of numbers. The
ordering of operations is not guaranteed for reduction operations, or in the
directly \swift-relevant case, for asynchronous communications in a
multi-threaded environment, unless the developers implemented explicit
mechanisms to force this (often slower) behaviour.}. Additionally, when bundling
random numbers in small tables, great care has to be taken to make sure the
indexing mechanism is sufficiently uniform so as to not bias the
results\footnote{A common mistake is to index the tables based on particle IDs
when these IDs themselves encode some information (e.g. only even numbers for
gas, or a position in the ICs).}.

In \swift, despite the intrinsic lack of ordering of the operations due to the
tasking, we decided to avoid these pitfalls by viewing the generation of random
numbers as a hashing of four unique quantities which are then used to construct
the mantissa of a number in the interval $[0,1)$. We combine the ID of the
  particle (64-bit), the current location on the integer timeline (64-bit), a
  unique identifier for this random process (64-bit), and a general seed
  (16-bit). By doing so, we always get the same random number for a given
  particle at the same point in simulation time. Since each process also gets a
  unique identifier, we can draw uncorrelated numbers between modules for the
  same particle in the same step. Finally, the global seed can be altered if one
  wanted to actually change the whole sequence to study the effect of a
  particular set of randoms \citep[see][for an example using \swift and the
    \eagle-like model]{BorrowStochastic}. The combined 144 bits thus generated
  are passed through a succession of XOR and random generator seed evolution
  functions to create a final source of entropy. We use this source as a seed
  for our last \techjargon{UNIX} random number call, \texttt{erand48()}, whose
  output bits are interpreted as the mantissa of our result.

We have thoroughly verified that this entire mechanism generates perfectly
uniform numbers. We also verified that there is no correlation between calls
using the same particle and time-step but varying the identifier of the random
process.

\section{Summary \& Conclusion}
\label{sec:conclusion}
\subsection{Summary}
\label{ssec:summary}

In this paper, we have presented the algorithms and numerical methods exploited
in the open-source astrophysical code, \swift. We have presented various test
problems performed with the code, as well as demonstrated its scaling capability
to reach the largest problems targeted by the community. In addition, we
described the sub-grid models and other features made available alongside the
code, and the various output strategies allowing the users to make the most
efficient use of their simulations.

The core design strategy of the \swift code was to focus on a
hydrodynamics-first approach, with a gravity solver added on top. In tandem with
this, the parallelisation strategy departs from traditional methods by
exploiting a task-based parallelism method with dependencies and conflicts. This
allows for the efficient load-balancing of problems by letting the runtime
scheduler dynamically shift work between the different compute units. This
approach, coupled to a domain decomposition method focusing on distributing work
and not data, is specifically designed to adhere to the best practices for
efficient use of modern hardware.

Various modern flavours of Smoothed Particle Hydrodynamics (SPH) are
implemented, alongside two sets of flexible sub-grid models for galaxy formation,
a modern way of evolving cosmological neutrinos, and extensions to handle
planetary simulations.  These additional components are presented and released
publicly along with the base code.

Besides testing and benchmarking (in simulations using more than
$2\times10^{12}$ particles), the \swift software package has already been
exploited to perform extremely challenging scientific calculations. These
include the very large dark-matter-only ``zoom-in'' ($>10^{11}$ particles in the
high resolution region) of the \simulationname{Sibelius} project \citep{McAlpine2022},
the large cosmological hydrodynamics runs (up to $2\times5040^3$ particles) of
the \flamingo project \citep{Schaye2022}, and the highest ever
resolution Moon-formation simulations \citep{Kegerreis2022}. We envision that
the public release of the code and its future developments will lead to more
projects adopting it as their backbone solver for the most difficult and largest
numerical astrophysics and cosmology problems.

\subsection{Future developments}
\label{ssec:future}

The \swift code is in constant development and we expect it to evolve
considerably in the future. This paper describes the first full public release
of the software and we expect improvements to the numerical aspects to be made,
new models to be added, as well as new computer architectures to be targeted in
the future. \\

One of the current grand challenges in high-performance computing is the jump
towards so-called exa-scale systems. It is widely believed that such computing
power can only be reached via the use of accelerators such as GPUs. This is a
challenge for methods such as SPH and generally for algorithms including deep
time-step hierarchies due to the low arithmetic intensity of these methods and
the use of largely irregular memory access patterns. In the context of \swift,
exploiting efficiently both CPUs and GPUs via a unified tasking approach is an
additional challenge. Some avenues and possible solutions are discussed by
\citet{Bower2021}, where some early work porting specific
computationally-intensive tasks to GPUs is also described.

In terms of physics models, we expect the public code to be soon expanded to
include the self-interacting dark matter model of \cite{Correa2022}. This will
expand the range of cosmological models that can be probed with the \swift
package. Work on other extensions beyond vanilla $\Lambda$CDM will likely
follow. Similarly, additional sub-grid models for galaxy formation and 
cosmological applications are in the process of being included in the main
code base and will be released in the future.

The code is also being expanded to include material strength models, as well as
further new equations of state, for planetary and other applications.

The various hydrodynamics solvers in the code are currently all variations of
SPH. This family of methods is known to have some limitations in the rate of
convergence towards analytic solutions in certain scenarios. In future releases
of the \swift package, we thus intend to supplement this with additional SPH
variations \citep[e.g.][]{Rosswog2020}, renormalised mesh-free methods
\citep[e.g.][]{Vila1999, Hopkins2015, Alonso2022}, and a moving mesh
implementation akin to \citet{shadowfax}. These methods all use unstructured
particles with neighbourhoods as their base algorithmic tool, which makes them
very suitable to fit within the framework currently existing in the \swift
code. Developments on top of the SPH flavours to include magneto-hydrodynamics
terms are also under way both using a direct induction formulation
\citep[e.g.][]{Price2018} and a vector-potential formulation
\citep[e.g.][]{Federico2015}.

The code is also being expanded to include radiative transfer modules, starting
with the SPH-based formalism of \citet{Chan2021} based on the M1-closure method
and a coupling to the \libraryname{CHIMES} non-equilibrium thermo-chemical solver
\citep{CHIMESa,CHIMESb}.  Developments to include sub-cycling steps, in an even
deeper hierarchy than in the gravity+hydro case \citep{Duncan1998}, for the
exchange of photons are also on-going, which coupled to the task-based approach
embraced by \swift should lead to significant gains over more classic methods
\citep{Ivkovic2022}.

Finally, an improved domain decomposition strategy for the special case of
zoom-in simulations with high-resolution regions small compared to the parent
box but too large to find in a single node's memory will be introduced by
\citet{Roper2024} (See also Chapter 2 of \citet{Roper2023} for a preliminary discussion).~\\

By publicly releasing the code and its extensions to the community, we also hope
to encourage external contributors to share their models built on top of the
version described here to other researchers by themselves making their work
public.

\section*{Acknowledgments}

The authors gratefully acknowledge the significant contribution to this project
and paper that the late Richard G. Bower made over the years. His unbounded
enthusiasm and immense expertise as well as his mentorship and guidance will be
sorely missed. \\

We are indebted to the support and admin staff running the DiRAC COSMA
facility at Durham, in particular to Lydia Heck and Alastair Basden. Their trust
in the project, as well as their help running, debugging, scaling, and testing
our code on the machines at scale, have been invaluable. \\

We thank Joop Schaye for the long-term support, the detailed discussions on
physics modelling, and the guidance this kind of large project requires. Adopting
\swift early on for multiple large projects has also been crucial to bring the
code to its current mature stage. We thank Carlos Frenk for support and
motivation in the early stages of this project.

We gratefully acknowledge useful discussions with Edoardo Altamura, Andres
Ar\'{a}mburo-Garcia, Stefan Arridge, Zorry Belcheva, Alejandro
Ben\'{i}tez-Llambay, Heinrich Bockhorst, Alexei Borissov, Peter Boyle, Joey
Braspenning, Jemima Briggs, Florian Cabot, Shaun Cole, Rob Crain, Claudio Dalla
Vecchia, Massimiliano Culpo, Vincent Eke, Pascal Elahi, Azadeh Fattahi,
Johnathan Frawley, Victor Forouhar, Daniel Giles, Cameron Grove, Oliver Hahn,
Patrick Hirling, Fabien Jeanquartier, Adrian Jenkins, Sarah Johnston, Orestis
Karapiperis, Ashley Kelly, Euthymios Kotsialos, Roi Kugel, Claudia Lagos, Angus
Lepper, B\`{a}rbara Levering, Aaron Ludlow, Ian McCarthy, Abouzied Nasar,
R\"{u}diger Pakmor, John Pennycook, Oliver Perks, Joel Pfeffer, Chris Power,
Daniel Price, Lieuwe de Regt, John Regan, Alex Richings, Darwin Roduit, Chris
Rowe, Jaime Salcido, Nikyta Shchutskyi, Volker Springel, Joachim Stadel,
Federico Stasyszyn, Lu\'{i}s Teodoro, Tom Theuns, Rodrigo Tobar, James Trayford,
and Tobias Weinzierl.

This work used the DiRAC@Durham facility managed by the Institute for
Computational Cosmology on behalf of the STFC DiRAC HPC Facility
(\url{www.dirac.ac.uk}). The equipment was funded by BEIS capital funding via
STFC capital grants ST/K00042X/1, ST/P002293/1, ST/R002371/1 and ST/S002502/1,
Durham University and STFC operations grant ST/R000832/1. DiRAC is part of the
National e-Infrastructure.  This work is supported by INTEL through the
establishment of the Institute for Computational Cosmology as an INTEL parallel
computing centre (IPCC). We acknowledge research software engineering support
for this project from the STFC DiRAC High Performance Computing Facility which
helped port the code on different architectures and performed thorough
benchmarking. This work was supported by the Swiss Federal Institute of
Technology in Lausanne (EPFL) through the use of the facilities of its
Scientific IT and Application Support Center (SCITAS) and the University of
Geneva through the usage of Yggdrasil. MS acknowledges support from NWO under
Veni grant number 639.041.749. PD is supported by STFC consolidated grant
ST/T000244/1. MI has been supported by EPSRC's Excalibur programme through its
cross-cutting project EX20-9 \textit{Exposing Parallelism: Task Parallelism}
(Grant ESA 10 CDEL) and the DDWG project \textit{PAX--HPC} (Gant
EP/W026775/1). YMB acknowledges support from NWO under Veni grant number
639.041.751. EC is supported by funding from the European Union's Horizon 2020
research and innovation programme under the Marie Sk\l{}odowska-Curie grant
agreement No 860744 (BiD4BESt). TKC is supported by the E. Margaret Burbidge
Prize Postdoctoral Fellowship from the Brinson Foundation at the Departments of
Astronomy and Astrophysics at the University of Chicago. CC acknowledges the
support of the Dutch Research Council (NWO Veni 192.020). FH is supported by the
STFC grant ST/P006744/1. JAK acknowledges support from a NASA Postdoctoral
Program Fellowship. SP acknowledges support by the Austrian Science Fund (FWF)
grant number V 982-N.S. TDS is supported by STFC grants ST/T506047/1 and
ST/V506643/1. WJR acknowledges funding from Sussex STFC Consolidated Grant
(ST/X001040/1).

\section*{Data availability}
The entirety of the software package presented in this paper, including all the
extensions and many examples, is fully publicly available.  It can be found
alongside an extensive documentation on the website of the project:
\url{www.swiftsim.com}.

\bibliographystyle{mnras}

\DeclareRobustCommand{\VAN}[3]{#3}
\bibliography{./bibliography.bib}

\begin{thebibliography}{}
\makeatletter
\relax
\def\mn@urlcharsother{\let\do\@makeother \do\$\do\&\do\#\do\^\do\_\do\%\do\~}
\def\mn@doi{\begingroup\mn@urlcharsother \@ifnextchar [ {\mn@doi@}
  {\mn@doi@[]}}
\def\mn@doi@[#1]#2{\def\@tempa{#1}\ifx\@tempa\@empty \href
  {http://dx.doi.org/#2} {doi:#2}\else \href {http://dx.doi.org/#2} {#1}\fi
  \endgroup}
\def\mn@eprint#1#2{\mn@eprint@#1:#2::\@nil}
\def\mn@eprint@arXiv#1{\href {http://arxiv.org/abs/#1} {{\tt arXiv:#1}}}
\def\mn@eprint@dblp#1{\href {http://dblp.uni-trier.de/rec/bibtex/#1.xml}
  {dblp:#1}}
\def\mn@eprint@#1:#2:#3:#4\@nil{\def\@tempa {#1}\def\@tempb {#2}\def\@tempc
  {#3}\ifx \@tempc \@empty \let \@tempc \@tempb \let \@tempb \@tempa \fi \ifx
  \@tempb \@empty \def\@tempb {arXiv}\fi \@ifundefined
  {mn@eprint@\@tempb}{\@tempb:\@tempc}{\expandafter \expandafter \csname
  mn@eprint@\@tempb\endcsname \expandafter{\@tempc}}}

\bibitem[\protect\citeauthoryear{{Abramowitz} \& {Stegun}}{{Abramowitz} \&
  {Stegun}}{1965}]{AS64}
{Abramowitz} M.,  {Stegun} I.~A.,  1965, {Handbook of mathematical functions
  with formulas, graphs, and mathematical tables}.
US Government printing office

\bibitem[\protect\citeauthoryear{{Adamek}, {Daverio}, {Durrer}  \&
  {Kunz}}{{Adamek} et~al.}{2016}]{gevolution}
{Adamek} J.,  {Daverio} D.,  {Durrer} R.,   {Kunz} M.,  2016, \mn@doi [\jcap]
  {10.1088/1475-7516/2016/07/053}, \href
  {https://ui.adsabs.harvard.edu/abs/2016JCAP...07..053A} {2016, 053}

\bibitem[\protect\citeauthoryear{{Agertz} et~al.,}{{Agertz}
  et~al.}{2007}]{Agertz2007}
{Agertz} O.,  et~al., 2007, \mn@doi [\mnras]
  {10.1111/j.1365-2966.2007.12183.x}, \href
  {https://ui.adsabs.harvard.edu/abs/2007MNRAS.380..963A} {380, 963}

\bibitem[\protect\citeauthoryear{{Ali-Ha{\"\i}moud} \&
  {Bird}}{{Ali-Ha{\"\i}moud} \& {Bird}}{2013}]{AliHaimoud2013}
{Ali-Ha{\"\i}moud} Y.,  {Bird} S.,  2013, \mn@doi [\mnras]
  {10.1093/mnras/sts286}, \href
  {https://ui.adsabs.harvard.edu/abs/2013MNRAS.428.3375A} {428, 3375}

\bibitem[\protect\citeauthoryear{{Almgren}, {Bell}, {Lijewski}, {Luki{\'c}}  \&
  {Van Andel}}{{Almgren} et~al.}{2013}]{Nyx}
{Almgren} A.~S.,  {Bell} J.~B.,  {Lijewski} M.~J.,  {Luki{\'c}} Z.,   {Van
  Andel} E.,  2013, \mn@doi [\apj] {10.1088/0004-637X/765/1/39}, \href
  {https://ui.adsabs.harvard.edu/abs/2013ApJ...765...39A} {765, 39}

\bibitem[\protect\citeauthoryear{{Alonso Asensio}, {Dalla Vecchia}, {Potter}
  \& {Stadel}}{{Alonso Asensio} et~al.}{2023}]{Alonso2022}
{Alonso Asensio} I.,  {Dalla Vecchia} C.,  {Potter} D.,   {Stadel} J.,  2023,
  \mn@doi [\mnras] {10.1093/mnras/stac3447}, \href
  {https://ui.adsabs.harvard.edu/abs/2023MNRAS.519..300A} {519, 300}

\bibitem[\protect\citeauthoryear{{Altamura}, {Kay}, {Bower}, {Schaller},
  {Bah{\'e}}, {Schaye}, {Borrow}  \& {Towler}}{{Altamura}
  et~al.}{2023}]{Altamura2022}
{Altamura} E.,  {Kay} S.~T.,  {Bower} R.~G.,  {Schaller} M.,  {Bah{\'e}} Y.~M.,
   {Schaye} J.,  {Borrow} J.,   {Towler} I.,  2023, \mn@doi [\mnras]
  {10.1093/mnras/stad342}, \href
  {https://ui.adsabs.harvard.edu/abs/2023MNRAS.520.3164A} {520, 3164}

\bibitem[\protect\citeauthoryear{{Angulo} \& {Hahn}}{{Angulo} \&
  {Hahn}}{2022}]{Angulo2022}
{Angulo} R.~E.,  {Hahn} O.,  2022, \mn@doi [Living Reviews in Computational
  Astrophysics] {10.1007/s41115-021-00013-z}, \href
  {https://ui.adsabs.harvard.edu/abs/2022LRCA....8....1A} {8, 1}

\bibitem[\protect\citeauthoryear{{Aubert} \& {Teyssier}}{{Aubert} \&
  {Teyssier}}{2010}]{Aubert2010}
{Aubert} D.,  {Teyssier} R.,  2010, \mn@doi [\apj]
  {10.1088/0004-637X/724/1/244}, \href
  {https://ui.adsabs.harvard.edu/abs/2010ApJ...724..244A} {724, 244}

\bibitem[\protect\citeauthoryear{Augonnet, Thibault, Namyst  \&
  Wacrenier}{Augonnet et~al.}{2011}]{StarPU}
Augonnet C.,  Thibault S.,  Namyst R.,   Wacrenier P.-A.,  2011, \mn@doi
  [Concurrency and Computation: Practice and Experience, Special Issue:
  Euro-Par 2009] {10.1002/cpe.1631}, 23, 187

\bibitem[\protect\citeauthoryear{{Bagla}}{{Bagla}}{2002}]{Bagla2002}
{Bagla} J.~S.,  2002, \mn@doi [Journal of Astrophysics and Astronomy]
  {10.1007/BF02702282}, \href
  {https://ui.adsabs.harvard.edu/abs/2002JApA...23..185B} {23, 185}

\bibitem[\protect\citeauthoryear{{Bagla} \& {Ray}}{{Bagla} \&
  {Ray}}{2003}]{Bagla2003}
{Bagla} J.~S.,  {Ray} S.,  2003, \mn@doi [\na] {10.1016/S1384-1076(03)00056-3},
  \href {http://adsabs.harvard.edu/abs/2003NewA....8..665B} {8, 665}

\bibitem[\protect\citeauthoryear{{Bah{\'e}} et~al.,}{{Bah{\'e}}
  et~al.}{2022}]{Bahe2022}
{Bah{\'e}} Y.~M.,  et~al., 2022, \mn@doi [\mnras] {10.1093/mnras/stac1339},
  \href {https://ui.adsabs.harvard.edu/abs/2022MNRAS.516..167B} {516, 167}

\bibitem[\protect\citeauthoryear{{Balsara}}{{Balsara}}{1989}]{Balsara1989}
{Balsara} D.~S.,  1989, PhD thesis, -

\bibitem[\protect\citeauthoryear{{Balsara}}{{Balsara}}{1995}]{Balsara1995}
{Balsara} D.~S.,  1995, \mn@doi [J. Comput. Phys.]
  {10.1016/S0021-9991(95)90221-X}, \href
  {https://ui.adsabs.harvard.edu/\#abs/1995JCoPh.121..357B} {121, 357}

\bibitem[\protect\citeauthoryear{{Barnes} \& {Hut}}{{Barnes} \&
  {Hut}}{1986}]{Barnes1986}
{Barnes} J.,  {Hut} P.,  1986, \mn@doi [\nat] {10.1038/324446a0}, \href
  {http://adsabs.harvard.edu/abs/1986Natur.324..446B} {324, 446}

\bibitem[\protect\citeauthoryear{{Bate} \& {Burkert}}{{Bate} \&
  {Burkert}}{1997}]{Bate1997}
{Bate} M.~R.,  {Burkert} A.,  1997, \mn@doi [\mnras]
  {10.1093/mnras/288.4.1060}, \href
  {https://ui.adsabs.harvard.edu/abs/1997MNRAS.288.1060B} {288, 1060}

\bibitem[\protect\citeauthoryear{Blumofe, Joerg, Kuszmaul, Leiserson, Randall
  \& Zhou}{Blumofe et~al.}{1995}]{Cilk}
Blumofe R.~D.,  Joerg C.~F.,  Kuszmaul B.~C.,  Leiserson C.~E.,  Randall K.~H.,
    Zhou Y.,  1995, Cilk: An efficient multithreaded runtime system.
~ Vol. 30, ACM New York, NY, USA

\bibitem[\protect\citeauthoryear{Boehm}{Boehm}{2000}]{cocomo}
Boehm B.,  2000, Software Cost Estimation with Cocomo II.
No. vol.~1 in Software Cost Estimation with Cocomo II, Prentice Hall

\bibitem[\protect\citeauthoryear{{Bondi}}{{Bondi}}{1952}]{Bondi1952}
{Bondi} H.,  1952, \mn@doi [\mnras] {10.1093/mnras/112.2.195}, \href
  {https://ui.adsabs.harvard.edu/abs/1952MNRAS.112..195B} {112, 195}

\bibitem[\protect\citeauthoryear{{Booth} \& {Schaye}}{{Booth} \&
  {Schaye}}{2009}]{Booth2009}
{Booth} C.~M.,  {Schaye} J.,  2009, \mn@doi [\mnras]
  {10.1111/j.1365-2966.2009.15043.x}, \href
  {https://ui.adsabs.harvard.edu/abs/2009MNRAS.398...53B} {398, 53}

\bibitem[\protect\citeauthoryear{Borrow \& Borrisov}{Borrow \&
  Borrisov}{2020}]{swiftsimio}
Borrow J.,  Borrisov A.,  2020, \mn@doi [Journal of Open Source Software]
  {10.21105/joss.02430}, 5, 2430

\bibitem[\protect\citeauthoryear{{Borrow}, {Bower}, {Draper}, {Gonnet}  \&
  {Schaller}}{{Borrow} et~al.}{2018}]{Borrow2018}
{Borrow} J.,  {Bower} R.~G.,  {Draper} P.~W.,  {Gonnet} P.,   {Schaller} M.,
  2018, in {Proceedings of the 13th SPHERIC International Workshop}. pp 44--51
  (\mn@eprint {arXiv} {1807.01341})

\bibitem[\protect\citeauthoryear{{Borrow}, {Schaller}  \& {Bower}}{{Borrow}
  et~al.}{2021}]{Borrow2021}
{Borrow} J.,  {Schaller} M.,   {Bower} R.~G.,  2021, \mn@doi [\mnras]
  {10.1093/mnras/stab1423}, \href
  {https://ui.adsabs.harvard.edu/abs/2021MNRAS.505.2316B} {505, 2316}

\bibitem[\protect\citeauthoryear{{Borrow}, {Schaller}, {Bower}  \&
  {Schaye}}{{Borrow} et~al.}{2022}]{Borrow2022}
{Borrow} J.,  {Schaller} M.,  {Bower} R.~G.,   {Schaye} J.,  2022, \mn@doi
  [\mnras] {10.1093/mnras/stab3166}, \href
  {https://ui.adsabs.harvard.edu/abs/2022MNRAS.511.2367B} {511, 2367}

\bibitem[\protect\citeauthoryear{{Borrow}, {Schaller}, {Bah{\'e}}, {Schaye},
  {Ludlow}, {Ploeckinger}, {Nobels}  \& {Altamura}}{{Borrow}
  et~al.}{2023a}]{Borrow2023random}
{Borrow} J.,  {Schaller} M.,  {Bah{\'e}} Y.~M.,  {Schaye} J.,  {Ludlow} A.~D.,
  {Ploeckinger} S.,  {Nobels} F. S.~J.,   {Altamura} E.,  2023a, \mn@doi
  [\mnras] {10.1093/mnras/stad2928}, \href
  {https://ui.adsabs.harvard.edu/abs/2023MNRAS.526.2441B} {526, 2441}

\bibitem[\protect\citeauthoryear{{Borrow}, {Schaller}, {Bah{\'e}}, {Schaye},
  {Ludlow}, {Ploeckinger}, {Nobels}  \& {Altamura}}{{Borrow}
  et~al.}{2023b}]{BorrowStochastic}
{Borrow} J.,  {Schaller} M.,  {Bah{\'e}} Y.~M.,  {Schaye} J.,  {Ludlow} A.~D.,
  {Ploeckinger} S.,  {Nobels} F. S.~J.,   {Altamura} E.,  2023b, \mn@doi
  [\mnras] {10.1093/mnras/stad2928}, \href
  {https://ui.adsabs.harvard.edu/abs/2023MNRAS.526.2441B} {526, 2441}

\bibitem[\protect\citeauthoryear{{Borrow} et~al.}{{Borrow}
  et~al.}{2024}]{Borrow2022calibration}
{Borrow} J.,  et~al., 2024, in prep.

\bibitem[\protect\citeauthoryear{Bower, Rogers  \& Schaller}{Bower
  et~al.}{2022}]{Bower2021}
Bower R.,  Rogers B.~D.,   Schaller M.,  2022, \mn@doi [Computing in Science &
  Engineering] {10.1109/MCSE.2021.3134604}, 24, 14

\bibitem[\protect\citeauthoryear{{Braspenning}, {Schaye}, {Borrow}  \&
  {Schaller}}{{Braspenning} et~al.}{2023}]{Braspenning2022}
{Braspenning} J.,  {Schaye} J.,  {Borrow} J.,   {Schaller} M.,  2023, \mn@doi
  [\mnras] {10.1093/mnras/stad1243}, \href
  {https://ui.adsabs.harvard.edu/abs/2023MNRAS.523.1280B} {523, 1280}

\bibitem[\protect\citeauthoryear{{Bryan} et~al.,}{{Bryan} et~al.}{2014}]{ENZO}
{Bryan} G.~L.,  et~al., 2014, \mn@doi [\apjs] {10.1088/0067-0049/211/2/19},
  \href {https://ui.adsabs.harvard.edu/abs/2014ApJS..211...19B} {211, 19}

\bibitem[\protect\citeauthoryear{{Chabrier} \& {Debras}}{{Chabrier} \&
  {Debras}}{2021}]{Chabrier+2021}
{Chabrier} G.,  {Debras} F.,  2021, \mn@doi [\apj] {10.3847/1538-4357/abfc48},
  \href {https://ui.adsabs.harvard.edu/abs/2021ApJ...917....4C} {917, 4}

\bibitem[\protect\citeauthoryear{{Chaikin}, {Schaye}, {Schaller}, {Bah{\'e}},
  {Nobels}  \& {Ploeckinger}}{{Chaikin} et~al.}{2022}]{Chaikin2022}
{Chaikin} E.,  {Schaye} J.,  {Schaller} M.,  {Bah{\'e}} Y.~M.,  {Nobels} F.
  S.~J.,   {Ploeckinger} S.,  2022, \mn@doi [\mnras] {10.1093/mnras/stac1132},
  \href {https://ui.adsabs.harvard.edu/abs/2022MNRAS.514..249C} {514, 249}

\bibitem[\protect\citeauthoryear{{Chaikin}, {Schaye}, {Schaller},
  {Ben{\'\i}tez-Llambay}, {Nobels}  \& {Ploeckinger}}{{Chaikin}
  et~al.}{2023}]{Chaikin2023}
{Chaikin} E.,  {Schaye} J.,  {Schaller} M.,  {Ben{\'\i}tez-Llambay} A.,
  {Nobels} F. S.~J.,   {Ploeckinger} S.,  2023, \mn@doi [\mnras]
  {10.1093/mnras/stad1626}, \href
  {https://ui.adsabs.harvard.edu/abs/2023MNRAS.523.3709C} {523, 3709}

\bibitem[\protect\citeauthoryear{{Chan}, {Theuns}, {Bower}  \& {Frenk}}{{Chan}
  et~al.}{2021}]{Chan2021}
{Chan} T.~K.,  {Theuns} T.,  {Bower} R.,   {Frenk} C.,  2021, \mn@doi [\mnras]
  {10.1093/mnras/stab1686}, \href
  {https://ui.adsabs.harvard.edu/abs/2021MNRAS.505.5784C} {505, 5784}

\bibitem[\protect\citeauthoryear{Cheng, Greengard  \& Rokhlin}{Cheng
  et~al.}{1999}]{Cheng1999}
Cheng H.,  Greengard L.,   Rokhlin V.,  1999, \mn@doi [Journal of Computational
  Physics] {http://dx.doi.org/10.1006/jcph.1999.6355}, 155, 468

\bibitem[\protect\citeauthoryear{{Colombi}, {Jaffe}, {Novikov}  \&
  {Pichon}}{{Colombi} et~al.}{2009}]{Colombi2009}
{Colombi} S.,  {Jaffe} A.,  {Novikov} D.,   {Pichon} C.,  2009, \mn@doi
  [\mnras] {10.1111/j.1365-2966.2008.14176.x}, \href
  {https://ui.adsabs.harvard.edu/abs/2009MNRAS.393..511C} {393, 511}

\bibitem[\protect\citeauthoryear{{Correa}, {Schaller}, {Ploeckinger}, {Anau
  Montel}, {Weniger}  \& {Ando}}{{Correa} et~al.}{2022}]{Correa2022}
{Correa} C.~A.,  {Schaller} M.,  {Ploeckinger} S.,  {Anau Montel} N.,
  {Weniger} C.,   {Ando} S.,  2022, \mn@doi [\mnras] {10.1093/mnras/stac2830},
  \href {https://ui.adsabs.harvard.edu/abs/2022MNRAS.517.3045C} {517, 3045}

\bibitem[\protect\citeauthoryear{{Couchman}, {Thomas}  \& {Pearce}}{{Couchman}
  et~al.}{1995}]{Couchman1995}
{Couchman} H.~M.~P.,  {Thomas} P.~A.,   {Pearce} F.~R.,  1995, \mn@doi [\apj]
  {10.1086/176348}, \href
  {https://ui.adsabs.harvard.edu/abs/1995ApJ...452..797C} {452, 797}

\bibitem[\protect\citeauthoryear{{Crain} \& {van de Voort}}{{Crain} \& {van de
  Voort}}{2023}]{Crain2023}
{Crain} R.~A.,  {van de Voort} F.,  2023, \mn@doi [\araa]
  {10.1146/annurev-astro-041923-043618}, \href
  {https://ui.adsabs.harvard.edu/abs/2023ARA&A..61..473C} {61, 473}

\bibitem[\protect\citeauthoryear{{Crain} et~al.,}{{Crain}
  et~al.}{2015}]{Crain2015}
{Crain} R.~A.,  et~al., 2015, \mn@doi [\mnras] {10.1093/mnras/stv725}, \href
  {https://ui.adsabs.harvard.edu/abs/2015MNRAS.450.1937C} {450, 1937}

\bibitem[\protect\citeauthoryear{{Creasey}}{{Creasey}}{2018}]{Creasey2018}
{Creasey} P.,  2018, \mn@doi [Astronomy and Computing]
  {10.1016/j.ascom.2018.09.010}, \href
  {https://ui.adsabs.harvard.edu/abs/2018A&C....25..159C} {25, 159}

\bibitem[\protect\citeauthoryear{{Croton}}{{Croton}}{2013}]{Croton2013}
{Croton} D.~J.,  2013, \mn@doi [\pasa] {10.1017/pasa.2013.31}, \href
  {https://ui.adsabs.harvard.edu/abs/2013PASA...30...52C} {30, e052}

\bibitem[\protect\citeauthoryear{{Cullen} \& {Dehnen}}{{Cullen} \&
  {Dehnen}}{2010}]{Cullen2010}
{Cullen} L.,  {Dehnen} W.,  2010, \mn@doi [\mnras]
  {10.1111/j.1365-2966.2010.17158.x}, \href
  {https://ui.adsabs.harvard.edu/abs/2010MNRAS.408..669C} {408, 669}

\bibitem[\protect\citeauthoryear{{Dalla Vecchia} \& {Schaye}}{{Dalla Vecchia}
  \& {Schaye}}{2012}]{DV2012}
{Dalla Vecchia} C.,  {Schaye} J.,  2012, \mn@doi [\mnras]
  {10.1111/j.1365-2966.2012.21704.x}, \href
  {https://ui.adsabs.harvard.edu/abs/2012MNRAS.426..140D} {426, 140}

\bibitem[\protect\citeauthoryear{{Dav{\'e}}, {Dubinski}  \&
  {Hernquist}}{{Dav{\'e}} et~al.}{1997}]{Dave1997}
{Dav{\'e}} R.,  {Dubinski} J.,   {Hernquist} L.,  1997, \mn@doi [\na]
  {10.1016/S1384-1076(97)00019-5}, \href
  {https://ui.adsabs.harvard.edu/abs/1997NewA....2..277D} {2, 277}

\bibitem[\protect\citeauthoryear{{Davis}, {Efstathiou}, {Frenk}  \&
  {White}}{{Davis} et~al.}{1985}]{Davis1985}
{Davis} M.,  {Efstathiou} G.,  {Frenk} C.~S.,   {White} S.~D.~M.,  1985,
  \mn@doi [\apj] {10.1086/163168}, \href
  {https://ui.adsabs.harvard.edu/abs/1985ApJ...292..371D} {292, 371}

\bibitem[\protect\citeauthoryear{{Dehnen}}{{Dehnen}}{2000}]{Dehnen2000}
{Dehnen} W.,  2000, \mn@doi [\apjl] {10.1086/312724}, \href
  {http://adsabs.harvard.edu/abs/2000ApJ...536L..39D} {536, L39}

\bibitem[\protect\citeauthoryear{{Dehnen}}{{Dehnen}}{2001}]{Dehnen2001}
{Dehnen} W.,  2001, \mn@doi [\mnras] {10.1046/j.1365-8711.2001.04237.x}, \href
  {http://adsabs.harvard.edu/abs/2001MNRAS.324..273D} {324, 273}

\bibitem[\protect\citeauthoryear{{Dehnen}}{{Dehnen}}{2002}]{Dehnen2002}
{Dehnen} W.,  2002, \mn@doi [Journal of Computational Physics]
  {10.1006/jcph.2002.7026}, \href
  {http://adsabs.harvard.edu/abs/2002JCoPh.179...27D} {179, 27}

\bibitem[\protect\citeauthoryear{{Dehnen}}{{Dehnen}}{2014}]{Dehnen2014}
{Dehnen} W.,  2014, \mn@doi [Computational Astrophysics and Cosmology]
  {10.1186/s40668-014-0001-7}, \href
  {http://adsabs.harvard.edu/abs/2014ComAC...1....1D} {1, 1}

\bibitem[\protect\citeauthoryear{{Dehnen} \& {Aly}}{{Dehnen} \&
  {Aly}}{2012}]{Dehnen2012}
{Dehnen} W.,  {Aly} H.,  2012, \mn@doi [\mnras]
  {10.1111/j.1365-2966.2012.21439.x}, \href
  {https://ui.adsabs.harvard.edu/abs/2012MNRAS.425.1068D} {425, 1068}

\bibitem[\protect\citeauthoryear{{Dehnen} \& {Read}}{{Dehnen} \&
  {Read}}{2011}]{Dehnen2011}
{Dehnen} W.,  {Read} J.~I.,  2011, \mn@doi [European Physical Journal Plus]
  {10.1140/epjp/i2011-11055-3}, \href
  {https://ui.adsabs.harvard.edu/abs/2011EPJP..126...55D} {126, 55}

\bibitem[\protect\citeauthoryear{{Driver} et~al.,}{{Driver}
  et~al.}{2022}]{Driver2022}
{Driver} S.~P.,  et~al., 2022, \mn@doi [\mnras] {10.1093/mnras/stac472}, \href
  {https://ui.adsabs.harvard.edu/abs/2022MNRAS.513..439D} {513, 439}

\bibitem[\protect\citeauthoryear{{Duncan}, {Levison}  \& {Lee}}{{Duncan}
  et~al.}{1998}]{Duncan1998}
{Duncan} M.~J.,  {Levison} H.~F.,   {Lee} M.~H.,  1998, \mn@doi [\aj]
  {10.1086/300541}, \href
  {https://ui.adsabs.harvard.edu/abs/1998AJ....116.2067D} {116, 2067}

\bibitem[\protect\citeauthoryear{{Durier} \& {Dalla Vecchia}}{{Durier} \&
  {Dalla Vecchia}}{2012}]{Durier2012}
{Durier} F.,  {Dalla Vecchia} C.,  2012, \mn@doi [\mnras]
  {10.1111/j.1365-2966.2011.19712.x}, \href
  {https://ui.adsabs.harvard.edu/abs/2012MNRAS.419..465D} {419, 465}

\bibitem[\protect\citeauthoryear{{Elahi}, {Thacker}  \& {Widrow}}{{Elahi}
  et~al.}{2011}]{Elahi2011}
{Elahi} P.~J.,  {Thacker} R.~J.,   {Widrow} L.~M.,  2011, \mn@doi [\mnras]
  {10.1111/j.1365-2966.2011.19485.x}, \href
  {https://ui.adsabs.harvard.edu/abs/2011MNRAS.418..320E} {418, 320}

\bibitem[\protect\citeauthoryear{{Elahi}, {Ca{\~n}as}, {Poulton}, {Tobar},
  {Willis}, {Lagos}, {Power}  \& {Robotham}}{{Elahi} et~al.}{2019}]{Elahi2019}
{Elahi} P.~J.,  {Ca{\~n}as} R.,  {Poulton} R. J.~J.,  {Tobar} R.~J.,  {Willis}
  J.~S.,  {Lagos} C. d.~P.,  {Power} C.,   {Robotham} A. S.~G.,  2019, \mn@doi
  [\pasa] {10.1017/pasa.2019.12}, \href
  {https://ui.adsabs.harvard.edu/abs/2019PASA...36...21E} {36, 21}

\bibitem[\protect\citeauthoryear{{Elbers}}{{Elbers}}{2022}]{Elbers2022b}
{Elbers} W.,  2022, \mn@doi [\jcap] {10.1088/1475-7516/2022/11/058}, \href
  {https://ui.adsabs.harvard.edu/abs/2022JCAP...11..058E} {2022, 058}

\bibitem[\protect\citeauthoryear{{Elbers}, {Frenk}, {Jenkins}, {Li}  \&
  {Pascoli}}{{Elbers} et~al.}{2021}]{Elbers2020}
{Elbers} W.,  {Frenk} C.~S.,  {Jenkins} A.,  {Li} B.,   {Pascoli} S.,  2021,
  \mn@doi [\mnras] {10.1093/mnras/stab2260}, \href
  {https://ui.adsabs.harvard.edu/abs/2021MNRAS.507.2614E} {507, 2614}

\bibitem[\protect\citeauthoryear{{Ewald}}{{Ewald}}{1921}]{Ewald1921}
{Ewald} P.~P.,  1921, \mn@doi [Annalen der Physik] {10.1002/andp.19213690304},
  \href {http://adsabs.harvard.edu/abs/1921AnP...369..253E} {369, 253}

\bibitem[\protect\citeauthoryear{{Faber}, {Stibbe}, {Portegies Zwart},
  {McMillan}  \& {Boily}}{{Faber} et~al.}{2010}]{Faber2010}
{Faber} N.~T.,  {Stibbe} D.,  {Portegies Zwart} S.,  {McMillan} S.~L.~W.,
  {Boily} C.~M.,  2010, \mn@doi [\mnras] {10.1111/j.1365-2966.2009.15775.x},
  \href {https://ui.adsabs.harvard.edu/abs/2010MNRAS.401.1898F} {401, 1898}

\bibitem[\protect\citeauthoryear{{Faucher-Gigu{\`e}re}}{{Faucher-Gigu{\`e}re}}{2020}]{FG2020}
{Faucher-Gigu{\`e}re} C.-A.,  2020, \mn@doi [\mnras] {10.1093/mnras/staa302},
  \href {https://ui.adsabs.harvard.edu/abs/2020MNRAS.493.1614F} {493, 1614}

\bibitem[\protect\citeauthoryear{{Ferland}, {Korista}, {Verner}, {Ferguson},
  {Kingdon}  \& {Verner}}{{Ferland} et~al.}{1998}]{Ferland1998}
{Ferland} G.~J.,  {Korista} K.~T.,  {Verner} D.~A.,  {Ferguson} J.~W.,
  {Kingdon} J.~B.,   {Verner} E.~M.,  1998, \mn@doi [\pasp] {10.1086/316190},
  \href {https://ui.adsabs.harvard.edu/abs/1998PASP..110..761F} {110, 761}

\bibitem[\protect\citeauthoryear{{Ferland} et~al.,}{{Ferland}
  et~al.}{2017}]{Ferland2017}
{Ferland} G.~J.,  et~al., 2017, \mn@doi [\rmxaa] {10.48550/arXiv.1705.10877},
  \href {https://ui.adsabs.harvard.edu/abs/2017RMxAA..53..385F} {53, 385}

\bibitem[\protect\citeauthoryear{Frigo \& Johnson}{Frigo \&
  Johnson}{2005}]{fftw}
Frigo M.,  Johnson S.~G.,  2005, Proceedings of the IEEE, 93, 216

\bibitem[\protect\citeauthoryear{{Frontiere}, {Raskin}  \& {Owen}}{{Frontiere}
  et~al.}{2017}]{Frontiere2017}
{Frontiere} N.,  {Raskin} C.~D.,   {Owen} J.~M.,  2017, \mn@doi [Journal of
  Computational Physics] {10.1016/j.jcp.2016.12.004}, \href
  {https://ui.adsabs.harvard.edu/abs/2017JCoPh.332..160F} {332, 160}

\bibitem[\protect\citeauthoryear{{Fryxell} et~al.,}{{Fryxell}
  et~al.}{2000}]{flash}
{Fryxell} B.,  et~al., 2000, \mn@doi [\apjs] {10.1086/317361}, \href
  {https://ui.adsabs.harvard.edu/abs/2000ApJS..131..273F} {131, 273}

\bibitem[\protect\citeauthoryear{{Gaburov} \& {Nitadori}}{{Gaburov} \&
  {Nitadori}}{2011}]{Gaburov2011}
{Gaburov} E.,  {Nitadori} K.,  2011, \mn@doi [\mnras]
  {10.1111/j.1365-2966.2011.18313.x}, \href
  {https://ui.adsabs.harvard.edu/abs/2011MNRAS.414..129G} {414, 129}

\bibitem[\protect\citeauthoryear{Galler \& Fisher}{Galler \&
  Fisher}{1964}]{UnionFind}
Galler B.~A.,  Fisher M.~J.,  1964, \mn@doi [Commun. ACM]
  {10.1145/364099.364331}, 7, 301–303

\bibitem[\protect\citeauthoryear{{Garrison}, {Eisenstein}  \&
  {Pinto}}{{Garrison} et~al.}{2019}]{Garrison2019}
{Garrison} L.~H.,  {Eisenstein} D.~J.,   {Pinto} P.~A.,  2019, \mn@doi [\mnras]
  {10.1093/mnras/stz634}, \href
  {https://ui.adsabs.harvard.edu/abs/2019MNRAS.485.3370G} {485, 3370}

\bibitem[\protect\citeauthoryear{{Garrison}, {Eisenstein}, {Ferrer},
  {Maksimova}  \& {Pinto}}{{Garrison} et~al.}{2021}]{Abacus}
{Garrison} L.~H.,  {Eisenstein} D.~J.,  {Ferrer} D.,  {Maksimova} N.~A.,
  {Pinto} P.~A.,  2021, \mn@doi [\mnras] {10.1093/mnras/stab2482}, \href
  {https://ui.adsabs.harvard.edu/abs/2021MNRAS.508..575G} {508, 575}

\bibitem[\protect\citeauthoryear{{Garzilli}, {Magalich}, {Theuns}, {Frenk},
  {Weniger}, {Ruchayskiy}  \& {Boyarsky}}{{Garzilli}
  et~al.}{2019}]{Garzilli2019}
{Garzilli} A.,  {Magalich} A.,  {Theuns} T.,  {Frenk} C.~S.,  {Weniger} C.,
  {Ruchayskiy} O.,   {Boyarsky} A.,  2019, \mn@doi [\mnras]
  {10.1093/mnras/stz2188}, \href
  {https://ui.adsabs.harvard.edu/abs/2019MNRAS.489.3456G} {489, 3456}

\bibitem[\protect\citeauthoryear{{Gingold} \& {Monaghan}}{{Gingold} \&
  {Monaghan}}{1977}]{Monaghan1977}
{Gingold} R.~A.,  {Monaghan} J.~J.,  1977, \mn@doi [\mnras]
  {10.1093/mnras/181.3.375}, \href
  {https://ui.adsabs.harvard.edu/abs/1977MNRAS.181..375G} {181, 375}

\bibitem[\protect\citeauthoryear{Goldbaum, ZuHone, Turk, Kowalik  \&
  Rosen}{Goldbaum et~al.}{2018}]{unyt}
Goldbaum N.~J.,  ZuHone J.~A.,  Turk M.~J.,  Kowalik K.,   Rosen A.~L.,  2018,
  \mn@doi [Journal of Open Source Software] {10.21105/joss.00809}, 3, 809

\bibitem[\protect\citeauthoryear{Gonnet}{Gonnet}{2013}]{Gonnet2013}
Gonnet P.,  2013, \mn@doi [Molecular Simulation]
  {10.1080/08927022.2012.762097}, 39, 721

\bibitem[\protect\citeauthoryear{Gonnet}{Gonnet}{2015}]{Gonnet2015}
Gonnet P.,  2015, \mn@doi [SIAM Journal on Scientific Computing]
  {10.1137/140964266}, 37, C95

\bibitem[\protect\citeauthoryear{{Gonnet}, {Chalk}  \& {Schaller}}{{Gonnet}
  et~al.}{2016}]{Gonnet2016}
{Gonnet} P.,  {Chalk} A. B.~G.,   {Schaller} M.,  2016, arXiv e-prints, \href
  {https://ui.adsabs.harvard.edu/abs/2016arXiv160105384G} {p. arXiv:1601.05384}

\bibitem[\protect\citeauthoryear{{G{\'o}rski}, {Hivon}, {Banday}, {Wandelt},
  {Hansen}, {Reinecke}  \& {Bartelmann}}{{G{\'o}rski}
  et~al.}{2005}]{Gorski2005}
{G{\'o}rski} K.~M.,  {Hivon} E.,  {Banday} A.~J.,  {Wandelt} B.~D.,  {Hansen}
  F.~K.,  {Reinecke} M.,   {Bartelmann} M.,  2005, \mn@doi [\apj]
  {10.1086/427976}, \href
  {https://ui.adsabs.harvard.edu/abs/2005ApJ...622..759G} {622, 759}

\bibitem[\protect\citeauthoryear{Gough}{Gough}{2009}]{GSL}
Gough B.,  2009, GNU Scientific Library Reference Manual - Third Edition, 3rd
  edn.
Network Theory Ltd.

\bibitem[\protect\citeauthoryear{{\VAN{Graaff}{de}{de}}~Graaff, {Trayford},
  {Franx}, {Schaller}, {Schaye}  \& {van der
  Wel}}{{\VAN{Graaff}{de}{de}}~Graaff et~al.}{2022}]{Graaff2022}
{\VAN{Graaff}{de}{de}}~Graaff A.,  {Trayford} J.,  {Franx} M.,  {Schaller} M.,
  {Schaye} J.,   {van der Wel} A.,  2022, \mn@doi [\mnras]
  {10.1093/mnras/stab3510}, \href
  {https://ui.adsabs.harvard.edu/abs/2022MNRAS.511.2544D} {511, 2544}

\bibitem[\protect\citeauthoryear{Greengard \& Rokhlin}{Greengard \&
  Rokhlin}{1987}]{Greengard1987}
Greengard L.,  Rokhlin V.,  1987, \mn@doi [Journal of Computational Physics]
  {http://dx.doi.org/10.1016/0021-9991(87)90140-9}, 73, 325

\bibitem[\protect\citeauthoryear{{Greif}, {Glover}, {Bromm}  \&
  {Klessen}}{{Greif} et~al.}{2009}]{Greif2009}
{Greif} T.~H.,  {Glover} S. C.~O.,  {Bromm} V.,   {Klessen} R.~S.,  2009,
  \mn@doi [\mnras] {10.1111/j.1365-2966.2008.14169.x}, \href
  {https://ui.adsabs.harvard.edu/abs/2009MNRAS.392.1381G} {392, 1381}

\bibitem[\protect\citeauthoryear{{Grove} et~al.,}{{Grove}
  et~al.}{2022}]{Grove2021}
{Grove} C.,  et~al., 2022, \mn@doi [\mnras] {10.1093/mnras/stac1947}, \href
  {https://ui.adsabs.harvard.edu/abs/2022MNRAS.515.1854G} {515, 1854}

\bibitem[\protect\citeauthoryear{{Haardt} \& {Madau}}{{Haardt} \&
  {Madau}}{2001}]{HM2001}
{Haardt} F.,  {Madau} P.,  2001, in {Neumann} D.~M.,  {Tran} J.~T.~V.,  eds,
  Clusters of Galaxies and the High Redshift Universe Observed in X-rays. p.~64
  (\mn@eprint {arXiv} {astro-ph/0106018}),
  \mn@doi{10.48550/arXiv.astro-ph/0106018}

\bibitem[\protect\citeauthoryear{{Haardt} \& {Madau}}{{Haardt} \&
  {Madau}}{2012}]{Haardt2012}
{Haardt} F.,  {Madau} P.,  2012, \mn@doi [\apj] {10.1088/0004-637X/746/2/125},
  \href {https://ui.adsabs.harvard.edu/abs/2012ApJ...746..125H} {746, 125}

\bibitem[\protect\citeauthoryear{{Habib} et~al.,}{{Habib} et~al.}{2016}]{HACC}
{Habib} S.,  et~al., 2016, \mn@doi [\na] {10.1016/j.newast.2015.06.003}, \href
  {https://ui.adsabs.harvard.edu/abs/2016NewA...42...49H} {42, 49}

\bibitem[\protect\citeauthoryear{{Hahn}, {Rampf}  \& {Uhlemann}}{{Hahn}
  et~al.}{2021}]{Hahn2021}
{Hahn} O.,  {Rampf} C.,   {Uhlemann} C.,  2021, \mn@doi [\mnras]
  {10.1093/mnras/staa3773}, \href
  {https://ui.adsabs.harvard.edu/abs/2021MNRAS.503..426H} {503, 426}

\bibitem[\protect\citeauthoryear{{Harnois-D{\'e}raps}, {Pen}, {Iliev}, {Merz},
  {Emberson}  \& {Desjacques}}{{Harnois-D{\'e}raps} et~al.}{2013}]{CUBEPM}
{Harnois-D{\'e}raps} J.,  {Pen} U.-L.,  {Iliev} I.~T.,  {Merz} H.,  {Emberson}
  J.~D.,   {Desjacques} V.,  2013, \mn@doi [\mnras] {10.1093/mnras/stt1591},
  \href {https://ui.adsabs.harvard.edu/abs/2013MNRAS.436..540H} {436, 540}

\bibitem[\protect\citeauthoryear{{Harvey}, {Revaz}, {Robertson}  \&
  {Hausammann}}{{Harvey} et~al.}{2018}]{Harvey2018}
{Harvey} D.,  {Revaz} Y.,  {Robertson} A.,   {Hausammann} L.,  2018, \mn@doi
  [\mnras] {10.1093/mnrasl/sly159}, \href
  {https://ui.adsabs.harvard.edu/abs/2018MNRAS.481L..89H} {481, L89}

\bibitem[\protect\citeauthoryear{Hausammann}{Hausammann}{2021}]{LoicThesis}
Hausammann L.,  2021, PhD thesis, Ecole Polytechnique Fédérale de Lausanne,
  Lausanne

\bibitem[\protect\citeauthoryear{{Hausammann}, {Revaz}  \&
  {Jablonka}}{{Hausammann} et~al.}{2019}]{Hausammann2019}
{Hausammann} L.,  {Revaz} Y.,   {Jablonka} P.,  2019, \mn@doi [\aap]
  {10.1051/0004-6361/201834871}, \href
  {https://ui.adsabs.harvard.edu/abs/2019A&A...624A..11H} {624, A11}

\bibitem[\protect\citeauthoryear{{Hausammann}, {Gonnet}  \&
  {Schaller}}{{Hausammann} et~al.}{2022}]{Hausammann2022}
{Hausammann} L.,  {Gonnet} P.,   {Schaller} M.,  2022, \mn@doi [Astronomy and
  Computing] {10.1016/j.ascom.2022.100659}, \href
  {https://ui.adsabs.harvard.edu/abs/2022A&C....4100659H} {41, 100659}

\bibitem[\protect\citeauthoryear{{Heitmann} et~al.,}{{Heitmann}
  et~al.}{2008}]{Heitmann2008}
{Heitmann} K.,  et~al., 2008, \mn@doi [Computational Science and Discovery]
  {10.1088/1749-4699/1/1/015003}, \href
  {https://ui.adsabs.harvard.edu/abs/2008CS&D....1a5003H} {1, 015003}

\bibitem[\protect\citeauthoryear{{Hernquist}}{{Hernquist}}{1990}]{hernquist1990}
{Hernquist} L.,  1990, \mn@doi [\apj] {10.1086/168845}, \href
  {https://ui.adsabs.harvard.edu/abs/1990ApJ...356..359H} {356, 359}

\bibitem[\protect\citeauthoryear{{Hernquist} \& {Katz}}{{Hernquist} \&
  {Katz}}{1989}]{Hernquist1989}
{Hernquist} L.,  {Katz} N.,  1989, \mn@doi [\apjs] {10.1086/191344}, \href
  {http://adsabs.harvard.edu/abs/1989ApJS...70..419H} {70, 419}

\bibitem[\protect\citeauthoryear{Hietel, Junk, Keck  \& Teleaga}{Hietel
  et~al.}{2001}]{Hietel2001}
Hietel D.,  Junk M.,  Keck R.,   Teleaga D.,  2001, in Proceedings of {{GAMM
  Workshop}} "{{Discrete Modelling}} and Discrete {{Algorithms}} in {{Continuum
  Mechanics}}". p.~10

\bibitem[\protect\citeauthoryear{Hietel, Junk, Kuhnert  \& Tiwari}{Hietel
  et~al.}{2005}]{Hietel2005}
Hietel D.,  Junk M.,  Kuhnert J.,   Tiwari S.,  2005, Analysis and Numerics for
  Conservation Laws (G. Warnecke Edt.), pp 339--362

\bibitem[\protect\citeauthoryear{{Hockney} \& {Eastwood}}{{Hockney} \&
  {Eastwood}}{1988}]{Hockney1988}
{Hockney} R.~W.,  {Eastwood} J.~W.,  1988, {Computer simulation using
  particles}.
{CRC Press}

\bibitem[\protect\citeauthoryear{{Hopkins}}{{Hopkins}}{2013}]{Hopkins2013}
{Hopkins} P.~F.,  2013, \mn@doi [\mnras] {10.1093/mnras/sts210}, \href
  {http://adsabs.harvard.edu/abs/2013MNRAS.428.2840H} {428, 2840}

\bibitem[\protect\citeauthoryear{{Hopkins}}{{Hopkins}}{2015}]{Hopkins2015}
{Hopkins} P.~F.,  2015, \mn@doi [\mnras] {10.1093/mnras/stv195}, \href
  {https://ui.adsabs.harvard.edu/abs/2015MNRAS.450...53H} {450, 53}

\bibitem[\protect\citeauthoryear{{Hopkins}, {Quataert}  \& {Murray}}{{Hopkins}
  et~al.}{2011}]{hopkins2011}
{Hopkins} P.~F.,  {Quataert} E.,   {Murray} N.,  2011, \mn@doi [\mnras]
  {10.1111/j.1365-2966.2011.19306.x}, \href
  {https://ui.adsabs.harvard.edu/abs/2011MNRAS.417..950H} {417, 950}

\bibitem[\protect\citeauthoryear{{Hopkins}, {Nadler}, {Grudi{\'c}}, {Shen},
  {Sands}  \& {Jiang}}{{Hopkins} et~al.}{2023}]{Hopkins2023}
{Hopkins} P.~F.,  {Nadler} E.~O.,  {Grudi{\'c}} M.~Y.,  {Shen} X.,  {Sands} I.,
    {Jiang} F.,  2023, \mn@doi [\mnras] {10.1093/mnras/stad2548}, \href
  {https://ui.adsabs.harvard.edu/abs/2023MNRAS.525.5951H} {525, 5951}

\bibitem[\protect\citeauthoryear{{Hu}, {Naab}, {Walch}, {Moster}  \&
  {Oser}}{{Hu} et~al.}{2014}]{Hu2014}
{Hu} C.-Y.,  {Naab} T.,  {Walch} S.,  {Moster} B.~P.,   {Oser} L.,  2014,
  \mn@doi [\mnras] {10.1093/mnras/stu1187}, \href
  {https://ui.adsabs.harvard.edu/abs/2014MNRAS.443.1173H} {443, 1173}

\bibitem[\protect\citeauthoryear{{Hubber}, {Batty}, {McLeod}  \&
  {Whitworth}}{{Hubber} et~al.}{2011}]{Hubber2011}
{Hubber} D.~A.,  {Batty} C.~P.,  {McLeod} A.,   {Whitworth} A.~P.,  2011,
  \mn@doi [\aap] {10.1051/0004-6361/201014949}, \href
  {http://adsabs.harvard.edu/abs/2011A%26A...529A..27H} {529, A27}

\bibitem[\protect\citeauthoryear{{Hu{\v{s}}ko} \& {Lacey}}{{Hu{\v{s}}ko} \&
  {Lacey}}{2023}]{Husko2022a}
{Hu{\v{s}}ko} F.,  {Lacey} C.~G.,  2023, \mn@doi [\mnras]
  {10.1093/mnras/stad450}, \href
  {https://ui.adsabs.harvard.edu/abs/2023MNRAS.520.5090H} {520, 5090}

\bibitem[\protect\citeauthoryear{{Hu{\v{s}}ko}, {Lacey}, {Schaye}, {Schaller}
  \& {Nobels}}{{Hu{\v{s}}ko} et~al.}{2022}]{Husko2022b}
{Hu{\v{s}}ko} F.,  {Lacey} C.~G.,  {Schaye} J.,  {Schaller} M.,   {Nobels} F.
  S.~J.,  2022, \mn@doi [\mnras] {10.1093/mnras/stac2278}, \href
  {https://ui.adsabs.harvard.edu/abs/2022MNRAS.516.3750H} {516, 3750}

\bibitem[\protect\citeauthoryear{{Hu{\v{s}}ko}, {Lacey}, {Schaye}, {Nobels}  \&
  {Schaller}}{{Hu{\v{s}}ko} et~al.}{2024}]{Husko2023}
{Hu{\v{s}}ko} F.,  {Lacey} C.~G.,  {Schaye} J.,  {Nobels} F. S.~J.,
  {Schaller} M.,  2024, \mn@doi [\mnras] {10.1093/mnras/stad3548}, \href
  {https://ui.adsabs.harvard.edu/abs/2024MNRAS.527.5988H} {527, 5988}

\bibitem[\protect\citeauthoryear{Ishiyama, Nitadori  \& Makino}{Ishiyama
  et~al.}{2012}]{Greem}
Ishiyama T.,  Nitadori K.,   Makino J.,  2012, in SC '12: Proceedings of the
  International Conference on High Performance Computing, Networking, Storage
  and Analysis. pp 1--10 (\mn@eprint {arXiv} {1211.4406}),
  \mn@doi{10.1109/SC.2012.3}

\bibitem[\protect\citeauthoryear{Ivanova et~al.,}{Ivanova
  et~al.}{2013}]{Ivanova2013}
Ivanova N.,  et~al., 2013, \mn@doi [The Astronomy and Astrophysics Review]
  {10.1007/s00159-013-0059-2}, 21

\bibitem[\protect\citeauthoryear{{Ivkovic}}{{Ivkovic}}{2023}]{Ivkovic2022}
{Ivkovic} M.,  2023, PhD thesis, EPFL (\mn@eprint {arXiv} {2302.12727}),
  \mn@doi{10.5075/epfl-thesis-9973}

\bibitem[\protect\citeauthoryear{{Jenkins} et~al.,}{{Jenkins}
  et~al.}{1998}]{Jenkins1998}
{Jenkins} A.,  et~al., 1998, \mn@doi [\apj] {10.1086/305615}, \href
  {https://ui.adsabs.harvard.edu/abs/1998ApJ...499...20J} {499, 20}

\bibitem[\protect\citeauthoryear{Karypis \& Kumar}{Karypis \&
  Kumar}{1998}]{metis}
Karypis G.,  Kumar V.,  1998, \mn@doi [SIAM Journal on Scientific Computing]
  {10.1137/S1064827595287997}, 20, 359

\bibitem[\protect\citeauthoryear{{Katz}}{{Katz}}{1992}]{Katz1992}
{Katz} N.,  1992, \mn@doi [\apj] {10.1086/171366}, \href
  {https://ui.adsabs.harvard.edu/abs/1992ApJ...391..502K} {391, 502}

\bibitem[\protect\citeauthoryear{{Katz}, {Weinberg}  \& {Hernquist}}{{Katz}
  et~al.}{1996}]{Katz1996}
{Katz} N.,  {Weinberg} D.~H.,   {Hernquist} L.,  1996, \mn@doi [\apjs]
  {10.1086/192305}, \href
  {https://ui.adsabs.harvard.edu/abs/1996ApJS..105...19K} {105, 19}

\bibitem[\protect\citeauthoryear{{Kegerreis}, {Eke}, {Gonnet}, {Korycansky},
  {Massey}, {Schaller}  \& {Teodoro}}{{Kegerreis} et~al.}{2019}]{Kegerreis2019}
{Kegerreis} J.~A.,  {Eke} V.~R.,  {Gonnet} P.,  {Korycansky} D.~G.,  {Massey}
  R.~J.,  {Schaller} M.,   {Teodoro} L.~F.~A.,  2019, \mn@doi [\mnras]
  {10.1093/mnras/stz1606}, \href
  {https://ui.adsabs.harvard.edu/abs/2019MNRAS.487.5029K} {487, 5029}

\bibitem[\protect\citeauthoryear{{Kegerreis}, {Ruiz-Bonilla}, {Eke}, {Massey},
  {Sandnes}  \& {Teodoro}}{{Kegerreis} et~al.}{2022}]{Kegerreis2022}
{Kegerreis} J.~A.,  {Ruiz-Bonilla} S.,  {Eke} V.~R.,  {Massey} R.~J.,
  {Sandnes} T.~D.,   {Teodoro} L.~F.~A.,  2022, \mn@doi [\apjl]
  {10.3847/2041-8213/ac8d96}, \href
  {https://ui.adsabs.harvard.edu/abs/2022ApJ...937L..40K} {937, L40}

\bibitem[\protect\citeauthoryear{{Kennicutt}}{{Kennicutt}}{1998}]{Kennicutt1998}
{Kennicutt} Robert~C. J.,  1998, \mn@doi [\apj] {10.1086/305588}, \href
  {https://ui.adsabs.harvard.edu/abs/1998ApJ...498..541K} {498, 541}

\bibitem[\protect\citeauthoryear{{Kim} et~al.,}{{Kim} et~al.}{2016}]{Kim2016}
{Kim} J.-h.,  et~al., 2016, \mn@doi [\apj] {10.3847/1538-4357/833/2/202}, \href
  {https://ui.adsabs.harvard.edu/abs/2016ApJ...833..202K} {833, 202}

\bibitem[\protect\citeauthoryear{{Klessen}}{{Klessen}}{1997}]{Klessen1997}
{Klessen} R.,  1997, \mn@doi [\mnras] {10.1093/mnras/292.1.11}, \href
  {http://adsabs.harvard.edu/abs/1997MNRAS.292...11K} {292, 11}

\bibitem[\protect\citeauthoryear{{Knebe} et~al.,}{{Knebe}
  et~al.}{2013}]{Knebe2013}
{Knebe} A.,  et~al., 2013, \mn@doi [\mnras] {10.1093/mnras/stt1403}, \href
  {https://ui.adsabs.harvard.edu/abs/2013MNRAS.435.1618K} {435, 1618}

\bibitem[\protect\citeauthoryear{{Kravtsov}, {Klypin}  \&
  {Khokhlov}}{{Kravtsov} et~al.}{1997}]{ART}
{Kravtsov} A.~V.,  {Klypin} A.~A.,   {Khokhlov} A.~M.,  1997, \mn@doi [\apjs]
  {10.1086/313015}, \href
  {https://ui.adsabs.harvard.edu/abs/1997ApJS..111...73K} {111, 73}

\bibitem[\protect\citeauthoryear{{Kugel} et~al.,}{{Kugel}
  et~al.}{2023}]{Kugel2023}
{Kugel} R.,  et~al., 2023, \mn@doi [\mnras] {10.1093/mnras/stad2540}, \href
  {https://ui.adsabs.harvard.edu/abs/2023MNRAS.526.6103K} {526, 6103}

\bibitem[\protect\citeauthoryear{{Lesgourgues} \& {Pastor}}{{Lesgourgues} \&
  {Pastor}}{2006}]{Lesgourgues2006}
{Lesgourgues} J.,  {Pastor} S.,  2006, \mn@doi [\physrep]
  {10.1016/j.physrep.2006.04.001}, \href
  {https://ui.adsabs.harvard.edu/abs/2006PhR...429..307L} {429, 307}

\bibitem[\protect\citeauthoryear{{Lesgourgues} \& {Tram}}{{Lesgourgues} \&
  {Tram}}{2011}]{Lesgourgues2011}
{Lesgourgues} J.,  {Tram} T.,  2011, \mn@doi [\jcap]
  {10.1088/1475-7516/2011/09/032}, \href
  {https://ui.adsabs.harvard.edu/abs/2011JCAP...09..032L} {2011, 032}

\bibitem[\protect\citeauthoryear{{Linder} \& {Jenkins}}{{Linder} \&
  {Jenkins}}{2003}]{Linder2003}
{Linder} E.~V.,  {Jenkins} A.,  2003, \mn@doi [\mnras]
  {10.1046/j.1365-2966.2003.07112.x}, \href
  {http://adsabs.harvard.edu/abs/2003MNRAS.346..573L} {346, 573}

\bibitem[\protect\citeauthoryear{{Lucy}}{{Lucy}}{1977}]{Lucy1977}
{Lucy} L.~B.,  1977, \mn@doi [\aj] {10.1086/112164}, \href
  {https://ui.adsabs.harvard.edu/abs/1977AJ.....82.1013L} {82, 1013}

\bibitem[\protect\citeauthoryear{{Ludlow}, {Schaye}  \& {Bower}}{{Ludlow}
  et~al.}{2019}]{Ludlow2019}
{Ludlow} A.~D.,  {Schaye} J.,   {Bower} R.,  2019, \mn@doi [\mnras]
  {10.1093/mnras/stz1821}, \href
  {https://ui.adsabs.harvard.edu/abs/2019MNRAS.488.3663L} {488, 3663}

\bibitem[\protect\citeauthoryear{{Mangano}, {Miele}, {Pastor}, {Pinto},
  {Pisanti}  \& {Serpico}}{{Mangano} et~al.}{2005}]{Mangano2005}
{Mangano} G.,  {Miele} G.,  {Pastor} S.,  {Pinto} T.,  {Pisanti} O.,
  {Serpico} P.~D.,  2005, \mn@doi [Nuclear Physics B]
  {10.1016/j.nuclphysb.2005.09.041}, \href
  {https://ui.adsabs.harvard.edu/abs/2005NuPhB.729..221M} {729, 221}

\bibitem[\protect\citeauthoryear{{McAlpine} et~al.,}{{McAlpine}
  et~al.}{2022}]{McAlpine2022}
{McAlpine} S.,  et~al., 2022, \mn@doi [\mnras] {10.1093/mnras/stac295}, \href
  {https://ui.adsabs.harvard.edu/abs/2022MNRAS.512.5823M} {512, 5823}

\bibitem[\protect\citeauthoryear{{Menon}, {Wesolowski}, {Zheng}, {Jetley},
  {Kale}, {Quinn}  \& {Governato}}{{Menon} et~al.}{2015}]{CHANGA}
{Menon} H.,  {Wesolowski} L.,  {Zheng} G.,  {Jetley} P.,  {Kale} L.,  {Quinn}
  T.,   {Governato} F.,  2015, \mn@doi [Computational Astrophysics and
  Cosmology] {10.1186/s40668-015-0007-9}, \href
  {https://ui.adsabs.harvard.edu/abs/2015ComAC...2....1M} {2, 1}

\bibitem[\protect\citeauthoryear{{Message Passing Interface Forum}}{{Message
  Passing Interface Forum}}{2021}]{mpi_standard}
{Message Passing Interface Forum} 2021, {MPI}: A Message-Passing Interface
  Standard Version 4.0.
\url {https://www.mpi-forum.org/docs/mpi-4.0/mpi40-report.pdf}

\bibitem[\protect\citeauthoryear{{Michaux}, {Hahn}, {Rampf}  \&
  {Angulo}}{{Michaux} et~al.}{2021}]{Michaux2021}
{Michaux} M.,  {Hahn} O.,  {Rampf} C.,   {Angulo} R.~E.,  2021, \mn@doi
  [\mnras] {10.1093/mnras/staa3149}, \href
  {https://ui.adsabs.harvard.edu/abs/2021MNRAS.500..663M} {500, 663}

\bibitem[\protect\citeauthoryear{{Mignone}, {Zanni}, {Tzeferacos}, {van
  Straalen}, {Colella}  \& {Bodo}}{{Mignone} et~al.}{2012}]{pluto}
{Mignone} A.,  {Zanni} C.,  {Tzeferacos} P.,  {van Straalen} B.,  {Colella} P.,
    {Bodo} G.,  2012, \mn@doi [\apjs] {10.1088/0067-0049/198/1/7}, \href
  {https://ui.adsabs.harvard.edu/abs/2012ApJS..198....7M} {198, 7}

\bibitem[\protect\citeauthoryear{{Monaghan}}{{Monaghan}}{1992}]{Monaghan1992}
{Monaghan} J.~J.,  1992, \mn@doi [\araa] {10.1146/annurev.aa.30.090192.002551},
  \href {https://ui.adsabs.harvard.edu/abs/1992ARA&A..30..543M} {30, 543}

\bibitem[\protect\citeauthoryear{{Monaghan} \& {Lattanzio}}{{Monaghan} \&
  {Lattanzio}}{1985}]{Monaghan1985}
{Monaghan} J.~J.,  {Lattanzio} J.~C.,  1985, \aap, \href
  {http://adsabs.harvard.edu/abs/1985A%26A...149..135M} {149, 135}

\bibitem[\protect\citeauthoryear{{Monaghan} \& {Price}}{{Monaghan} \&
  {Price}}{2001}]{Monaghan2001}
{Monaghan} J.~J.,  {Price} D.~J.,  2001, \mn@doi [\mnras]
  {10.1046/j.1365-8711.2001.04742.x}, \href
  {https://ui.adsabs.harvard.edu/abs/2001MNRAS.328..381M} {328, 381}

\bibitem[\protect\citeauthoryear{Morris \& Monaghan}{Morris \&
  Monaghan}{1997}]{Morris1997}
Morris J.~P.,  Monaghan J.~J.,  1997, Journal of Computational Physics, 136, 41

\bibitem[\protect\citeauthoryear{{Naab} \& {Ostriker}}{{Naab} \&
  {Ostriker}}{2017}]{Naab2017}
{Naab} T.,  {Ostriker} J.~P.,  2017, \mn@doi [\araa]
  {10.1146/annurev-astro-081913-040019}, \href
  {https://ui.adsabs.harvard.edu/abs/2017ARA&A..55...59N} {55, 59}

\bibitem[\protect\citeauthoryear{{Narayan} \& {Yi}}{{Narayan} \&
  {Yi}}{1994}]{Narayan1994}
{Narayan} R.,  {Yi} I.,  1994, \mn@doi [\apjl] {10.1086/187381}, \href
  {https://ui.adsabs.harvard.edu/abs/1994ApJ...428L..13N} {428, L13}

\bibitem[\protect\citeauthoryear{{Navarro}, {Frenk}  \& {White}}{{Navarro}
  et~al.}{1997}]{navarro1997}
{Navarro} J.~F.,  {Frenk} C.~S.,   {White} S. D.~M.,  1997, \mn@doi [\apj]
  {10.1086/304888}, \href
  {https://ui.adsabs.harvard.edu/abs/1997ApJ...490..493N} {490, 493}

\bibitem[\protect\citeauthoryear{{Nelson} \& {Papaloizou}}{{Nelson} \&
  {Papaloizou}}{1994}]{Nelson1994}
{Nelson} R.~P.,  {Papaloizou} J.~C.~B.,  1994, \mn@doi [\mnras]
  {10.1093/mnras/270.1.1}, \href
  {https://ui.adsabs.harvard.edu/abs/1994MNRAS.270....1N} {270, 1}

\bibitem[\protect\citeauthoryear{{Nobels}, {Schaye}, {Schaller}, {Bah{\'e}}  \&
  {Chaikin}}{{Nobels} et~al.}{2022}]{nobels2022}
{Nobels} F. S.~J.,  {Schaye} J.,  {Schaller} M.,  {Bah{\'e}} Y.~M.,   {Chaikin}
  E.,  2022, \mn@doi [\mnras] {10.1093/mnras/stac2061}, \href
  {https://ui.adsabs.harvard.edu/abs/2022MNRAS.515.4838N} {515, 4838}

\bibitem[\protect\citeauthoryear{{Nobels}, {Schaye}, {Schaller}, {Ploeckinger},
  {Chaikin}  \& {Richings}}{{Nobels} et~al.}{2023}]{Nobels2023}
{Nobels} F. S.~J.,  {Schaye} J.,  {Schaller} M.,  {Ploeckinger} S.,  {Chaikin}
  E.,   {Richings} A.~J.,  2023, \mn@doi [arXiv e-prints]
  {10.48550/arXiv.2309.13750}, \href
  {https://ui.adsabs.harvard.edu/abs/2023arXiv230913750N} {p. arXiv:2309.13750}

\bibitem[\protect\citeauthoryear{{Okamoto}, {Eke}, {Frenk}  \&
  {Jenkins}}{{Okamoto} et~al.}{2005}]{Okamoto2005}
{Okamoto} T.,  {Eke} V.~R.,  {Frenk} C.~S.,   {Jenkins} A.,  2005, \mn@doi
  [\mnras] {10.1111/j.1365-2966.2005.09525.x}, \href
  {http://adsabs.harvard.edu/abs/2005MNRAS.363.1299O} {363, 1299}

\bibitem[\protect\citeauthoryear{{Owen} \& {Villumsen}}{{Owen} \&
  {Villumsen}}{1997}]{Owen1997}
{Owen} J.~M.,  {Villumsen} J.~V.,  1997, \mn@doi [\apj] {10.1086/304018}, \href
  {https://ui.adsabs.harvard.edu/abs/1997ApJ...481....1O} {481, 1}

\bibitem[\protect\citeauthoryear{{Peebles}}{{Peebles}}{1980}]{Peebles1980}
{Peebles} P.~J.~E.,  1980, {The large-scale structure of the universe}.
Princeton University Press

\bibitem[\protect\citeauthoryear{Perez, Badia  \& Labarta}{Perez
  et~al.}{2008}]{SMPSs}
Perez J.~M.,  Badia R.~M.,   Labarta J.,  2008, in 2008 IEEE international
  conference on cluster computing. pp 142--151

\bibitem[\protect\citeauthoryear{{Ploeckinger} \& {Schaye}}{{Ploeckinger} \&
  {Schaye}}{2020}]{Ploeckinger2020}
{Ploeckinger} S.,  {Schaye} J.,  2020, \mn@doi [\mnras]
  {10.1093/mnras/staa2172}, \href
  {https://ui.adsabs.harvard.edu/abs/2020MNRAS.497.4857P} {497, 4857}

\bibitem[\protect\citeauthoryear{{Plummer}}{{Plummer}}{1911}]{Plummer1911}
{Plummer} H.~C.,  1911, \mn@doi [\mnras] {10.1093/mnras/71.5.460}, \href
  {https://ui.adsabs.harvard.edu/abs/1911MNRAS..71..460P} {71, 460}

\bibitem[\protect\citeauthoryear{{Portegies Zwart}}{{Portegies
  Zwart}}{2020}]{SPZ2020}
{Portegies Zwart} S.,  2020, \mn@doi [Nature Astronomy]
  {10.1038/s41550-020-1208-y}, \href
  {https://ui.adsabs.harvard.edu/abs/2020NatAs...4..819P} {4, 819}

\bibitem[\protect\citeauthoryear{{Potter}, {Stadel}  \& {Teyssier}}{{Potter}
  et~al.}{2017}]{pkdGrav3}
{Potter} D.,  {Stadel} J.,   {Teyssier} R.,  2017, \mn@doi [Computational
  Astrophysics and Cosmology] {10.1186/s40668-017-0021-1}, \href
  {https://ui.adsabs.harvard.edu/abs/2017ComAC...4....2P} {4, 2}

\bibitem[\protect\citeauthoryear{{Power}, {Navarro}, {Jenkins}, {Frenk},
  {White}, {Springel}, {Stadel}  \& {Quinn}}{{Power} et~al.}{2003}]{Power2003}
{Power} C.,  {Navarro} J.~F.,  {Jenkins} A.,  {Frenk} C.~S.,  {White} S.~D.~M.,
   {Springel} V.,  {Stadel} J.,   {Quinn} T.,  2003, \mn@doi [\mnras]
  {10.1046/j.1365-8711.2003.05925.x}, \href
  {https://ui.adsabs.harvard.edu/abs/2003MNRAS.338...14P} {338, 14}

\bibitem[\protect\citeauthoryear{{Price}}{{Price}}{2008}]{Price2008}
{Price} D.~J.,  2008, \mn@doi [Journal of Computational Physics]
  {10.1016/j.jcp.2008.08.011}, \href
  {https://ui.adsabs.harvard.edu/abs/2008JCoPh.22710040P} {227, 10040}

\bibitem[\protect\citeauthoryear{{Price}}{{Price}}{2012}]{Price2012}
{Price} D.~J.,  2012, \mn@doi [Journal of Computational Physics]
  {10.1016/j.jcp.2010.12.011}, \href
  {http://adsabs.harvard.edu/abs/2012JCoPh.231..759P} {231, 759}

\bibitem[\protect\citeauthoryear{{Price} \& {Monaghan}}{{Price} \&
  {Monaghan}}{2007}]{Price2007}
{Price} D.~J.,  {Monaghan} J.~J.,  2007, \mn@doi [\mnras]
  {10.1111/j.1365-2966.2006.11241.x}, \href
  {http://adsabs.harvard.edu/abs/2007MNRAS.374.1347P} {374, 1347}

\bibitem[\protect\citeauthoryear{{Price} et~al.,}{{Price}
  et~al.}{2018}]{Price2018}
{Price} D.~J.,  et~al., 2018, \mn@doi [\pasa] {10.1017/pasa.2018.25}, \href
  {https://ui.adsabs.harvard.edu/abs/2018PASA...35...31P} {35, e031}

\bibitem[\protect\citeauthoryear{{Quinn}, {Katz}, {Stadel}  \& {Lake}}{{Quinn}
  et~al.}{1997}]{Quinn1997}
{Quinn} T.,  {Katz} N.,  {Stadel} J.,   {Lake} G.,  1997, arXiv e-prints, \href
  {https://ui.adsabs.harvard.edu/abs/1997astro.ph.10043Q} {pp
  astro--ph/9710043}

\bibitem[\protect\citeauthoryear{{Rahmati}, {Schaye}, {Pawlik}  \&
  {Rai{\v{c}}evi{\'c}}}{{Rahmati} et~al.}{2013}]{Rahmati2013}
{Rahmati} A.,  {Schaye} J.,  {Pawlik} A.~H.,   {Rai{\v{c}}evi{\'c}} M.,  2013,
  \mn@doi [\mnras] {10.1093/mnras/stt324}, \href
  {https://ui.adsabs.harvard.edu/abs/2013MNRAS.431.2261R} {431, 2261}

\bibitem[\protect\citeauthoryear{{Ramsey}, {Haugb{\o}lle}  \&
  {Nordlund}}{{Ramsey} et~al.}{2018}]{dispatch}
{Ramsey} J.~P.,  {Haugb{\o}lle} T.,   {Nordlund} {\r{A}}.,  2018, in Journal of
  Physics Conference Series. p. 012021 (\mn@eprint {arXiv} {1806.10098}),
  \mn@doi{10.1088/1742-6596/1031/1/012021}

\bibitem[\protect\citeauthoryear{{Regan}, {Haehnelt}  \& {Viel}}{{Regan}
  et~al.}{2007}]{Regan2007}
{Regan} J.~A.,  {Haehnelt} M.~G.,   {Viel} M.,  2007, \mn@doi [\mnras]
  {10.1111/j.1365-2966.2006.11132.x}, \href
  {https://ui.adsabs.harvard.edu/abs/2007MNRAS.374..196R} {374, 196}

\bibitem[\protect\citeauthoryear{{Rein} \& {Tamayo}}{{Rein} \&
  {Tamayo}}{2017}]{Rein2017}
{Rein} H.,  {Tamayo} D.,  2017, \mn@doi [\mnras] {10.1093/mnras/stx232}, \href
  {https://ui.adsabs.harvard.edu/abs/2017MNRAS.467.2377R} {467, 2377}

\bibitem[\protect\citeauthoryear{Reinders}{Reinders}{2007}]{TBB}
Reinders J.,  2007, Intel Threading Building Blocks: Outfitting C++ for
  Multi-core Processor Parallelism.
O'Reilly Media, \url {https://books.google.nl/books?id=do86P6kb0msC}

\bibitem[\protect\citeauthoryear{{Revaz}}{{Revaz}}{2013}]{pNbody}
{Revaz} Y.,  2013, {pNbody: A python parallelized N-body reduction toolbox},
  Astrophysics Source Code Library, record ascl:1302.004 (\mn@eprint {ascl}
  {1302.004})

\bibitem[\protect\citeauthoryear{{Revaz} \& {Jablonka}}{{Revaz} \&
  {Jablonka}}{2012}]{Revaz2012}
{Revaz} Y.,  {Jablonka} P.,  2012, \mn@doi [\aap]
  {10.1051/0004-6361/201117402}, \href
  {https://ui.adsabs.harvard.edu/abs/2012A&A...538A..82R} {538, A82}

\bibitem[\protect\citeauthoryear{{Revaz} \& {Jablonka}}{{Revaz} \&
  {Jablonka}}{2018}]{Revaz2018}
{Revaz} Y.,  {Jablonka} P.,  2018, \mn@doi [\aap]
  {10.1051/0004-6361/201832669}, \href
  {https://ui.adsabs.harvard.edu/abs/2018A&A...616A..96R} {616, A96}

\bibitem[\protect\citeauthoryear{{Revaz}, {Arnaudon}, {Nichols}, {Bonvin}  \&
  {Jablonka}}{{Revaz} et~al.}{2016}]{Revaz2016}
{Revaz} Y.,  {Arnaudon} A.,  {Nichols} M.,  {Bonvin} V.,   {Jablonka} P.,
  2016, \mn@doi [\aap] {10.1051/0004-6361/201526438}, \href
  {https://ui.adsabs.harvard.edu/abs/2016A&A...588A..21R} {588, A21}

\bibitem[\protect\citeauthoryear{{Richings}, {Schaye}  \&
  {Oppenheimer}}{{Richings} et~al.}{2014a}]{CHIMESa}
{Richings} A.~J.,  {Schaye} J.,   {Oppenheimer} B.~D.,  2014a, \mn@doi [\mnras]
  {10.1093/mnras/stu525}, \href
  {https://ui.adsabs.harvard.edu/abs/2014MNRAS.440.3349R} {440, 3349}

\bibitem[\protect\citeauthoryear{{Richings}, {Schaye}  \&
  {Oppenheimer}}{{Richings} et~al.}{2014b}]{CHIMESb}
{Richings} A.~J.,  {Schaye} J.,   {Oppenheimer} B.~D.,  2014b, \mn@doi [\mnras]
  {10.1093/mnras/stu1046}, \href
  {https://ui.adsabs.harvard.edu/abs/2014MNRAS.442.2780R} {442, 2780}

\bibitem[\protect\citeauthoryear{{Robertson} \& {Kravtsov}}{{Robertson} \&
  {Kravtsov}}{2008}]{Robertson2008}
{Robertson} B.~E.,  {Kravtsov} A.~V.,  2008, \mn@doi [\apj] {10.1086/587796},
  \href {https://ui.adsabs.harvard.edu/abs/2008ApJ...680.1083R} {680, 1083}

\bibitem[\protect\citeauthoryear{{Roca-F{\`a}brega} et~al.,}{{Roca-F{\`a}brega}
  et~al.}{2021}]{Roca-Fabrega2021}
{Roca-F{\`a}brega} S.,  et~al., 2021, \mn@doi [\apj]
  {10.3847/1538-4357/ac088a}, \href
  {https://ui.adsabs.harvard.edu/abs/2021ApJ...917...64R} {917, 64}

\bibitem[\protect\citeauthoryear{{Rogers} et~al.,}{{Rogers}
  et~al.}{2022}]{Rogers2022}
{Rogers} B.,  et~al., 2022, in {G. Bilotta} ed., {Proceedings of the 16th
  SPHERIC International Workshop}. pp 391--398

\bibitem[\protect\citeauthoryear{{Roper}}{{Roper}}{2023}]{Roper2023}
{Roper} W.,  2023, PhD thesis, University of Sussex, UK, \url
  {https://hdl.handle.net/10779/uos.24131940.v1}

\bibitem[\protect\citeauthoryear{{Roper} et~al.}{{Roper}
  et~al.}{2024}]{Roper2024}
{Roper} W.,  et~al., 2024, in prep.

\bibitem[\protect\citeauthoryear{{Rosas-Guevara} et~al.,}{{Rosas-Guevara}
  et~al.}{2015}]{Rosas2015}
{Rosas-Guevara} Y.~M.,  et~al., 2015, \mn@doi [\mnras] {10.1093/mnras/stv2056},
  \href {https://ui.adsabs.harvard.edu/abs/2015MNRAS.454.1038R} {454, 1038}

\bibitem[\protect\citeauthoryear{{Rosswog}}{{Rosswog}}{2020}]{Rosswog2020}
{Rosswog} S.,  2020, \mn@doi [\mnras] {10.1093/mnras/staa2591}, \href
  {https://ui.adsabs.harvard.edu/abs/2020MNRAS.498.4230R} {498, 4230}

\bibitem[\protect\citeauthoryear{{Ruiz-Bonilla}, {Eke}, {Kegerreis}, {Massey}
  \& {Teodoro}}{{Ruiz-Bonilla} et~al.}{2021}]{RuizBonilla2021}
{Ruiz-Bonilla} S.,  {Eke} V.~R.,  {Kegerreis} J.~A.,  {Massey} R.~J.,
  {Teodoro} L.~F.~A.,  2021, \mn@doi [\mnras] {10.1093/mnras/staa3385}, \href
  {https://ui.adsabs.harvard.edu/abs/2021MNRAS.500.2861R} {500, 2861}

\bibitem[\protect\citeauthoryear{{Ruiz-Bonilla}, {Borrow}, {Eke}, {Kegerreis},
  {Massey}, {Sandnes}  \& {Teodoro}}{{Ruiz-Bonilla}
  et~al.}{2022}]{RuizBonilla2022}
{Ruiz-Bonilla} S.,  {Borrow} J.,  {Eke} V.~R.,  {Kegerreis} J.~A.,  {Massey}
  R.~J.,  {Sandnes} T.~D.,   {Teodoro} L.~F.~A.,  2022, \mn@doi [\mnras]
  {10.1093/mnras/stac857}, \href
  {https://ui.adsabs.harvard.edu/abs/2022MNRAS.512.4660R} {512, 4660}

\bibitem[\protect\citeauthoryear{{Saitoh} \& {Makino}}{{Saitoh} \&
  {Makino}}{2009}]{Saitoh2009}
{Saitoh} T.~R.,  {Makino} J.,  2009, \mn@doi [\apjl]
  {10.1088/0004-637X/697/2/L99}, \href
  {https://ui.adsabs.harvard.edu/abs/2009ApJ...697L..99S} {697, L99}

\bibitem[\protect\citeauthoryear{{Saitoh} \& {Makino}}{{Saitoh} \&
  {Makino}}{2013}]{Saitoh2013}
{Saitoh} T.~R.,  {Makino} J.,  2013, \mn@doi [\apj]
  {10.1088/0004-637X/768/1/44}, \href
  {https://ui.adsabs.harvard.edu/abs/2013ApJ...768...44S} {768, 44}

\bibitem[\protect\citeauthoryear{{Salmon} \& {Warren}}{{Salmon} \&
  {Warren}}{1994}]{Salmon1994}
{Salmon} J.~K.,  {Warren} M.~S.,  1994, \mn@doi [Journal of Computational
  Physics] {10.1006/jcph.1994.1050}, \href
  {https://ui.adsabs.harvard.edu/abs/1994JCoPh.111..136S} {111, 136}

\bibitem[\protect\citeauthoryear{{Sanati}, {Revaz}, {Schober}, {Kunze}  \&
  {Jablonka}}{{Sanati} et~al.}{2020}]{Sanati2020}
{Sanati} M.,  {Revaz} Y.,  {Schober} J.,  {Kunze} K.~E.,   {Jablonka} P.,
  2020, \mn@doi [\aap] {10.1051/0004-6361/202038382}, \href
  {https://ui.adsabs.harvard.edu/abs/2020A&A...643A..54S} {643, A54}

\bibitem[\protect\citeauthoryear{{Sanati}, {Jeanquartier}, {Revaz}  \&
  {Jablonka}}{{Sanati} et~al.}{2023}]{Sanati2022}
{Sanati} M.,  {Jeanquartier} F.,  {Revaz} Y.,   {Jablonka} P.,  2023, \mn@doi
  [\aap] {10.1051/0004-6361/202244309}, \href
  {https://ui.adsabs.harvard.edu/abs/2023A&A...669A..94S} {669, A94}

\bibitem[\protect\citeauthoryear{{Sandnes} et~al.}{{Sandnes}
  et~al.}{2024}]{Sandnes2024}
{Sandnes} T.~D.,  et~al., 2024, in prep.

\bibitem[\protect\citeauthoryear{{Schaller}, {Dalla Vecchia}, {Schaye},
  {Bower}, {Theuns}, {Crain}, {Furlong}  \& {McCarthy}}{{Schaller}
  et~al.}{2015}]{Schaller2015}
{Schaller} M.,  {Dalla Vecchia} C.,  {Schaye} J.,  {Bower} R.~G.,  {Theuns} T.,
   {Crain} R.~A.,  {Furlong} M.,   {McCarthy} I.~G.,  2015, \mn@doi [\mnras]
  {10.1093/mnras/stv2169}, \href
  {https://ui.adsabs.harvard.edu/abs/2015MNRAS.454.2277S} {454, 2277}

\bibitem[\protect\citeauthoryear{{Schaller}, {Gonnet}, {Chalk}  \&
  {Draper}}{{Schaller} et~al.}{2016}]{Schaller2016}
{Schaller} M.,  {Gonnet} P.,  {Chalk} A. B.~G.,   {Draper} P.~W.,  2016, in
  Proceedings of the PASC Conference. PASC '16.
ACM, New York, NY, USA (\mn@eprint {arXiv} {1606.02738}),
  \mn@doi{10.1145/2929908.2929916}

\bibitem[\protect\citeauthoryear{{Schaye}}{{Schaye}}{2004}]{Schaye2004}
{Schaye} J.,  2004, \mn@doi [\apj] {10.1086/421232}, \href
  {https://ui.adsabs.harvard.edu/abs/2004ApJ...609..667S} {609, 667}

\bibitem[\protect\citeauthoryear{{Schaye} \& {Dalla Vecchia}}{{Schaye} \&
  {Dalla Vecchia}}{2008}]{Schaye2008}
{Schaye} J.,  {Dalla Vecchia} C.,  2008, \mn@doi [\mnras]
  {10.1111/j.1365-2966.2007.12639.x}, \href
  {https://ui.adsabs.harvard.edu/abs/2008MNRAS.383.1210S} {383, 1210}

\bibitem[\protect\citeauthoryear{{Schaye}, {Aguirre}, {Kim}, {Theuns}, {Rauch}
  \& {Sargent}}{{Schaye} et~al.}{2003}]{SpecWizard1}
{Schaye} J.,  {Aguirre} A.,  {Kim} T.-S.,  {Theuns} T.,  {Rauch} M.,
  {Sargent} W. L.~W.,  2003, \mn@doi [\apj] {10.1086/378044}, \href
  {https://ui.adsabs.harvard.edu/abs/2003ApJ...596..768S} {596, 768}

\bibitem[\protect\citeauthoryear{{Schaye} et~al.,}{{Schaye}
  et~al.}{2015}]{Schaye2015}
{Schaye} J.,  et~al., 2015, \mn@doi [\mnras] {10.1093/mnras/stu2058}, \href
  {https://ui.adsabs.harvard.edu/abs/2015MNRAS.446..521S} {446, 521}

\bibitem[\protect\citeauthoryear{{Schaye} et~al.,}{{Schaye}
  et~al.}{2023}]{Schaye2022}
{Schaye} J.,  et~al., 2023, \mn@doi [\mnras] {10.1093/mnras/stad2419}, \href
  {https://ui.adsabs.harvard.edu/abs/2023MNRAS.526.4978S} {526, 4978}

\bibitem[\protect\citeauthoryear{{Schmidt}}{{Schmidt}}{1959}]{Schmidt1959}
{Schmidt} M.,  1959, \mn@doi [\apj] {10.1086/146614}, \href
  {https://ui.adsabs.harvard.edu/abs/1959ApJ...129..243S} {129, 243}

\bibitem[\protect\citeauthoryear{{Schneider} et~al.,}{{Schneider}
  et~al.}{2016}]{Schneider2016}
{Schneider} A.,  et~al., 2016, \mn@doi [\jcap] {10.1088/1475-7516/2016/04/047},
  \href {https://ui.adsabs.harvard.edu/abs/2016JCAP...04..047S} {2016, 047}

\bibitem[\protect\citeauthoryear{{Sembolini} et~al.,}{{Sembolini}
  et~al.}{2016}]{Sembolini2016}
{Sembolini} F.,  et~al., 2016, \mn@doi [\mnras] {10.1093/mnras/stw250}, \href
  {https://ui.adsabs.harvard.edu/abs/2016MNRAS.457.4063S} {457, 4063}

\bibitem[\protect\citeauthoryear{{Shakura} \& {Sunyaev}}{{Shakura} \&
  {Sunyaev}}{1973}]{Shakura1973}
{Shakura} N.~I.,  {Sunyaev} R.~A.,  1973, \aap, \href
  {https://ui.adsabs.harvard.edu/abs/1973A&A....24..337S} {24, 337}

\bibitem[\protect\citeauthoryear{{Smith} et~al.,}{{Smith}
  et~al.}{2017}]{Smith2017}
{Smith} B.~D.,  et~al., 2017, \mn@doi [\mnras] {10.1093/mnras/stw3291}, \href
  {https://ui.adsabs.harvard.edu/abs/2017MNRAS.466.2217S} {466, 2217}

\bibitem[\protect\citeauthoryear{{Somerville} \& {Dav{\'e}}}{{Somerville} \&
  {Dav{\'e}}}{2015}]{Somerville2015}
{Somerville} R.~S.,  {Dav{\'e}} R.,  2015, \mn@doi [\araa]
  {10.1146/annurev-astro-082812-140951}, \href
  {https://ui.adsabs.harvard.edu/abs/2015ARA&A..53...51S} {53, 51}

\bibitem[\protect\citeauthoryear{{Springel}}{{Springel}}{2005}]{Springel2005}
{Springel} V.,  2005, \mn@doi [\mnras] {10.1111/j.1365-2966.2005.09655.x},
  \href {http://adsabs.harvard.edu/abs/2005MNRAS.364.1105S} {364, 1105}

\bibitem[\protect\citeauthoryear{{Springel}}{{Springel}}{2010a}]{SpringelSPHreview}
{Springel} V.,  2010a, \mn@doi [\araa] {10.1146/annurev-astro-081309-130914},
  \href {https://ui.adsabs.harvard.edu/abs/2010ARA&A..48..391S} {48, 391}

\bibitem[\protect\citeauthoryear{{Springel}}{{Springel}}{2010b}]{Springel2010}
{Springel} V.,  2010b, \mn@doi [\mnras] {10.1111/j.1365-2966.2009.15715.x},
  \href {http://adsabs.harvard.edu/abs/2010MNRAS.401..791S} {401, 791}

\bibitem[\protect\citeauthoryear{{Springel} \& {Hernquist}}{{Springel} \&
  {Hernquist}}{2002}]{Springel2002}
{Springel} V.,  {Hernquist} L.,  2002, \mn@doi [\mnras]
  {10.1046/j.1365-8711.2002.05445.x}, \href
  {http://adsabs.harvard.edu/abs/2002MNRAS.333..649S} {333, 649}

\bibitem[\protect\citeauthoryear{{Springel} \& {Hernquist}}{{Springel} \&
  {Hernquist}}{2003}]{Springel2003}
{Springel} V.,  {Hernquist} L.,  2003, \mn@doi [\mnras]
  {10.1046/j.1365-8711.2003.06206.x}, \href
  {https://ui.adsabs.harvard.edu/abs/2003MNRAS.339..289S} {339, 289}

\bibitem[\protect\citeauthoryear{{Springel}, {Yoshida}  \& {White}}{{Springel}
  et~al.}{2001}]{Springel2001}
{Springel} V.,  {Yoshida} N.,   {White} S. D.~M.,  2001, \mn@doi [\na]
  {10.1016/S1384-1076(01)00042-2}, \href
  {https://ui.adsabs.harvard.edu/abs/2001NewA....6...79S} {6, 79}

\bibitem[\protect\citeauthoryear{{Springel}, {Di Matteo}  \&
  {Hernquist}}{{Springel} et~al.}{2005a}]{springel2005b}
{Springel} V.,  {Di Matteo} T.,   {Hernquist} L.,  2005a, \mn@doi [\mnras]
  {10.1111/j.1365-2966.2005.09238.x}, \href
  {https://ui.adsabs.harvard.edu/abs/2005MNRAS.361..776S} {361, 776}

\bibitem[\protect\citeauthoryear{{Springel} et~al.,}{{Springel}
  et~al.}{2005b}]{Millennium}
{Springel} V.,  et~al., 2005b, \mn@doi [\nat] {10.1038/nature03597}, \href
  {https://ui.adsabs.harvard.edu/abs/2005Natur.435..629S} {435, 629}

\bibitem[\protect\citeauthoryear{{Springel}, {Pakmor}, {Zier}  \&
  {Reinecke}}{{Springel} et~al.}{2021}]{Springel2021}
{Springel} V.,  {Pakmor} R.,  {Zier} O.,   {Reinecke} M.,  2021, \mn@doi
  [\mnras] {10.1093/mnras/stab1855}, \href
  {https://ui.adsabs.harvard.edu/abs/2021MNRAS.506.2871S} {506, 2871}

\bibitem[\protect\citeauthoryear{{Stasyszyn} \& {Elstner}}{{Stasyszyn} \&
  {Elstner}}{2015}]{Federico2015}
{Stasyszyn} F.~A.,  {Elstner} D.,  2015, \mn@doi [Journal of Computational
  Physics] {10.1016/j.jcp.2014.11.011}, \href
  {https://ui.adsabs.harvard.edu/abs/2015JCoPh.282..148S} {282, 148}

\bibitem[\protect\citeauthoryear{{Stevens}, {Bellstedt}, {Elahi}  \&
  {Murphy}}{{Stevens} et~al.}{2020}]{Stevens2020}
{Stevens} A. R.~H.,  {Bellstedt} S.,  {Elahi} P.~J.,   {Murphy} M.~T.,  2020,
  \mn@doi [Nature Astronomy] {10.1038/s41550-020-1169-1}, \href
  {https://ui.adsabs.harvard.edu/abs/2020NatAs...4..843S} {4, 843}

\bibitem[\protect\citeauthoryear{{Stewart} et~al.,}{{Stewart}
  et~al.}{2020}]{Stewart+2020}
{Stewart} S.,  et~al., 2020, in American Institute of Physics Conference
  Series. p. 080003 (\mn@eprint {arXiv} {1910.04687}),
  \mn@doi{10.1063/12.0000946}

\bibitem[\protect\citeauthoryear{{Stinson}, {Seth}, {Katz}, {Wadsley},
  {Governato}  \& {Quinn}}{{Stinson} et~al.}{2006}]{Stinson2006}
{Stinson} G.,  {Seth} A.,  {Katz} N.,  {Wadsley} J.,  {Governato} F.,   {Quinn}
  T.,  2006, \mn@doi [\mnras] {10.1111/j.1365-2966.2006.11097.x}, \href
  {https://ui.adsabs.harvard.edu/abs/2006MNRAS.373.1074S} {373, 1074}

\bibitem[\protect\citeauthoryear{{Stone}, {Tomida}, {White}  \&
  {Felker}}{{Stone} et~al.}{2020}]{athena}
{Stone} J.~M.,  {Tomida} K.,  {White} C.~J.,   {Felker} K.~G.,  2020, \mn@doi
  [\apjs] {10.3847/1538-4365/ab929b}, \href
  {https://ui.adsabs.harvard.edu/abs/2020ApJS..249....4S} {249, 4}

\bibitem[\protect\citeauthoryear{{Tepper-Garc{\'\i}a}, {Richter}, {Schaye},
  {Booth}, {Dalla Vecchia}, {Theuns}  \& {Wiersma}}{{Tepper-Garc{\'\i}a}
  et~al.}{2011}]{SpecWizard2}
{Tepper-Garc{\'\i}a} T.,  {Richter} P.,  {Schaye} J.,  {Booth} C.~M.,  {Dalla
  Vecchia} C.,  {Theuns} T.,   {Wiersma} R. P.~C.,  2011, \mn@doi [\mnras]
  {10.1111/j.1365-2966.2010.18123.x}, \href
  {https://ui.adsabs.harvard.edu/abs/2011MNRAS.413..190T} {413, 190}

\bibitem[\protect\citeauthoryear{{Teyssier}}{{Teyssier}}{2002}]{Ramses}
{Teyssier} R.,  2002, \mn@doi [\aap] {10.1051/0004-6361:20011817}, \href
  {https://ui.adsabs.harvard.edu/abs/2002A&A...385..337T} {385, 337}

\bibitem[\protect\citeauthoryear{{The HDF Group}}{{The HDF Group}}{2022}]{hdf5}
{The HDF Group} 1997-2022, {Hierarchical Data Format, version 5}, \url
  {https://www.hdfgroup.org/HDF5/}

\bibitem[\protect\citeauthoryear{Tillotson}{Tillotson}{1962}]{Tillotson1962}
Tillotson J.~H.,  1962, General Atomic Report, GA-3216, 141

\bibitem[\protect\citeauthoryear{{Tinker}, {Robertson}, {Kravtsov}, {Klypin},
  {Warren}, {Yepes}  \& {Gottl{\"o}ber}}{{Tinker} et~al.}{2010}]{Tinker2010}
{Tinker} J.~L.,  {Robertson} B.~E.,  {Kravtsov} A.~V.,  {Klypin} A.,  {Warren}
  M.~S.,  {Yepes} G.,   {Gottl{\"o}ber} S.,  2010, \mn@doi [\apj]
  {10.1088/0004-637X/724/2/878}, \href
  {https://ui.adsabs.harvard.edu/abs/2010ApJ...724..878T} {724, 878}

\bibitem[\protect\citeauthoryear{{Tornatore}, {Borgani}, {Dolag}  \&
  {Matteucci}}{{Tornatore} et~al.}{2007}]{Tornatore2007}
{Tornatore} L.,  {Borgani} S.,  {Dolag} K.,   {Matteucci} F.,  2007, \mn@doi
  [\mnras] {10.1111/j.1365-2966.2007.12070.x}, \href
  {http://adsabs.harvard.edu/abs/2007MNRAS.382.1050T} {382, 1050}

\bibitem[\protect\citeauthoryear{{Trayford} et~al.,}{{Trayford}
  et~al.}{2015}]{Trayford2015}
{Trayford} J.~W.,  et~al., 2015, \mn@doi [\mnras] {10.1093/mnras/stv1461},
  \href {https://ui.adsabs.harvard.edu/abs/2015MNRAS.452.2879T} {452, 2879}

\bibitem[\protect\citeauthoryear{{Truelove}, {Klein}, {McKee}, {Holliman},
  {Howell}  \& {Greenough}}{{Truelove} et~al.}{1997}]{Truelove1997}
{Truelove} J.~K.,  {Klein} R.~I.,  {McKee} C.~F.,  {Holliman} John~H. I.,
  {Howell} L.~H.,   {Greenough} J.~A.,  1997, \mn@doi [\apjl] {10.1086/310975},
  \href {https://ui.adsabs.harvard.edu/abs/1997ApJ...489L.179T} {489, L179}

\bibitem[\protect\citeauthoryear{{Turk}, {Smith}, {Oishi}, {Skory}, {Skillman},
  {Abel}  \& {Norman}}{{Turk} et~al.}{2011}]{yt}
{Turk} M.~J.,  {Smith} B.~D.,  {Oishi} J.~S.,  {Skory} S.,  {Skillman} S.~W.,
  {Abel} T.,   {Norman} M.~L.,  2011, \mn@doi [The Astrophysical Journal
  Supplement Series] {10.1088/0067-0049/192/1/9}, \href
  {https://ui.adsabs.harvard.edu/abs/2011ApJS..192....9T} {192, 9}

\bibitem[\protect\citeauthoryear{{Vandenbroucke} \& {De
  Rijcke}}{{Vandenbroucke} \& {De Rijcke}}{2016}]{shadowfax}
{Vandenbroucke} B.,  {De Rijcke} S.,  2016, \mn@doi [Astronomy and Computing]
  {10.1016/j.ascom.2016.05.001}, \href
  {https://ui.adsabs.harvard.edu/abs/2016A&C....16..109V} {16, 109}

\bibitem[\protect\citeauthoryear{Verlet}{Verlet}{1967}]{Verlet1967}
Verlet L.,  1967, \mn@doi [Phys. Rev.] {10.1103/PhysRev.159.98}, 159, 98

\bibitem[\protect\citeauthoryear{{Viel}, {Haehnelt}  \& {Springel}}{{Viel}
  et~al.}{2004}]{qla}
{Viel} M.,  {Haehnelt} M.~G.,   {Springel} V.,  2004, \mn@doi [\mnras]
  {10.1111/j.1365-2966.2004.08224.x}, \href
  {https://ui.adsabs.harvard.edu/abs/2004MNRAS.354..684V} {354, 684}

\bibitem[\protect\citeauthoryear{{Vila}}{{Vila}}{1999}]{Vila1999}
{Vila} J.~P.,  1999, \mn@doi [Mathematical Models and Methods in Applied
  Sciences] {10.1142/S0218202599000117}, 09, 161

\bibitem[\protect\citeauthoryear{{Vogelsberger}, {Marinacci}, {Torrey}  \&
  {Puchwein}}{{Vogelsberger} et~al.}{2020}]{Vogelsberger2020}
{Vogelsberger} M.,  {Marinacci} F.,  {Torrey} P.,   {Puchwein} E.,  2020,
  \mn@doi [Nature Reviews Physics] {10.1038/s42254-019-0127-2}, \href
  {https://ui.adsabs.harvard.edu/abs/2020NatRP...2...42V} {2, 42}

\bibitem[\protect\citeauthoryear{{Wadsley}, {Stadel}  \& {Quinn}}{{Wadsley}
  et~al.}{2004}]{Wadsley2004}
{Wadsley} J.~W.,  {Stadel} J.,   {Quinn} T.,  2004, \mn@doi [\na]
  {10.1016/j.newast.2003.08.004}, \href
  {https://ui.adsabs.harvard.edu/abs/2004NewA....9..137W} {9, 137}

\bibitem[\protect\citeauthoryear{{Wadsley}, {Veeravalli}  \&
  {Couchman}}{{Wadsley} et~al.}{2008}]{Wadsley2008}
{Wadsley} J.~W.,  {Veeravalli} G.,   {Couchman} H.~M.~P.,  2008, \mn@doi
  [\mnras] {10.1111/j.1365-2966.2008.13260.x}, \href
  {https://ui.adsabs.harvard.edu/abs/2008MNRAS.387..427W} {387, 427}

\bibitem[\protect\citeauthoryear{{Wadsley}, {Keller}  \& {Quinn}}{{Wadsley}
  et~al.}{2017}]{Wadsley2017}
{Wadsley} J.~W.,  {Keller} B.~W.,   {Quinn} T.~R.,  2017, \mn@doi [\mnras]
  {10.1093/mnras/stx1643}, \href
  {https://ui.adsabs.harvard.edu/abs/2017MNRAS.471.2357W} {471, 2357}

\bibitem[\protect\citeauthoryear{Warren}{Warren}{2013}]{2HOT}
Warren M.~S.,  2013, in Proceedings of the International Conference on High
  Performance Computing, Networking, Storage and Analysis. SC '13.
ACM, New York, NY, USA, \mn@doi{10.1145/2503210.2503220}, \url
  {https://doi.org/10.1145/2503210.2503220}

\bibitem[\protect\citeauthoryear{{Warren} \& {Salmon}}{{Warren} \&
  {Salmon}}{1995}]{Warren1995}
{Warren} M.~S.,  {Salmon} J.~K.,  1995, \mn@doi [Computer Physics
  Communications] {10.1016/0010-4655(94)00177-4}, \href
  {http://adsabs.harvard.edu/abs/1995CoPhC..87..266W} {87, 266}

\bibitem[\protect\citeauthoryear{Wendland}{Wendland}{1995}]{Wendland1995}
Wendland H.,  1995, \mn@doi [Advances in Computational Mathematics]
  {10.1007/BF02123482}, 4, 389

\bibitem[\protect\citeauthoryear{{Wiersma}, {Schaye}  \& {Smith}}{{Wiersma}
  et~al.}{2009a}]{Wiersma2009cooling}
{Wiersma} R. P.~C.,  {Schaye} J.,   {Smith} B.~D.,  2009a, \mn@doi [\mnras]
  {10.1111/j.1365-2966.2008.14191.x}, \href
  {https://ui.adsabs.harvard.edu/abs/2009MNRAS.393...99W} {393, 99}

\bibitem[\protect\citeauthoryear{{Wiersma}, {Schaye}, {Theuns}, {Dalla Vecchia}
   \& {Tornatore}}{{Wiersma} et~al.}{2009b}]{Wiersma2009enrichment}
{Wiersma} R. P.~C.,  {Schaye} J.,  {Theuns} T.,  {Dalla Vecchia} C.,
  {Tornatore} L.,  2009b, \mn@doi [\mnras] {10.1111/j.1365-2966.2009.15331.x},
  \href {https://ui.adsabs.harvard.edu/abs/2009MNRAS.399..574W} {399, 574}

\bibitem[\protect\citeauthoryear{{Willis}, {Schaller}, {Gonnet}, {Bower}  \&
  {Draper}}{{Willis} et~al.}{2018}]{Willis2018}
{Willis} J.~S.,  {Schaller} M.,  {Gonnet} P.,  {Bower} R.~G.,   {Draper} P.~W.,
   2018, in Parallel Computing is Everywhere. {IOS} Press, pp 507 -- 516
  (\mn@eprint {arXiv} {1804.06231}), \mn@doi{10.3233/978-1-61499-843-3-507}

\bibitem[\protect\citeauthoryear{Willis, Schaller, Gonnet  \& Helly}{Willis
  et~al.}{2020}]{Willis2020}
Willis J.~S.,  Schaller M.,  Gonnet P.,   Helly J.~C.,  2020, in Parallel
  Computing: Technology Trends. {IOS} Press, pp 263 -- 274 (\mn@eprint {arXiv}
  {2003.11468}), \mn@doi{10.3233/apc200050}

\bibitem[\protect\citeauthoryear{{Wright}}{{Wright}}{2006}]{Wright2006}
{Wright} E.~L.,  2006, \mn@doi [\pasp] {10.1086/510102}, \href
  {http://adsabs.harvard.edu/abs/2006PASP..118.1711W} {118, 1711}

\bibitem[\protect\citeauthoryear{{Xu}}{{Xu}}{1995}]{Xu1995}
{Xu} G.,  1995, \mn@doi [\apjs] {10.1086/192166}, \href
  {https://ui.adsabs.harvard.edu/abs/1995ApJS...98..355X} {98, 355}

\bibitem[\protect\citeauthoryear{{Zennaro}, {Bel}, {Villaescusa-Navarro},
  {Carbone}, {Sefusatti}  \& {Guzzo}}{{Zennaro} et~al.}{2017}]{Zennaro2016}
{Zennaro} M.,  {Bel} J.,  {Villaescusa-Navarro} F.,  {Carbone} C.,  {Sefusatti}
  E.,   {Guzzo} L.,  2017, \mn@doi [\mnras] {10.1093/mnras/stw3340}, \href
  {https://ui.adsabs.harvard.edu/abs/2017MNRAS.466.3244Z} {466, 3244}

\makeatother
\end{thebibliography}

\section*{Author Affiliations}
\noindent
{\it \small
$^{1}$Lorentz Institute for Theoretical Physics, Leiden University, PO Box 9506, NL-2300 RA Leiden, The Netherlands\\
$^{2}$Leiden Observatory, Leiden University, PO Box 9513, NL-2300 RA Leiden, The Netherlands\\
$^{3}$Department of Physics and Astronomy, University of Pennsylvania, 209 South 33rd Street, Philadelphia, PA, USA 19104\\
$^{4}$Department of Physics and Kavli Institute for Astrophysics and Space Research, Massachusetts Institute of Technology, Cambridge, MA 02139, USA\\
$^{5}$Institute for Computational Cosmology, Department of Physics, Durham University, South Road, Durham DH1 3LE, UK\\
$^{6}$Laboratoire d'astrophysique, \'{E}cole Polytechnique F\'{e}d\'{e}rale de Lausanne (EPFL), 1290 Sauverny, Switzerland\\
$^{7}$Observatoire de Gen\`{e}ve, Universit\'{e} de Gen\`{e}ve, Chemin Pegasi 51, 1290 Versoix, Switzerland\\
$^{8}$Department of Computer Science, Durham University, Upper Mountjoy Campus, Stockton Road,  Durham\\
$^{9}$Department of Physics, University of Helsinki, Gustaf H\"{a}llstr\"{o}min katu 2, FI-00014 Helsinki, Finland\\
$^{10}$The Oskar Klein Centre, Department of Physics, Stockholm University, Albanova University Center, 106 91 Stockholm, Sweden\\
$^{11}$Sterrenkundig Observatorium, Universiteit Gent, Krijgslaan 281, B-9000 Gent, Belgium\\
$^{12}$STFC Hartree Centre, Sci-Tech Daresbury, Warrington, WA4 4AD, UK\\
$^{13}$Department of Physics, The Chinese University of Hong Kong, Shatin, Hong Kong, China\\
$^{14}$Department of Astronomy and Astrophysics, The University of Chicago, Chicago, IL60637, USA\\
$^{15}$Universit\'{e} Paris-Saclay, Universit\'{e} Paris Cit\'{e}, CEA, CNRS, AIM, 91191, Gif-sur-Yvette, France\\
$^{16}$GRAPPA Institute, University of Amsterdam, Science Park 904, 1098 XH Amsterdam, The Netherland\\
$^{17}$Google AI Perception, Google Switzerland, CH-8002 Zurich, Switzerland\\
$^{18}$ITS High Performance Computing, Eidgen\"{o}ssische Technische Hochschule Z\"{u}rich, 8092 Z\"{u}rich, Switzerland\\
$^{19}$NASA Ames Research Center, Moffett Field, CA 94035, USA\\
$^{20}$Department of Astrophysics, University of Vienna, T\"{u}rkenschanzstrasse 17, 1180 Vienna, Austria\\
$^{21}$Astronomy Centre, University of Sussex, Falmer, Brighton BN1 9QH, UK\\
$^{22}$SciNet HPC Consortium, University of Toronto, Toronto, Ontario, Canada\\
$^{23}$Space Research and Planetary Sciences, Physikalisches Institut, University of Bern, Bern, Switzerland\\
$^{24}$Institute for Astronomy, University of Edinburgh, Royal Observatory, Blackford Hill, Edinburgh EH9 3HJ, UK
}

\appendix
\section{Additional SPH schemes}
\label{appendix:SPH}
For completeness, we summarise here the equations of motion for the
the additional modern SPH schemes present in \swift. These are
re-implementation of schemes from the literature and can be used to
perform comparisons between models in a framework where all the rest
of the solver's infrastructure is kept exactly fixed.

\subsection{Pressure-smoothed SPH}
\label{ssec:sph:psph}

Pressure-smoothed SPH solves the same generic equation of motion as described in
eq. \ref{eqn:genericeom}, but with a different choice of fundamental variables
$a$ and $b$. In general, instead of smoothing the density $\hat{\rho}$, we
introduce a smoothed pressure $\hat{P}$ which is generated through loops over
neighbours (as described below).  This approach is commonplace in astrophysics,
with it described and used in \citet{Saitoh2013}, \citet{Hopkins2013}, and
\citet{Hu2014}, amongst many others.\\

\noindent For the two choices of thermodynamic variable, internal energy (per
unit mass) $u$, or entropy $A$, we generate two different (but equivalent)
smoothed pressures,
\begin{align}
    \hat{P}_i =& (\gamma - 1) \sum_j m_j u_j W_{ij}, \\
    \hat{P}_i =& \left[\sum_j m_j A_j^{1/\gamma} W_{ij} \right]^\gamma,
\end{align}
respectively. As described by \citet{Borrow2021}, this then leads to issues
integrating the pressure in simulations with multiple time-stepping, especially
in scenarios where there is a high $\dot{u}$ (for instance in the presence of a
strong cooling term in the sub-grid physics), as we should use
\begin{equation}
    \frac{\mathrm{d}\hat{P}_i}{\mathrm{d}t} =
    (\gamma - 1)\sum_j m_j \left(W_{ij}\frac{\mathrm{d}u_j}{\mathrm{d}t} 
                                 + u_j \mathbf{v}_{ij} \cdot \nabla_j W_{ij}\right)
\end{equation}
for the evolution of $\hat{P}_i$, which would formally require an extra loop
over the neighbours. As such, we do not recommend these schemes for practical
use, but we implement them in \swift for cross-compatibility with the original
\gadget-based \eagle code.

The changes in the smoothed variable give rise to a different equation of motion,
\begin{align}
    \frac{\mathrm{d}\mathbf{v}_i}{\mathrm{d}t} = - u_i (\gamma -1)^2 \sum_j
       m_j u_j  \left[
          \frac{f_{ij}}{\hat{P}_i} \nabla_i W_{ij} + \frac{f_{ji}}{\hat{P}_j} \nabla_j W_{ji}
       \right],
\end{align}
shown for the internal energy variant (Pressure--Energy) only for
brevity\footnote{Expanded derivations and definitions are available in the
theory documentation provided with the \swift code.}. The factors $f_{ij}$ read
\begin{equation}
    f_{ij} = 1 - \frac{1}{m_j u_j} \left[\frac{\partial \hat{P}_i}{\partial h_i} \frac{h_i}{(\gamma - 1) n_{\rm d} \hat{n}_i}\right]\left[1 + \frac{h_i}{n_{\rm d} \hat{n}_i}\frac{\partial \hat{n}_i}{\partial h_i}\right]^{-1}
\end{equation}
As, in practice, we do not make an additional loop over neighbours to calculate
the derivative in the smoothed pressure, we use a simple chain rule,
\begin{align}
    \frac{\mathrm{d}\hat{P}_i}{\mathrm{d}t} = \rho_i \frac{\mathrm{d}u_i}{\mathrm{d}t} +
    u_i \frac{\mathrm{d}\rho_i}{\mathrm{d}t},
\end{align}
to integrate the smoothed pressure with time. This is commonplace amongst
Pressure-SPH schemes implemented in real codes, as it is impractical from a
performance perspective to require an additional loop solely for the
reconstruction of the smoothed pressure time differential.

There are base Pressure--Entropy and Pressure--Energy schemes available in
\swift that use the same equations of motion for artificial viscosity as the
Density-based schemes (eq. \ref{eqn:basicartvisc}).

\subsection{{\bfseries \scshape Anarchy}-SPH}
\label{ssec:sph:anarchy}

In addition to these base schemes, we implement `\sphflavour{Anarchy-PU}', which is a
Pressure--Energy-based variant of the original \anarchy scheme used for \eagle
(see \citet{Schaller2015} and Appendix A of \citet{Schaye2015}) which used
entropy as the thermodynamic variable to evolve. We reformulate the base
equations of motions in terms of internal energy in \swift as described in the
previous section.

\sphflavour{Anarchy-PU} uses the same artificial viscosity implementation as \sphenix
(eq.\ref{eq:sphenix_visc1}-\ref{eqn:av_ij}) but uses a slightly different
value of decay length $\ell = 0.25$.

The artificial conduction differs more markedly. The base equation
(eq. \ref{eqn:art_cond} and \ref{eqn:artconddt}) remain unchanged w.r.t
\sphenix but three of the ingredients are altered. Firstly, \sphflavour{Anarchy-PU} does not
pressure-weight the contributions of both interacting particles and thus
\begin{equation}
    \alpha_{ij} = \frac{\alpha_{{\rm c}, i} + \alpha_{{\rm c}, j}}{2}.
\end{equation}
Secondly, the conduction velocity is changed to
\begin{equation}
    v_{{\rm c}, ij} = c_{{\rm s},i} + v_{{\rm c},j} + \mu_{ij},
\end{equation}
which is similar to the signal velocity entering viscosity but with the sign of
$\mu$ reversed. Thirdly, the dimensionless constant $\beta_{\rm c}$ entering the
time evolution of the conduction parameter (eq. \ref{eqn:artconddt}) is lowered
to $\beta_{\rm c}=0.01$. This is because \sphflavour{Anarchy-PU} uses a smoothed-pressure
implementation and thus a lower amount of conduction is required.

Finally, the conduction limiter in strong shocks
(eq. \ref{eqn:condshocklimiter}) is not used. Our implementation is consistent
with the original \anarchy scheme.

\subsection{{\bfseries \scshape Phantom}-like flavour}
\label{ssec:sph:phantom}

\swift includes a reduced, and slightly modified, version of the \phantomSPH SPH
scheme, from \citep{Price2018}. It employs the same Density--Energy SPH scheme
as \sphenix, and also implements variable artificial conduction and viscosity
parameters. At present, our implementation in \swift is hydrodynamics only, but
an extension to include magnetohydrodynamical effects is planned for the future.

Our \phantomSPH artificial viscosity implementation is the same as \sphenix and
\anarchy, with $\ell = 0.25$. This differs slightly from the original
\phantomSPH description, where a modified version of the \citet{Balsara1989}
switch is also used. For artificial conduction, a fixed $\alpha_{\rm c}=1$ is
used for all particles, effectively removing the need for
eq.~\ref{eqn:artconddt}.  The conduction speed is given as
\begin{align}
    v_{{\rm c}, i} = \sqrt{2\frac{|P_i - P_j|}{\hat{\rho}_i + \hat{\rho}_j}},
\end{align}
with the \phantomSPH implementation only designed for use with purely
hydrodynamical simulations. \cite{Price2018} recommend a different conduction
speed in simulations involving self-gravity.

\subsection{{\bfseries \scshape Gasoline-2}-like (GDF-like) flavour}
\label{ssec:sph:gasoline}

\swift also includes a re-implementation of the equations of the \gasoline-2
model presented by \citet{Wadsley2017}. The implementation and default
parameters follow the paper closely, though there are minor differences. We give
the equations here for completeness but refer the reader to the original
\citet{Wadsley2017} work for the motivation behind their derivation.

\noindent The equation of motion in Gasoline uses the so-called `Geometric
Density Force' (GDF) formulation, and is as follows:
\begin{align}
    \frac{\mathrm{d} \mathbf{v}_{i}}{\mathrm{d} t}&=-\sum_{j} m_{j}\left(\frac{P_{i}+P_{j}}{\hat{\rho}_{i} \hat{\rho}_{j}}\right) \nabla_{i} \bar{W}_{i j}, \\
    \frac{\mathrm{d} u_{i}}{\mathrm{d} t}&=\sum_{j} m_{j}\left(\frac{P_{i}}{\hat{\rho}_{i} \hat{\rho}_{j}} \right) \mathbf{v}_{i j} \cdot \nabla_{i} \bar{W}_{i j},
\end{align}
where
\begin{equation}
    \nabla_{i} \bar{W}_{i j}=\frac{1}{2} f_{i} \nabla_{i} W\left(r_{i j}, h_{i}\right)+\frac{1}{2} f_{j} \nabla_{j} W\left(r_{i j}, h_{j}\right),
\end{equation}  
is the symmetric average of both usual kernel contributions, and the variable
smoothing length correction terms read:
\begin{equation}
    f_{i}=\frac{\sum_{j} \frac{m_{j}}{\hat{\rho}_{i}} \mathbf{r}_{i j}^{2} W_{ij}}{\sum_{j} \frac{m_{j}}{\hat{\rho}_{j}} \mathbf{r}_{i j}^{2} W_{ij}}.
\end{equation}
~\\

\noindent The artificial viscosity and conduction implementations use matrix
calculations based on local pressure gradients. Here,
\begin{align}
    \nabla P_i=&(\gamma-1) \sum_{j} m_{j} u_{j} \nabla_{i} W_{ij}, \\
\mathbf{n}_i=&\frac{\nabla P_i}{|\nabla P_i \, |}, \\
\frac{\mathrm{d} v_i}{\mathrm{d} n_i}=&\sum_{\alpha, \beta} \mathbf{n}_{i, \alpha} \mathbf{V}_{\alpha \beta, i} \mathbf{n}_{ i, \beta},
\end{align}
with the velocity gradient tensor
\begin{align}
    \mathbf{V}_{\alpha \beta, i}=\frac{\sum_{j}\left(\mathbf{v}_{\alpha i}-
    \mathbf{v}_{\alpha j}\right)\left(\mathbf{r}_{\beta i}-\mathbf{r}_{\beta j}\right) m_{j} W_{ij}}{\frac{1}{3} \sum_{j} \mathbf{r}_{i j}^{2} m_{j} W_{ij}},
\end{align}
and the shock detector
\begin{align}
    D_i=\frac{3}{2}\left[\frac{\mathrm{d} v_i}{\mathrm{d} n_i}+\max \left(-\frac{1}{3} \nabla \cdot \mathbf{v}_i, 0\right)\right]
\end{align}
with $\alpha$ and $\beta$ indices along the Cartesian axes in our case. These
give rise to the evolution equation for the artificial viscosity parameter,
which is evolved in a similar manner to \anarchy, \sphenix, and \phantomSPH:
\begin{align}
    \alpha_{\mathrm{V, loc}, i} &= \alpha_{\rm{V, max}} \frac{A_{i}}{A_{i}+v_{\mathrm{sig}, i}^{2}} \\
A_{i} &= 2 h_{i}^{2} B_{i} \max \left(-\frac{\mathrm{d} D_i}{\mathrm{~d} t}, 0\right) \\
\frac{\mathrm{d} \alpha_{i}}{\mathrm{d} t} &= 0.2 c_{{\rm s}, i} \left(\alpha_{\mathrm{V, loc}, i}-\alpha_{{\rm V}, i}\right) / h_{i}.
\end{align}
We note that the \swift implementation again uses the \citet{Balsara1989} switch
(the $B_i$ term), rather than the \citet{Cullen2010} style limiter used in the
original \gasoline-2 paper. \\

\noindent Artificial conduction is implemented using the trace-free shear tensor,
\begin{align}
    \mathbf{S}^2_{\alpha, \beta, i} = \frac{\mathbf{V}_{\alpha, \beta, i} + \mathbf{V}_{\beta, \alpha,i}}{2} - \frac{\delta_{\alpha, \beta} \nabla \cdot \mathbf{v}_i}{3},
\end{align}
and the conduction parameter: 
\begin{align}
    \alpha_{{\rm c},i} &= C |\mathbf{S}| h_i^2, \\
|\mathbf{S}| &= \sum_{\alpha, \beta} \mathbf{S}^2_{\alpha, \beta},
\end{align}
with the fixed parameter $C=0.03$. Note that unlike the other
schemes $\alpha_{{\rm c},i}$ is not dimensionless. These then get
added to the equation of motion for thermal energy using
\begin{align}
    \frac{\mathrm{d} u_{i}}{\mathrm{d} t}=-\sum_{j} m_{j} \frac{\left(\alpha_{{\rm c}, i} + \alpha_{{\rm c}, j}\right)\left(u_{i}-u_{j}\right)\left(\mathbf{r}_{i j} \cdot \nabla_{i} \bar{W}_{i j}\right)}{\frac{1}{2}\left(\rho_{i}+\rho_{j}\right) \mathbf{r}_{i j}^{2}},
\end{align}
which is very similar to the other schemes presented above.

\section{Multi-index notation}
\label{sec:multi_index_notation}

Following \cite{Dehnen2014}, we define a multi-index $\mathbf{n}$ as a triplet of 
non-negative integers:
\begin{equation}
  \mathbf{n} \equiv \left(n_x, n_y, n_z\right), \qquad n_i \in \mathbb{N},
\end{equation}
with a norm $n$ given by
\begin{equation}
  n = |\mathbf{n}| \equiv n_x + n_y + n_z. 
\end{equation}
We also define the exponentiation of a vector
$\mathbf{r}=(r_x,r_y,r_z)$ by a multi-index $\mathbf{n}$ as
\begin{equation}
  \mathbf{r}^\mathbf{n} \equiv r_x^{n_x} \cdot r_y^{n_y} \cdot r_z^{n_z},
\end{equation}
which for a scalar $\alpha$ reduces to
\begin{equation}
  \alpha^\mathbf{n} = \alpha^{n}.
\end{equation}
Finally, the factorial of a multi-index is defined to be
\begin{equation}
  \mathbf{n}! \equiv n_x! \cdot n_y! \cdot n_z!,
\end{equation}
which leads to a simple expression for the binomial coefficients of
two multi-indices entering Taylor expansions:
\begin{equation}
  \binom{\mathbf{n}}{\mathbf{k}} = \binom{n_x}{k_x}\binom{n_y}{k_y}\binom{n_z}{k_z}.
\end{equation}
When appearing as the index in a sum, a multi-index represents all
values that the triplet can take up to a given norm. For instance,
$\sum_{\mathbf{n}}^{p}$ indicates that the sum runs over all possible
multi-indices whose norm is $\leq p$.

\label{lastpage}

\end{document}